\documentclass[a4paper,fleqn,usenatbib]{mnras}

\usepackage[T1]{fontenc}
\usepackage{ae,aecompl}

\usepackage{soul}
\usepackage{xcolor}

\usepackage{graphicx}	
\usepackage{amsmath}	
\usepackage{amssymb}	
\usepackage{lscape} 
\usepackage{subfigure}
\usepackage{caption}
\usepackage{multirow}
\usepackage{float}
\usepackage{lineno}

\makeatletter
\newcommand{\labtext}[2]{%
  \@bsphack
  \csname phantomsection\endcsname 
  \def\@currentlabel{#1}{\label{#2}}%
  \@esphack
}
\makeatother

\newcommand\explvar{$\sigma^2_{exp}$}
\usepackage{newtxtext,newtxmath}

\title[Optical spectral variability of $\gamma$-ray blazars]{A statistical study of the optical spectral variability in gamma-ray blazars}
 
\author[J. Otero-Santos et al.]{
J. Otero-Santos,$^{1,2}$\thanks{E-mail: joteros@iac.es (JOS)}
J. A. Acosta-Pulido,$^{1,2}$
J. Becerra Gonz\'alez,$^{1,2}$
A. Luashvili,$^{3}$ 
\newauthor
N. Castro Segura,$^{2,4}$
O. González-Martín,$^{5}$ 
C. M. Raiteri$^{6}$ and
M. I. Carnerero$^{6}$
\\
$^{1}$Instituto de Astrof\'isica de Canarias (IAC), E-38200 La Laguna, Tenerife, Spain\\
$^{2}$Universidad de La Laguna (ULL), Departamento de Astrof\'isica, E-38206 La Laguna, Tenerife, Spain\\
$^{3}$Laboratoire Univers et Théories, Observatoire de Paris, Unversité PSL, CNRS, Université de Paris, 92190 Meudon, France\\
$^{4}$School of Physics \& Astronomy, University of Southampton, Highfield, Southampton SO17 1BJ, UK\\
$^{5}$Instituto de Radioastronomía y Astrofísica (IRyA-UNAM), 3-72 (Xangari), 8701, Morelia, Mexico\\
$^{6}$INAF-Osservatorio Astrofisico di Torino, via Osservatorio 20, 10025 Pino Torinese, Italy\\
}

\date{Accepted XXX. Received YYY; in original form ZZZ}

\pubyear{2020}

\begin{document}
\label{firstpage}
\pagerange{\pageref{firstpage}--\pageref{lastpage}}
\maketitle

\begin{abstract}

Blazars optical emission is generally dominated by relativistic jets, although the host galaxy, accretion disk and broad line region (BLR) may also contribute significantly. Disentangling their contributions has been challenging for years due to the dominance of the jet. To quantify the contributions to the spectral variability, we use the statistical technique for dimensionality reduction Non-Negative Matrix Factorization on a spectroscopic data set of 26 $\gamma$-ray blazars. This technique allows to model large numbers of spectra in terms of a reduced number of components. We use \textit{a priori} knowledge to obtain components associated to meaningful physical processes. The sources are classified according to their optical spectrum as host-galaxy dominated BL Lac objects (BL Lacs), BL Lacs, or Flat Spectrum Radio Quasars (FSRQs). Host-galaxy sources show less variability, as expected, and bluer-when-brighter trends, as the other BL Lacs. For FSRQs, more complicated colour-flux behaviours are observed: redder-when-brighter for low states saturating above a certain level and, in some cases, turning to bluer-when-brighter. We are able to reproduce the variability observed during 10 years using only 2 to 4 components, depending on the type. The simplest scenario corresponds to host-galaxy blazars, whose spectra are reconstructed using the stellar population and a power law for the jet. BL Lac spectra are reproduced using from 2 to 4 power laws. Different components can be associated to acceleration/cooling processes taking place in the jet. The reconstruction of FSRQs also incorporates a QSO-like component to account for the BLR, plus a very steep power law, associated to the accretion disk. 
\end{abstract}

\begin{keywords}
BL Lacertae objects: general --  galaxies: active -- galaxies: nuclei -- galaxies: jets -- methods: statistical 
\end{keywords}





\section{Introduction}
Blazars are a subclass of Active Galactic Nuclei (AGNs) whose relativistic jets are closely aligned with the line of vision, characterized by a strong flux variability. Their emission can extend from radio wavelengths up to very-high-energy (VHE) $\gamma$ rays with energies of a few TeVs. The spectral energy distribution (SED) of these sources shows a double bump structure \citep[e.g.][]{abdo2010}. The first bump is caused by the synchrotron emission of relativistic electrons coming from the jet \citep{ghisellini2010}. The second one is commonly explained with a leptonic scenario via Inverse Compton (IC) scattering of the synchrotron photons of the same electron population \citep[synchrotron self Compton, SSC,][]{maraschi1992}, and/or IC of seed photons from outside the jet \citep[external Compton, EC, see e.g.][]{dermer1994}. Alternatively, models based on hadronic processes have been proposed to explain and reproduce the high-energy bump of these objects \citep[e.g.][]{cerruti2015}. Blazars can be subdivided in two groups \citep{urry1995}: BL Lac objects and Flat Spectrum Radio Quasars (FSRQs). Both classes can be distinguished by their optical spectra. BL Lacs show an optical spectrum with (almost) no features, whereas FSRQs have narrow and broad lines in their spectrum with a rest-frame equivalent width EW~$>5$~\AA \ \citep[e.g.][]{stickel1991,stocke1991}. Moreover, they also differ in the frequency $\nu$ of the SED peaks and their intensity ratio \citep[Compton dominance,][]{finke2013}. Furthermore, they can also be classified according to the frequency of the synchrotron peak of the SED as low-, intermediate-, and high-synchrotron peaked blazars (LSPs, ISPs and HSPs, respectively). LSPs display a synchrotron peak located at $\nu_{sync}<10^{14}$ Hz. ISPs show their synchrotron peak between frequencies of $10^{14} \ \text{Hz} < \nu_{sync}<10^{15}$ Hz. Finally, HSPs have a synchrotron peak frequency $\nu_{sync}>10^{15}$ Hz.

Variability in these sources has been detected in a wide range of time scales \citep[see e.g.][]{fan2005}: in scales as short as a few minutes or several hours \citep[intraday variability, for instance][]{vovk2015}, longer time scale variations on the order of days to weeks \citep[short-term variability, e.g.][]{abdo2010b}, and variations in scales of months to years \citep[long-term variability, e.g.][]{sandrinelli2014}. Variability has a crucial importance in the understanding of these violent sources, since it has been interpreted within several scenarios, depending on the nature and time scale of the variation. For instance, intraday variability has been explained in terms of flare-like events or phenomena such as the lighthouse effect presented by \cite{camenzind1992}. Another interpretation involves the combination of chromatic variations linked to substructures in the jet, shocks propagating in turbulent plasma, and stochastic processes related to the injection and acceleration of particles, and/or changes of the Doppler factor most likely due to variations of the viewing angle \citep{raiteri2021a,raiteri2021b}. Moreover, short-term variability may be caused by instabilities or variations in the accretion rate \citep{mangalam1993}. Long-term variations in blazars have been interpreted with several models, e.g. a binary black hole system with a precession period of the secondary black hole around the primary \citep{lehto1996}, helical jets \citep{raiteri2010}, shocks \citep{marscher1985} or changes in the viewing angle \citep{raiteri2017}. These long-term variations have been detected in the past showing periodic patterns \citep[see for instance][]{ackermann2015,fan2018a,sandrinelli2018,covino2019,otero2020,penil2020,zhang2020}. These flux variations are frequently associated with spectral variability of the source \citep[see e.g.][]{bonning2012,zhang2015,isler2017,li2018}.

The synchrotron emission coming from the relativistic jets is the main contribution to their optical emission \citep{blandford1978}. Nevertheless, other regions (e.g. host galaxy, broad line region (BLR) in FSRQs or accretion disk) may contribute significantly to the emission \citep[e.g.][]{massaro2015,raiteri2019}. In such cases, all these components must be taken into account when studying the variability of the source. However, disentangling these contributions to the emission of blazars has been challenging due to the high dominance of the emission from the jet.

Understanding the origin of this variability in blazars is crucial to comprehend the processes taking place in the most extreme Universe. Several previous studies have tried to explain and reproduce the variability in AGNs taking into account the different contributions. For this purpose, statistical tools for dimensionality reduction have been used, such as the Principal Component Analysis \citep[PCA, see][]{pearson1901,hotelling1933,ivezic2014} or the Non-negative Matrix Factorization \citep[NMF, see][]{paatero1994,ivezic2014}. These techniques allow the reconstruction of a large data set by making use of a reduced number of components. They have been applied before to other types of AGNs, in which the jet does not outshine the other contributions, due to a larger viewing angle \citep{hurley2014, parker2015, zhu2016, gallant2018}. However, the powerful jets of blazars make the application of such technique challenging. Nevertheless, understanding the impact of other physical regions such as the accretion disk or the BLR to the global variability observed in blazars is a matter of great importance to comprehend the processes taking place in the environment of these extreme sources. For this purpose, the combination of statistical tools such as the NMF together with some physical assumptions based on our knowledge of the different components, can help to shed light to the overall emission of blazars.



This work is focused on the characterization of the overall optical variability of a blazar sample based on the minimum number of spectral components. The global variability of the sources and the contribution of each component to the overall flux variation are studied. For this purpose, the NMF is used to decompose the optical spectra of each source and identify the different components contributing to the emission. A first approach of applying the NMF to model and reproduce the variability observed in blazars was performed in the past by \cite{acosta2017}. For this analysis, we make use of the observations performed by the Steward Observatory in Arizona, as a support program for the \textit{Fermi} satellite observations of bright $\gamma$-ray blazars. Photometric and spectropolarimetric data of a large sample of blazars are available, with a time coverage of $\sim$10 years. The paper is structured as follows: in Sections \ref{sec2} and \ref{sec3} we introduce the data set used in the analysis, the data reduction and the variability observed in the data sample; in Section \ref{sec4} the analysis tools and the methodology followed for this work are presented; in Sections \ref{sec5} and \ref{sec6} we present the results of the analysis and the physical interpretation and implications of these results. The conclusions and final remarks of this work are summarized in Section \ref{sec7}. Finally, a detailed discussion on some sources of interest is included in the Appendix \ref{appendix}.

\section{Observations and data reduction}\label{sec2}
We have analysed the optical spectral data taken by the Steward Observatory program in support of the \textit{Fermi}-LAT $\gamma$-ray observations\footnote{\url{http://james.as.arizona.edu/~psmith/Fermi/}} of a sample of blazars. The Steward Observatory has monitored a sample of 80 blazars from October 2008 until July 2018 \citep{smith2009}. We have selected those sources with more than 15 spectra available. This selection criteria led us to a final data sample of 26 sources, listed in Table \ref{table_observations}. A total of 11 sources have been classified as BL Lac objects and 12 are FSRQs. The 3 remaining ones are BL Lac objects whose optical spectra are dominated by the stellar emission from the host galaxy (hereafter, galaxy-dominated blazars). This classification was made based on the EW of the emission lines observed in the mean spectrum. In Table \ref{table_observations} we also include the classification according to the frequency of the synchrotron peak in the SED.

\begin{table}
\begin{center}
\caption{Sample of blazars from the Steward Observatory monitoring program included in this work.}
\begin{tabular}{ccccc} \hline
Source & Sp. Type & $z$ & SED Type$^{*}$ & $\nu_{sync}$$^{*}$  \\ \hline \hline
3C 66A & BL Lac & 0.444$^{**}$ & ISP & 7.1$\cdot$10$^{14}$ \\ \hline
AO 0235+164 & BL Lac & 0.940 & LSP & 2.5$\cdot$10$^{14}$ \\ \hline
S5 0716+714 & BL Lac & 0.30$^{***}$ & ISP & 1.5$\cdot$10$^{14}$ \\ \hline
PKS 0735+178 & BL Lac & 0.424 & LSP & 2.7$\cdot$10$^{13}$ \\ \hline
OJ 287 & BL Lac & 0.306 & LSP & 1.8$\cdot$10$^{13}$  \\ \hline
Mkn 421 & BL Lac & 0.031 & HSP & 1.7$\cdot$10$^{16}$ \\ \hline
W Comae & BL Lac & 0.103 & ISP & 4.5$\cdot$10$^{14}$  \\ \hline
H 1219+305 & BL Lac & 0.184 & HSP & 1.9$\cdot$10$^{16}$ \\ \hline
1ES 1959+650 & BL Lac & 0.047 & HSP & 9.0$\cdot$10$^{15}$ \\ \hline
PKS 2155-304 & BL Lac & 0.116 & HSP & 5.7$\cdot$10$^{15}$ \\ \hline
BL Lacertae & BL Lac & 0.069 & LSP & 3.9$\cdot$10$^{13}$ \\ \hline \hline
PKS 0420-014 & FSRQ &  0.916 & LSP & 5.9$\cdot$10$^{12}$ \\ \hline
PKS 0736+017 & FSRQ &  0.189 & LSP & 1.5$\cdot$10$^{13}$ \\ \hline
OJ 248 & FSRQ & 0.941 & LSP & 3.4$\cdot$10$^{12}$ \\ \hline
Ton 599 & FSRQ & 0.725 & LSP & 1.2$\cdot$10$^{13}$ \\ \hline
PKS 1222+216 & FSRQ &  0.434 & LSP & 2.9$\cdot$10$^{13}$ \\ \hline
3C 273 & FSRQ & 0.158 & LSP & 7.0$\cdot$10$^{13}$ \\ \hline
3C 279 & FSRQ & 0.536 & LSP & 5.2$\cdot$10$^{12}$ \\ \hline
PKS 1510-089 & FSRQ &  0.360 & LSP & 1.1$\cdot$10$^{13}$ \\ \hline
B2 1633+38 & FSRQ & 1.814 & LSP & 3.0$\cdot$10$^{12}$ \\ \hline
3C 345 & FSRQ & 0.593 & LSP & 1.0$\cdot$10$^{13}$ \\ \hline
CTA 102 & FSRQ & 1.032 & LSP & 3.0$\cdot$10$^{12}$ \\ \hline
3C 454.3 & FSRQ & 0.859 & LSP & 1.3$\cdot$10$^{13}$ \\ \hline \hline
H 1426+428 & \begin{tabular}[c]{@{}c@{}}Galaxy\\dominated\end{tabular} & 0.129 & HSP & 1.0$\cdot$10$^{18}$ \\ \hline
Mkn 501 & \begin{tabular}[c]{@{}c@{}}Galaxy\\dominated\end{tabular} & 0.033 & HSP & 2.8$\cdot$10$^{15}$ \\ \hline
1ES 2344+514 & \begin{tabular}[c]{@{}c@{}}Galaxy\\dominated\end{tabular} &  0.044 & HSP & 1.6$\cdot$10$^{16}$ \\ \hline \hline
\end{tabular}
\label{table_observations}
\end{center}
\footnotesize{$^{*}$Extracted from the 4LAC-DR2 catalog of the \textit{Fermi}-LAT satellite \citep{ajello2020}.

$^{**}$Redshift value still under debate. Lower limit of $z \geqslant 0.33$ determined by \cite{torres-zafra2018}. 

$^{***}$Redshift still uncertain. Upper limit of $z < 0.322$ estimated by \cite{danforth2013}.}
\end{table}

The available spectra cover a wavelength range from 4000~\AA \ to 7550~\AA, with typical resolutions between 16-24~\AA, depending on the width of the slit used. They present telluric absorption features due to atmospheric O$_{2}$ at 6980~\AA, and H$_{2}$O atmospheric absorption at 7200-7300 \AA. To avoid introducing systematic effects due to these features, we cut the spectra at 6650~\AA. We have also corrected them by their corresponding redshift in order to perform the analysis in the rest frame of each source. A rebinning of the spectra was performed to correct small shifts in $\lambda$ between different exposures or observations due to small differences in their spectral resolution.
Finally, the spectra have also been dereddened, i.e. corrected of galactic extinction. For this we have made use of the \textit{python} package \texttt{dust\_extinction}\footnote{\url{https://dust-extinction.readthedocs.io/en/stable/}} and the extinction model from \cite{fitzpatrick1999}. A value of $R_{V}=3.1$ was assumed for the extinction law, and the extinction values, $A_{\lambda}$, for each source were extracted from \cite{schlafly2011}, using the tool provided by NED\footnote{The NASA/IPAC Extragalactic Database (NED) is operated by the Jet Propulsion Laboratory, California Institute of Technology under contract with the National Aeronautics and Space Administration.}. 


\section{Variability}\label{sec3}
\begin{figure}
	\includegraphics[width=0.99\columnwidth]{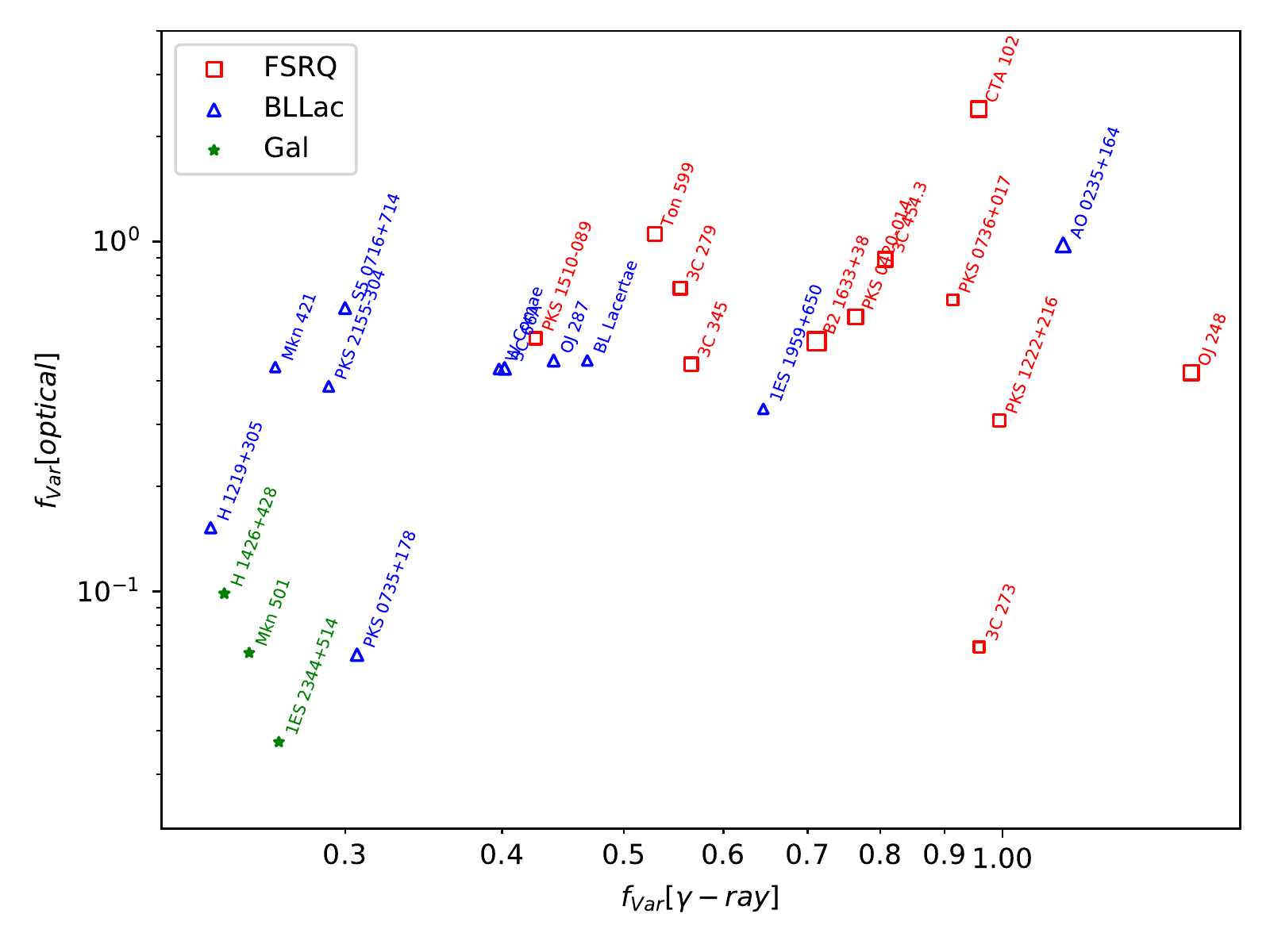}
    \caption{Optical fractional variability vs. HE $\gamma$-ray fractional variability of the blazar sample. Green stars represent those blazars dominated by the stellar emission of the host galaxy, blue triangles correspond to BL Lac objects and red squares to FSRQs. The size of the symbols is proportional to the redshift of the object.}
    \label{fvar_plot}
\end{figure}

One of the key signatures of blazars is their extreme variability at all wavelengths. In order to quantify the intensity of the variability, we computed the optical fractional variability of each object. This quantity is estimated as follows:

\begin{equation}
F_{var}=\sqrt{\frac{S^2-\langle\sigma^2_{err}\rangle}{\langle x \rangle^2}},
\label{fractional_variability_equation}
\end{equation}

\noindent where $S^{2}$ is the variance of the data sample, $\langle \sigma^2_{err} \rangle$ is the mean square error and $\langle x \rangle$ is the mean value of the sample. A detailed discussion on the fractional variability can be found in \cite{vaughan2003}. The estimation of the fractional variability in the optical range for our sample is reported in Table \ref{table_results}. The uncertainties have been estimated using the expression B2 from \cite{vaughan2003}. In all cases except for the galaxy-dominated targets, the $F_{var}$ is much larger than its uncertainty, indicating that the intrinsic variability is well detected. Additionally, Fig. \ref{fvar_plot} shows the optical fractional variability w.r.t. the fractional variability in high-energy (HE) $\gamma$ rays estimated in the 4LAC-DR2 catalog of the Large Area Telescope (LAT) on board the \textit{Fermi} satellite \citep{ajello2020}. A clear difference can be seen between the different AGN types. While galaxy-dominated objects are much less variable in the optical and $\gamma$-ray bands, with $F_{var} \lesssim 0.1$, BL Lacs show a relatively low optical fractional variability, typically ranging from values close to 0.1 up to $\sim$0.5. Finally, FSRQs display the highest values of $F_{var}$, from $\sim$0.5 up to values close to 1.0. A more clear separation is observed for the $F_{var}$ in the $\gamma$-ray band. We note however that the $F_{var}$ in the HE regime can be affected by the frequency of the high-energy peak of the SED. If the peak is located at higher energies, it may happen that the LAT instrument does not have enough sensitivity to detect significant variability (for instance Mkn 501 is highly variable in VHE $\gamma$ rays but its $F_{var}$ in HE $\gamma$ rays is <0.1).

\begin{figure*}
    \includegraphics[width=0.5\textwidth]{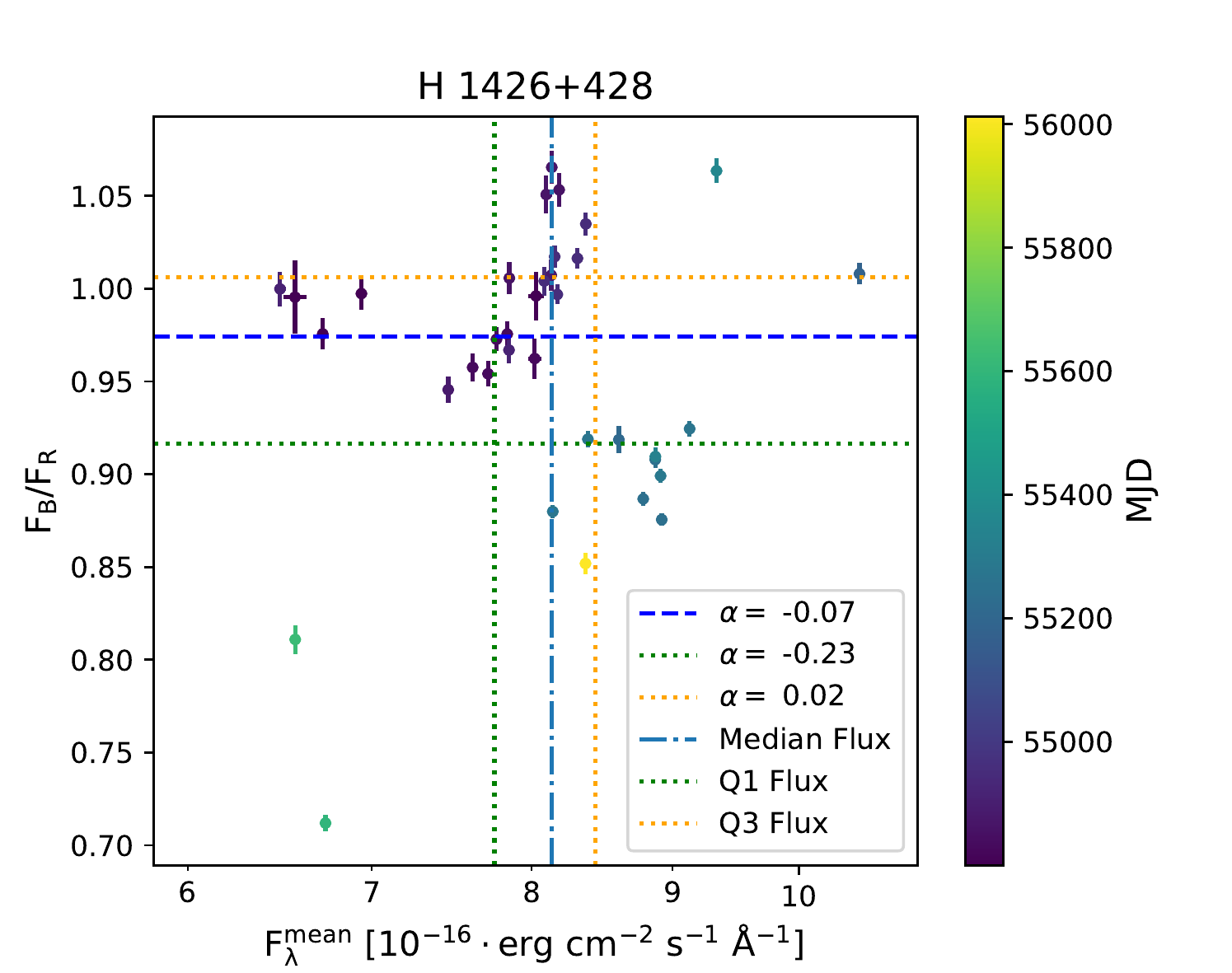}\hfill
    \includegraphics[width=.5\textwidth]{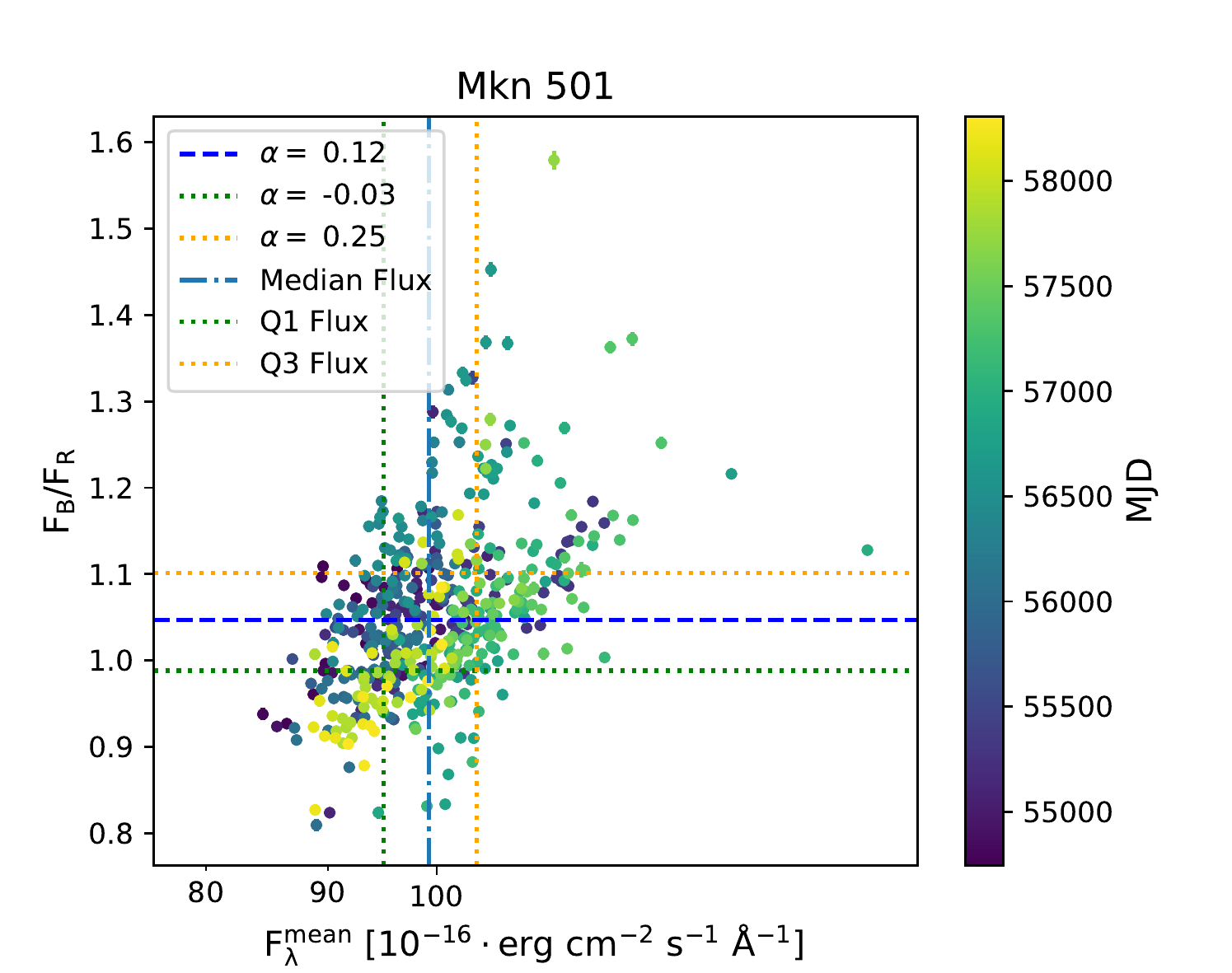}\hfill
    \\[\smallskipamount]
    \begin{center}
    \includegraphics[width=.5\textwidth]{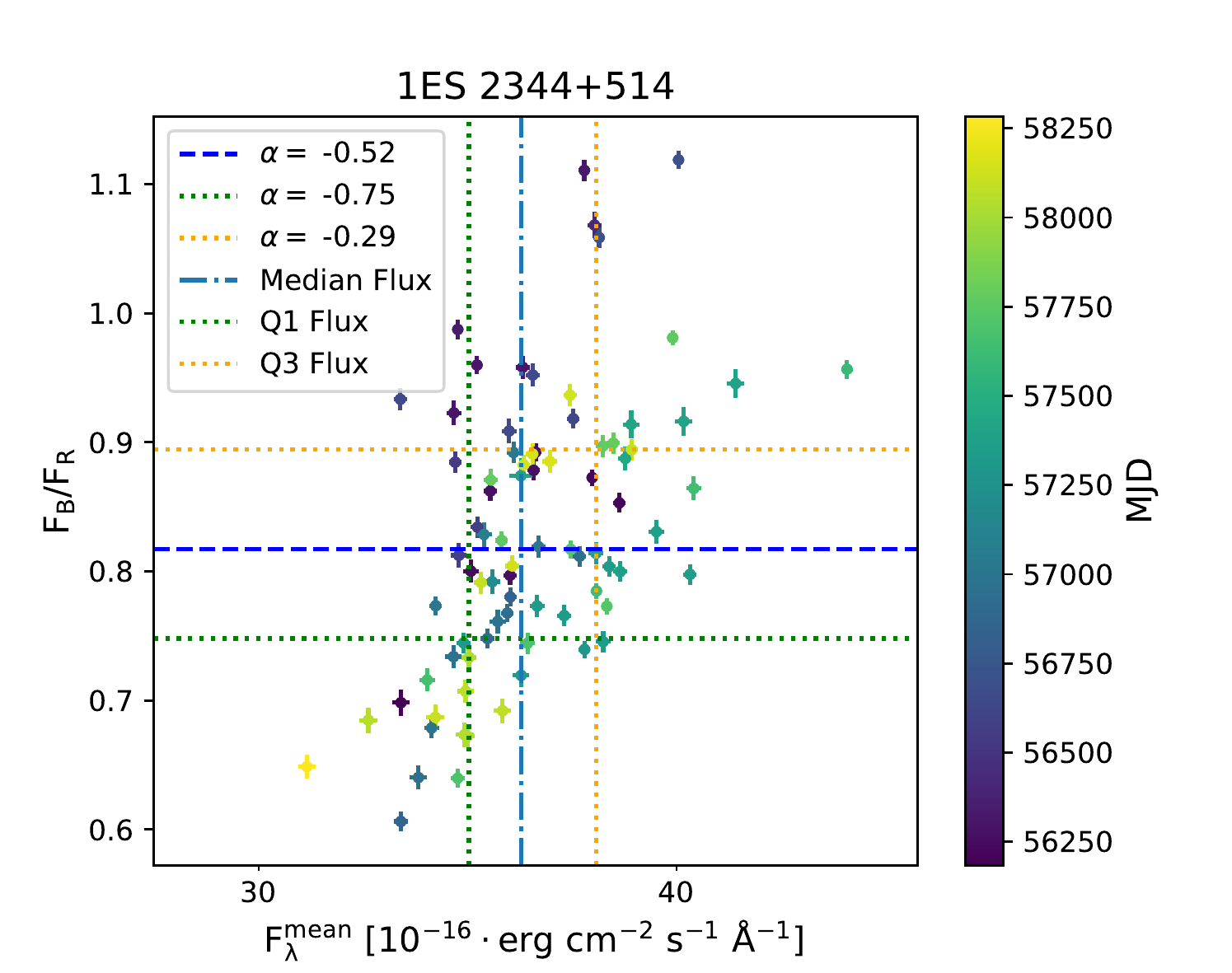}\hfill
    \end{center}
    \caption{Colour-flux diagrams for galaxy-dominated blazars. The colour is estimated as the ratio between the blue and red parts of the spectra. Thus, higher colour values represent bluer spectra. The vertical dashed-dotted line corresponds to the median flux. Vertical dotted lines represent the first and third quartiles for the flux values. The horizontal dashed line denotes the median colour value, and its corresponding spectral index is noted in the legend. Horizontal dotted lines represent the first and third quartiles of the colour and their corresponding spectral index are noted in the legend.}
    \label{variability_galdom}
\end{figure*}

Moreover, this variability comes not only in form of flux variations, but also as spectral variability and changes in different properties of the source (e.g. polarization). The spectral variability is usually associated with changes in the colour of the spectra. Traditionally, different colour evolutions have been associated to the different blazar types \citep[see for instance][]{zhang2015,li2018}: bluer-when-brighter (BWB) for BL Lac objects and redder-when-brighter (RWB) for FSRQs. However, for the latter, more complicated behaviours have been found in the past \cite[e.g.][]{bonning2012,zhang2015,isler2017}. The colour-flux diagrams showing the evolution over time of the colour with respect to the flux for the sources included in this analysis are shown in Figs. \ref{variability_galdom}, \ref{variability_bllacs} and \ref{variability_fsrqs} for galaxy-dominated sources, BL Lac objects and FSRQs, respectively.
We note that for this study, the colour is estimated as the ratio between the blue (4000-4500~\AA) and red (6000-6500~\AA) bands of the spectrum in the observed frame. Thus, higher (lower) colour values represent bluer (redder) spectra. Assuming a power-law (PL) shape as $F_{\lambda}\propto\lambda^{-\alpha}$, we compute the equivalent spectral index as
\begin{equation}
\alpha=-\frac{\log{F_B/F_R}}{\log{\lambda_B/\lambda_R}}\simeq 6\, \log{F_B/F_R}
\end{equation}
With this definition, higher flux ratios (bluer colours) correspond to higher spectral indices (steeper spectra). These diagrams can be of great use to identify different trends or spectral behaviours of the sources, associated to their flux variations. Studying and understanding these behaviours can reveal different acceleration and cooling mechanisms taking place in the jets of these active galaxies, helping to shed light to the origin and the physics involved in their emission.

Figs. \ref{variability_galdom} and \ref{variability_bllacs} reveal that almost all the BL Lac objects of this sample agree with the historical BWB trends observed in this blazar type, as well as the three galaxy-dominated blazars studied here. However, some atypical cases are found in these objects: AO~0235+164 shows different trends, more similar to those found in FSRQs; OJ~287 shows no clear behaviour in the colour except for an achromatic brightening; and 3C~66A displays a subset of spectra deviating from the general BWB trend of the rest of the data sample. These behaviours will be discussed in more detail in Section \ref{sec5}.

FSRQs show the RWB trend mentioned before (see Fig. \ref{variability_fsrqs}). However, more complicated colour behaviours can be seen in this blazar type, with colour flattening and even BWB trends during the highest emission states for some sources. In almost all the FSRQs we observe this RWB trend, followed by a flatter-when-brighter (FWB) behaviour during bright outbursts, once the flux rises over a certain saturation value. The RWB trends in these objects show a large colour change with small flux changes. This is associated with flux changes during low states of these sources, which also imply large changes in their spectral slopes \citep{dai2009}. Moreover, only three sources of the data set (3C~454.3, CTA~102 and PKS~0420-014) display a clear BWB evolution during a high flux state. This behaviour will be discussed in Sections \ref{sec5} and \ref{sec6} in the framework of their more complicated morphology, the different acceleration/cooling processes taking place in the jet, and the relative contributions of the different components.

\begin{figure*}
    \includegraphics[width=0.5\textwidth]{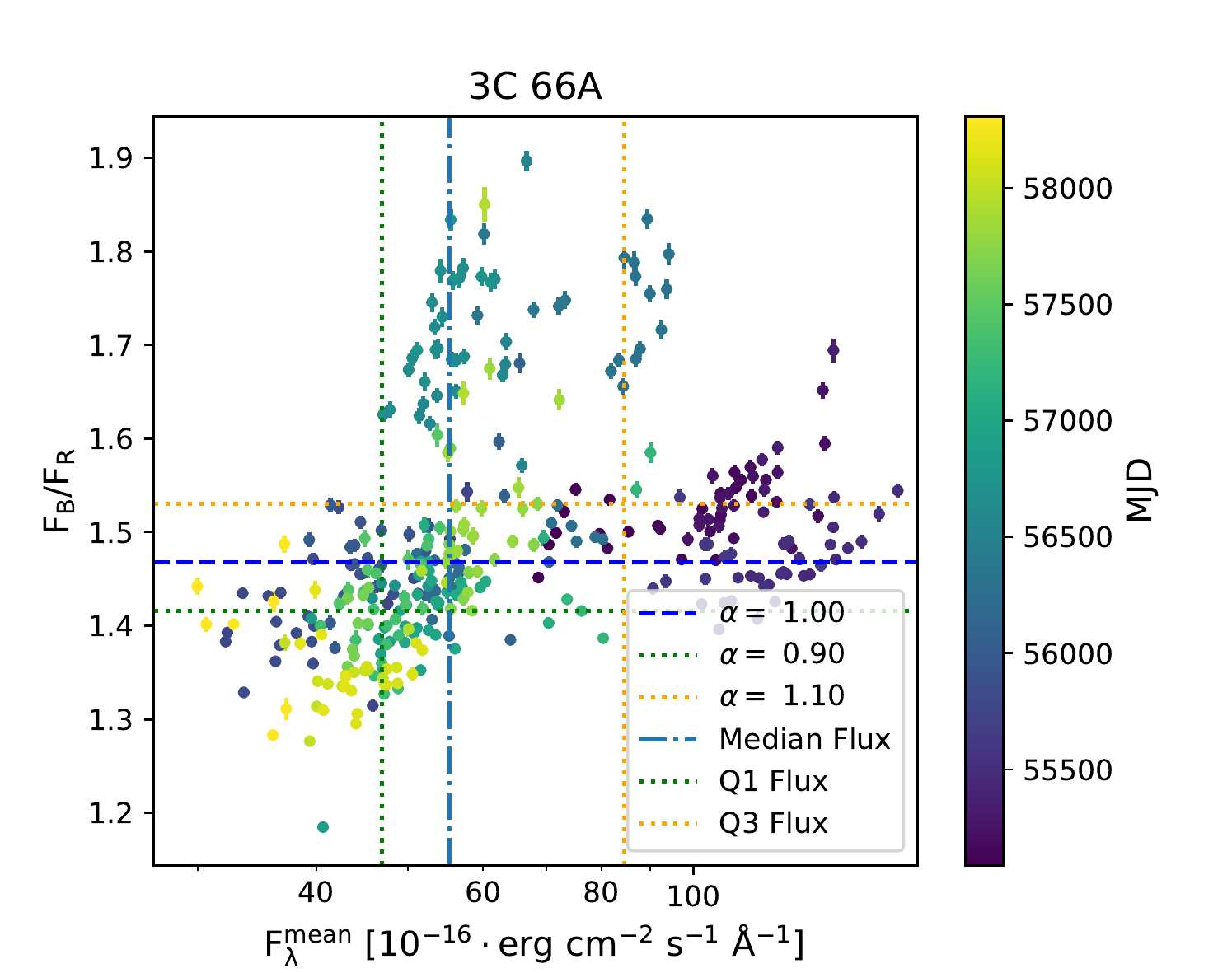}\hfill
    \includegraphics[width=.5\textwidth]{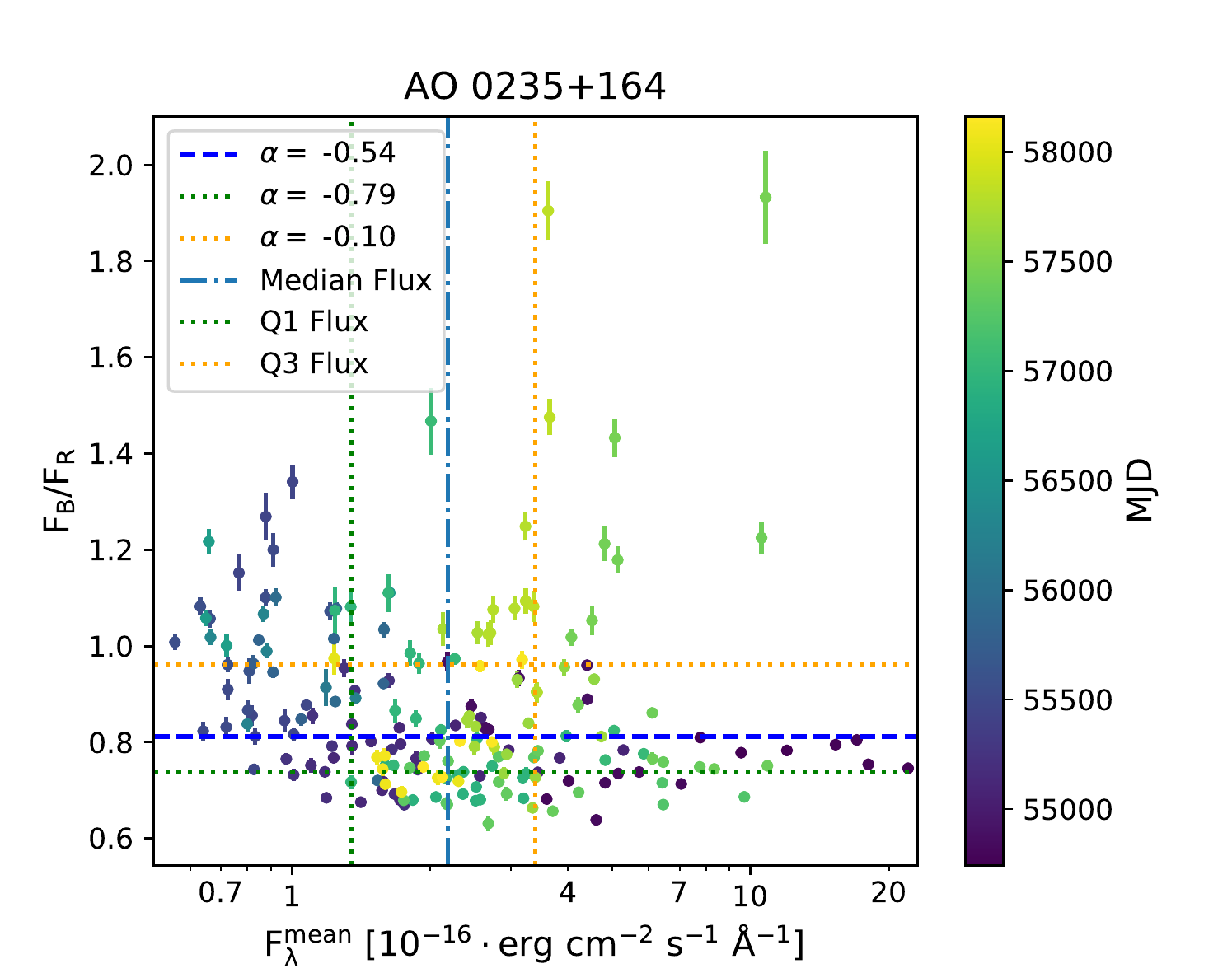}\hfill
    \\[\smallskipamount]
    \includegraphics[width=.5\textwidth]{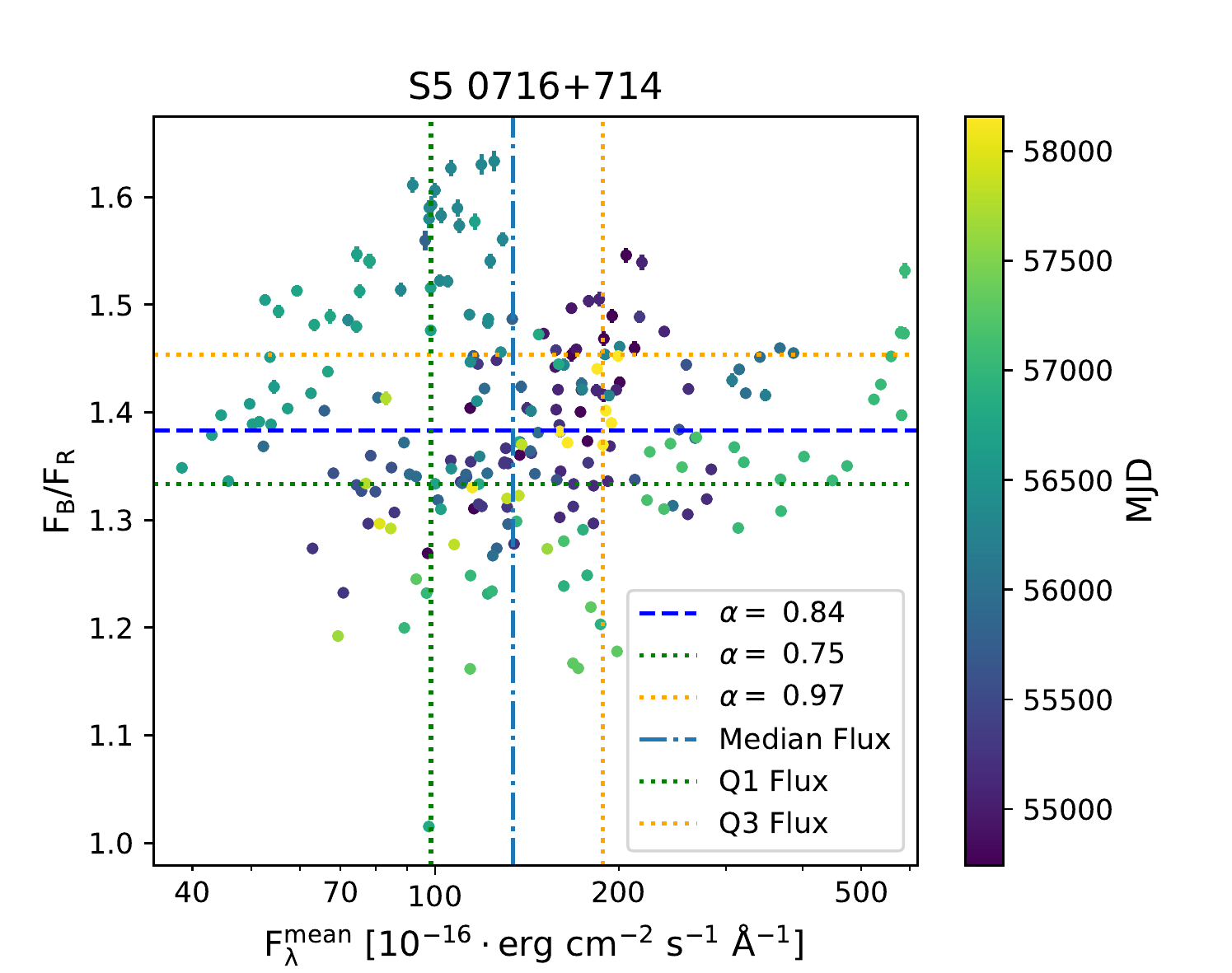}\hfill
    \includegraphics[width=.5\textwidth]{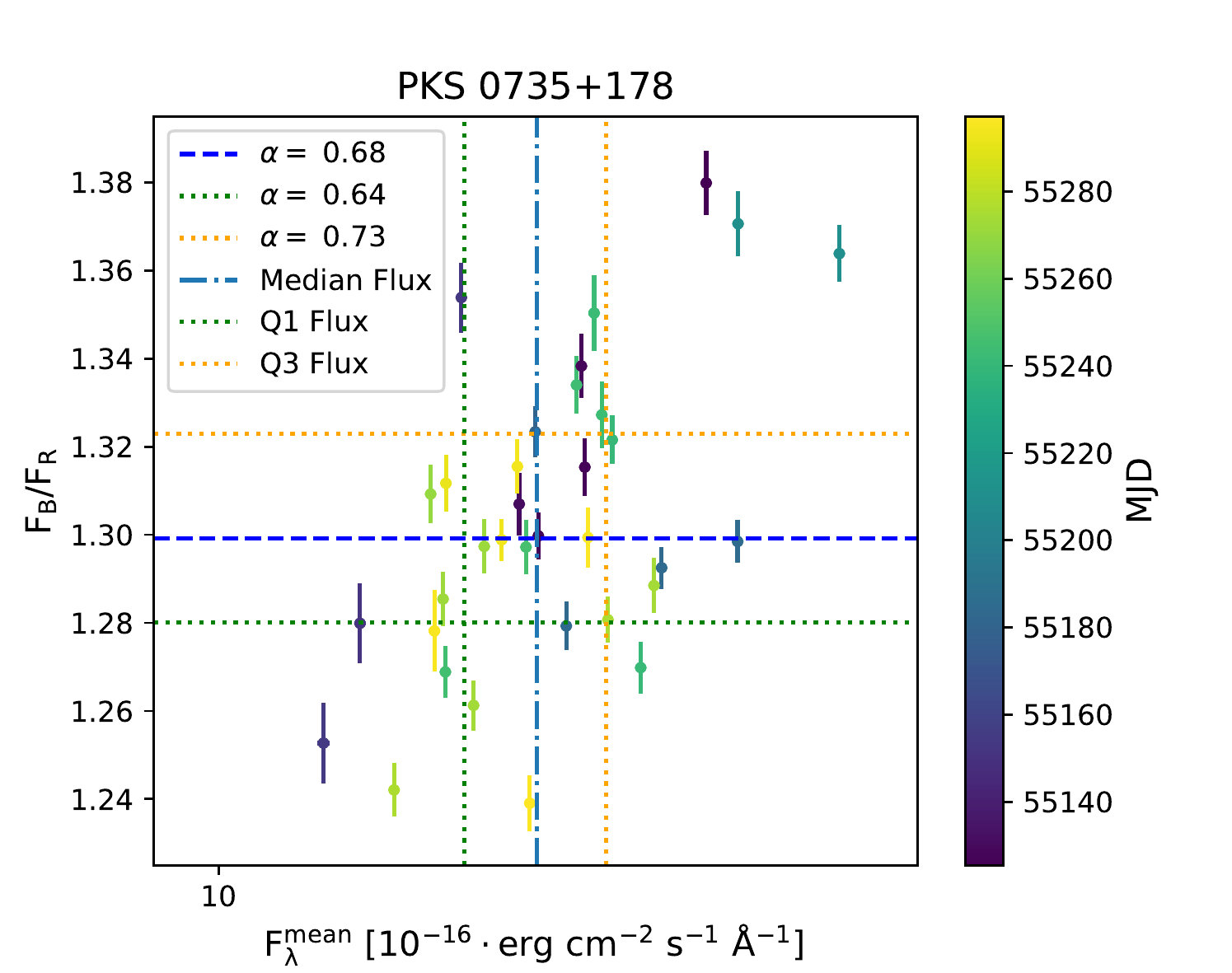}
    \\[\smallskipamount]
    \includegraphics[width=.5\textwidth]{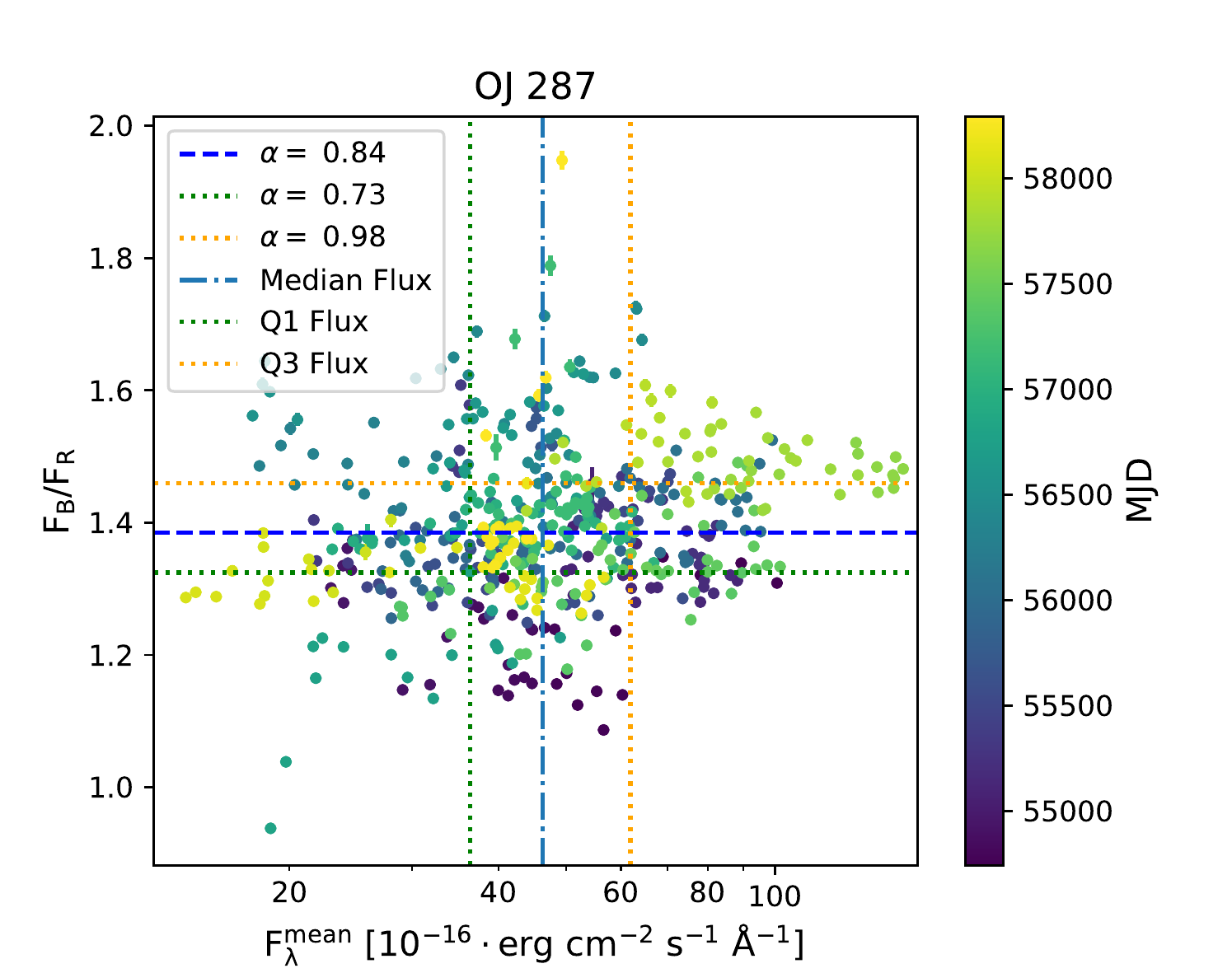}\hfill
    \includegraphics[width=.5\textwidth]{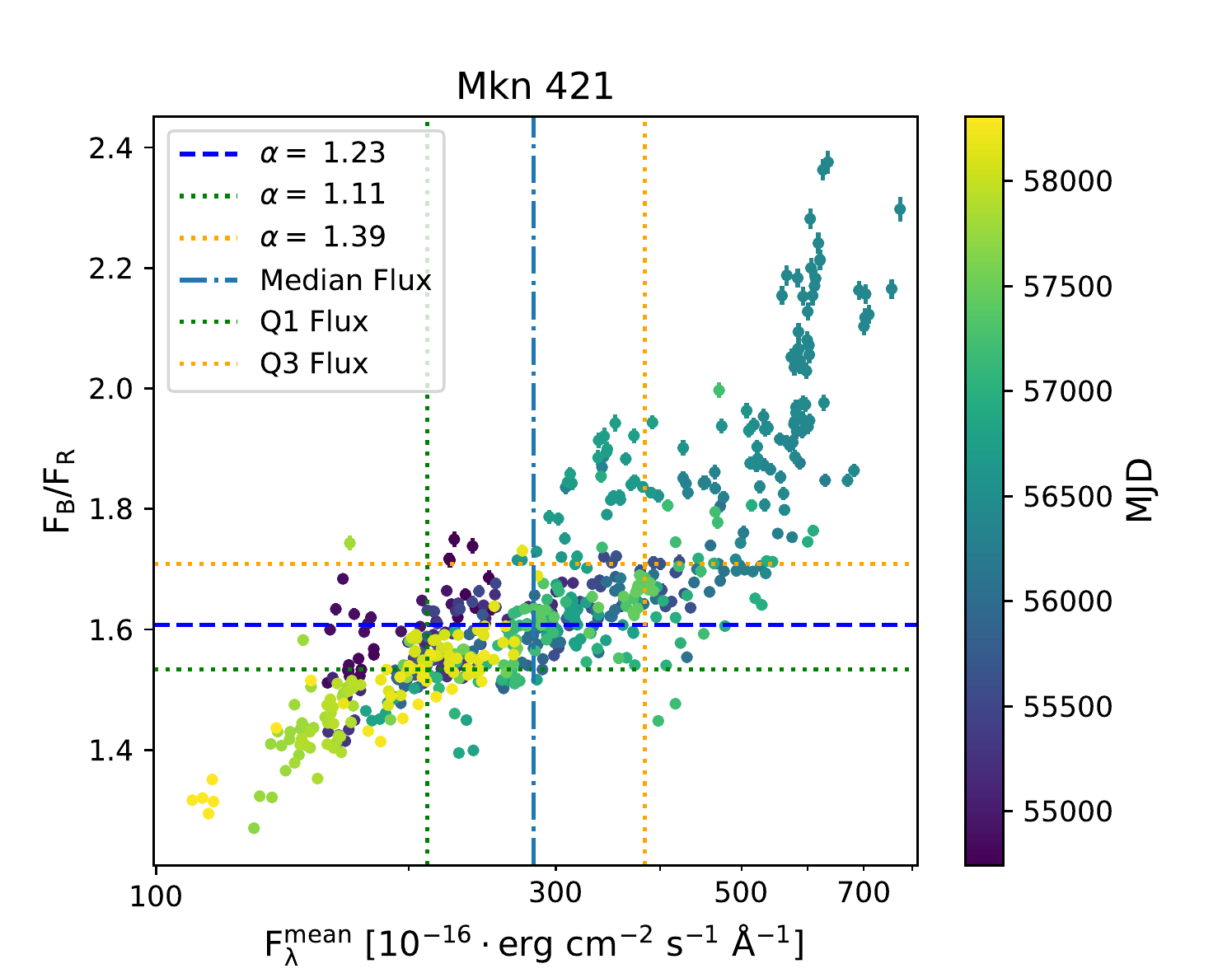}\hfill
    \caption{Colour-flux diagrams for BL Lac objects. The colour is estimated as the ratio between the blue and red parts of the spectra. Thus, higher colour values represent bluer spectra. The vertical dashed-dotted line corresponds to the median flux. Vertical dotted lines represent the first and third quartiles for the flux values. The horizontal dashed line denotes the median colour value, and its corresponding spectral index is noted in the legend. Horizontal dotted lines represent the first and third quartiles of the colour and their corresponding spectral index are noted in the legend.}
    \label{fig:variability_bllacs}
\end{figure*}
\begin{figure*}\ContinuedFloat
    \includegraphics[width=0.5\textwidth]{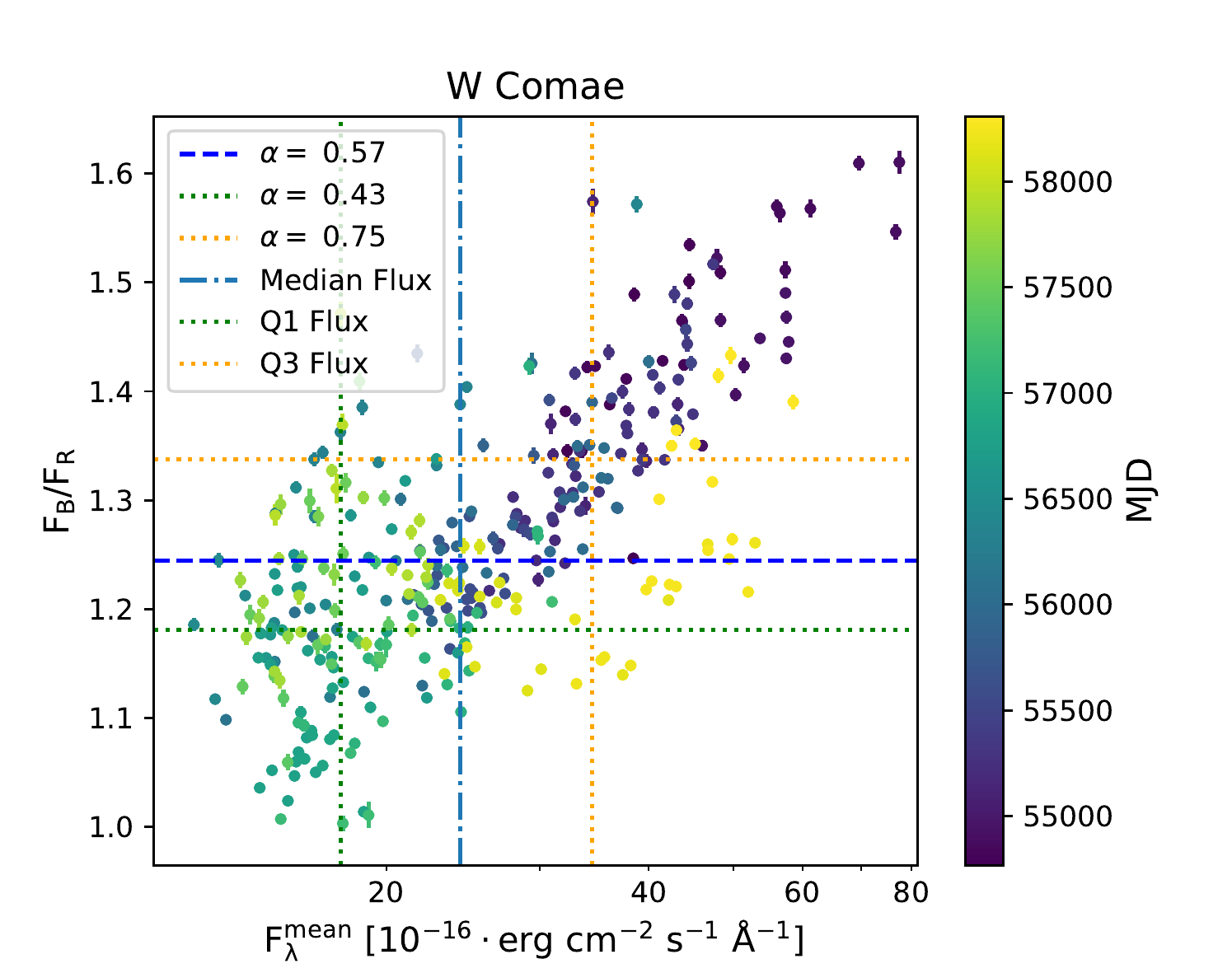}\hfill
    \includegraphics[width=.5\textwidth]{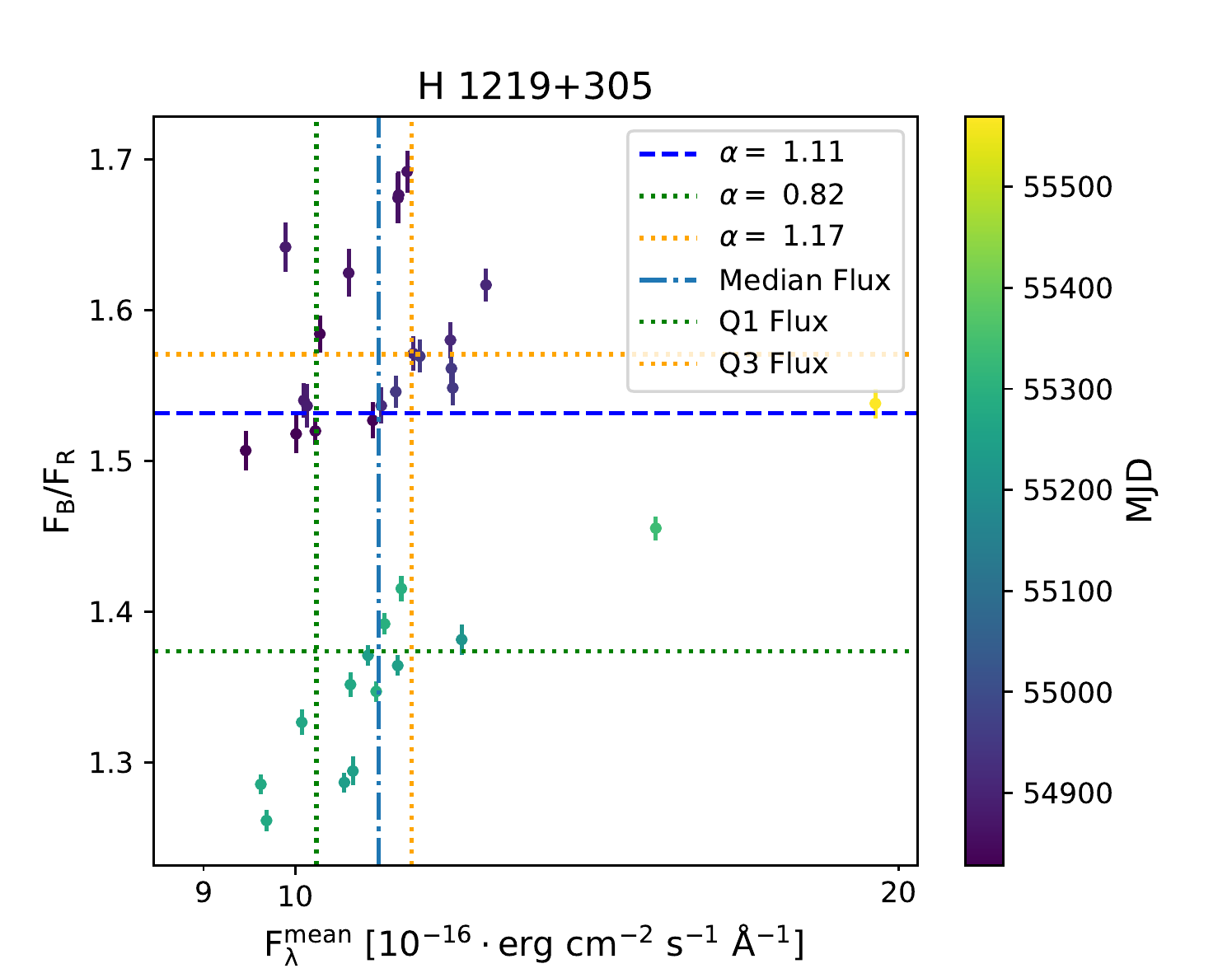}\hfill
    \\[\smallskipamount]
    \includegraphics[width=.5\textwidth]{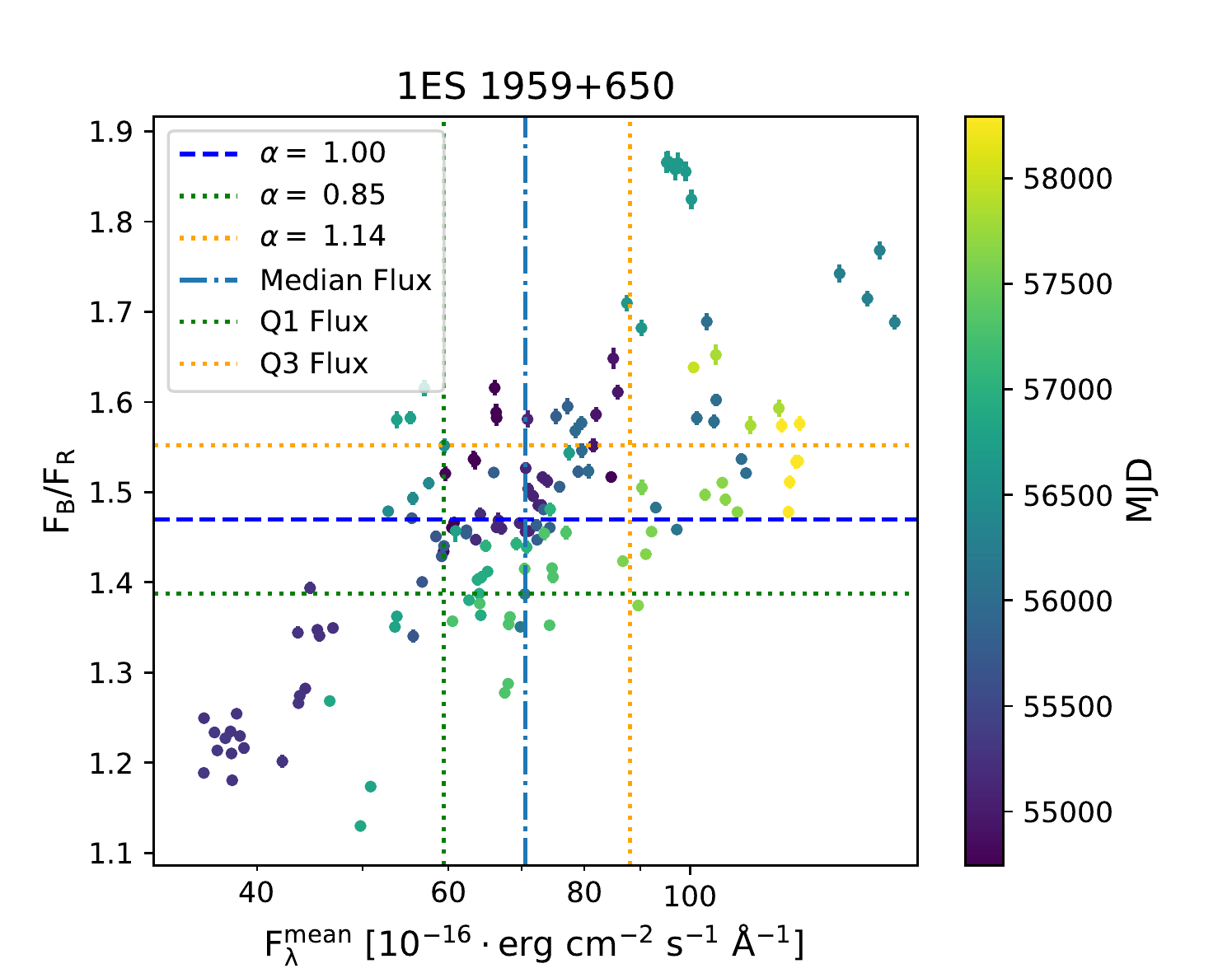}\hfill
    \includegraphics[width=.5\textwidth]{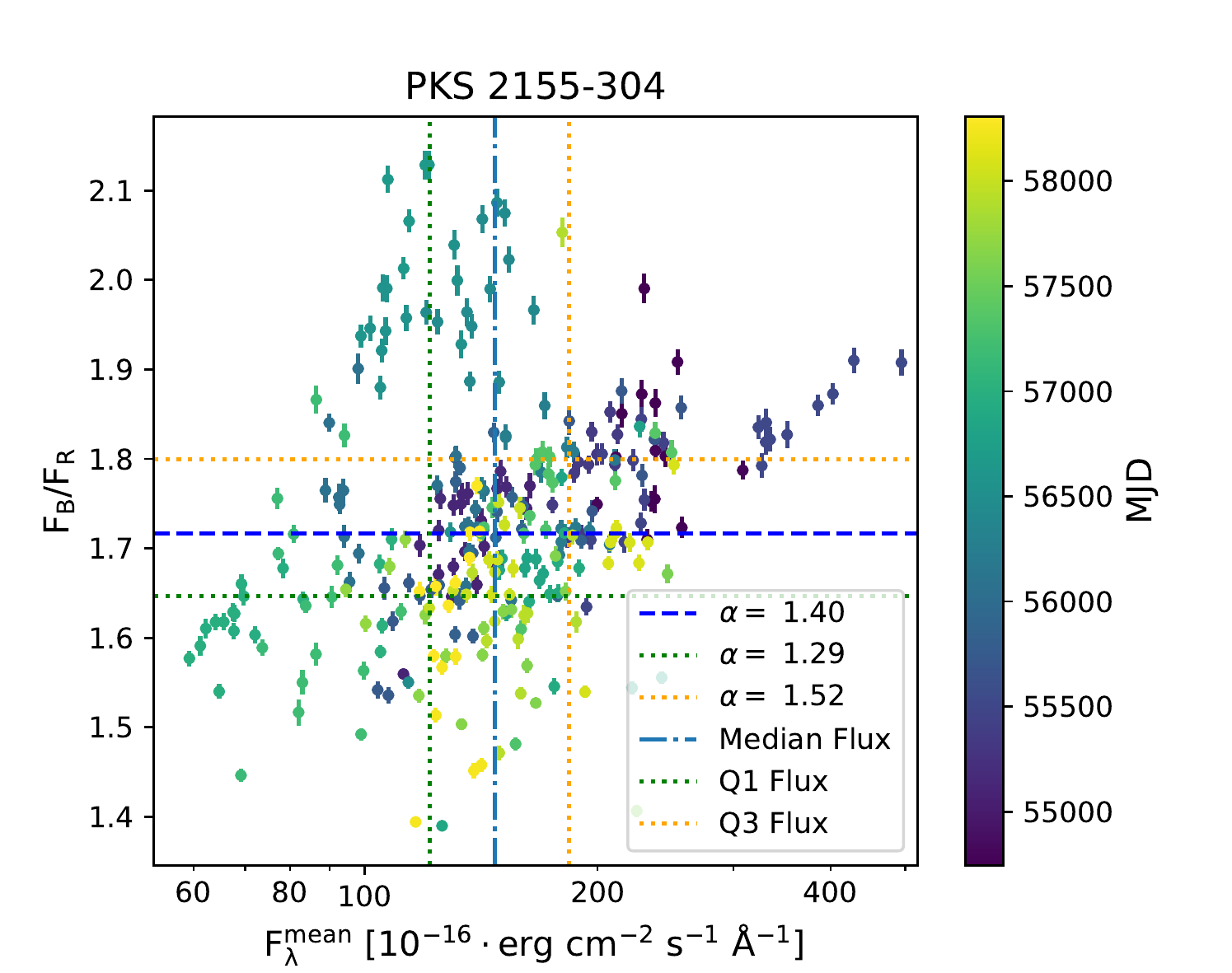}
    \\[\smallskipamount]
    \begin{center}
    \includegraphics[width=.5\textwidth]{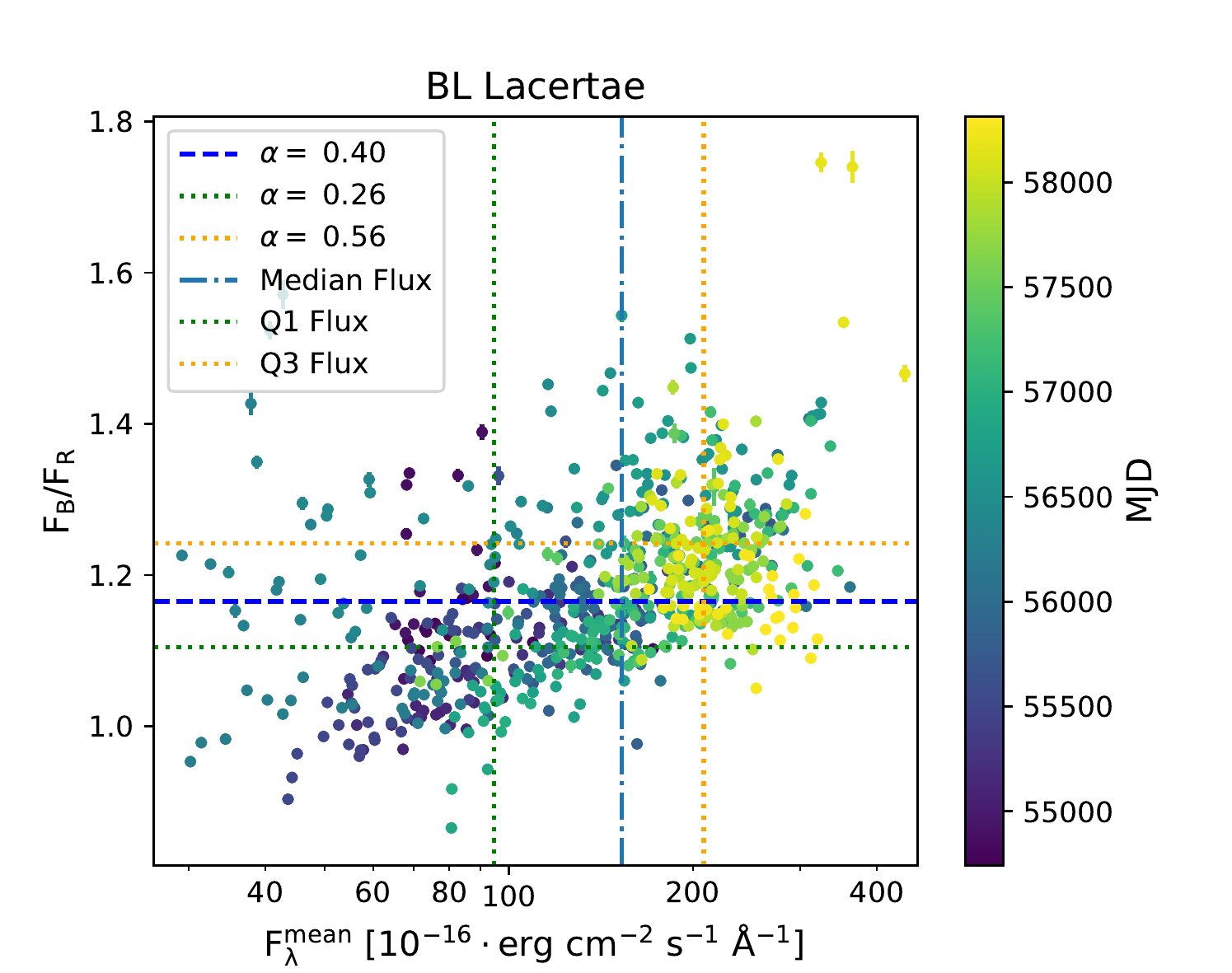}\hfill
    \end{center}
    \caption{Colour-flux diagrams for BL Lac objects (cont.).}
    \label{variability_bllacs}
\end{figure*}
\begin{figure*}
    \includegraphics[width=0.5\textwidth]{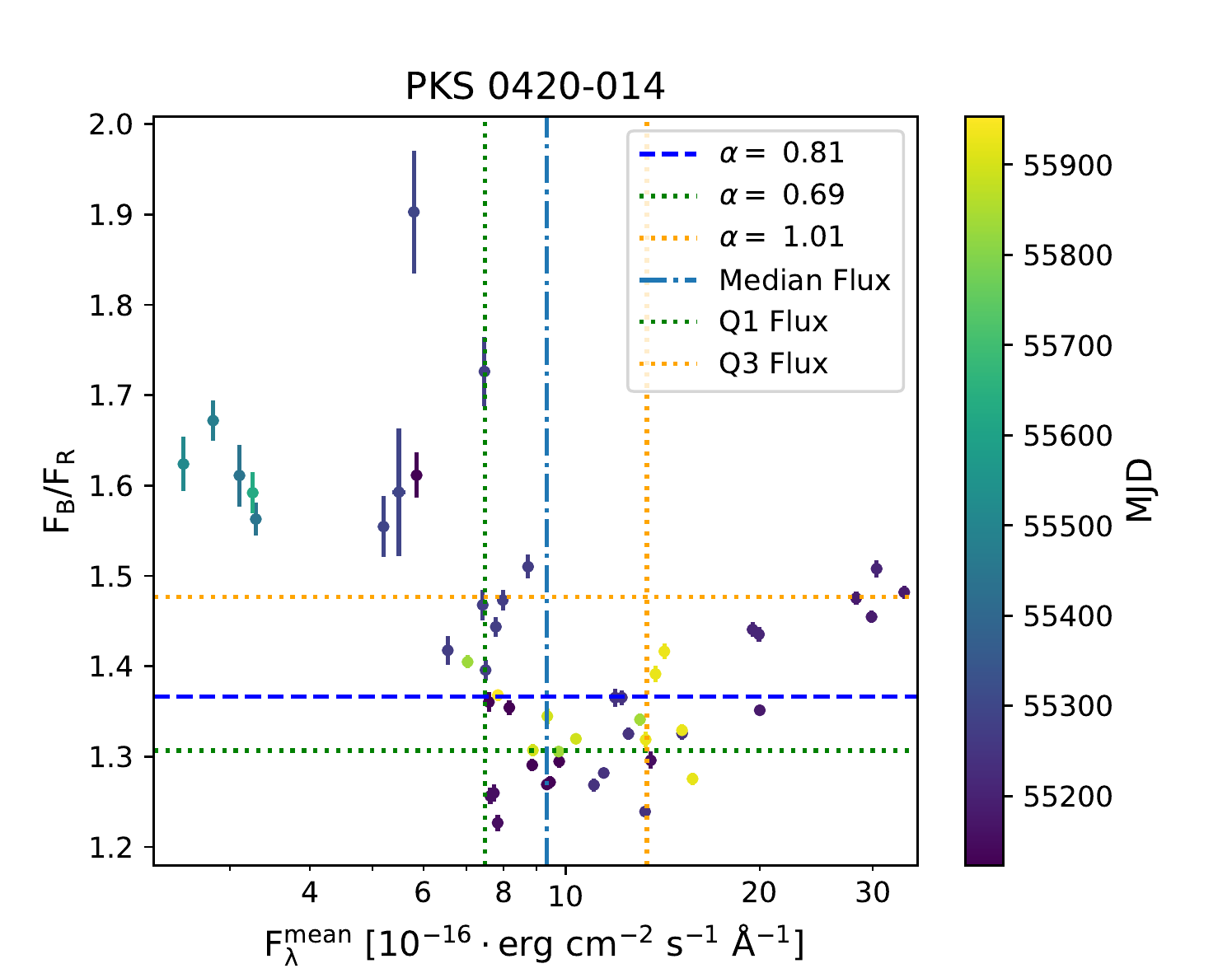}\hfill
    \includegraphics[width=.5\textwidth]{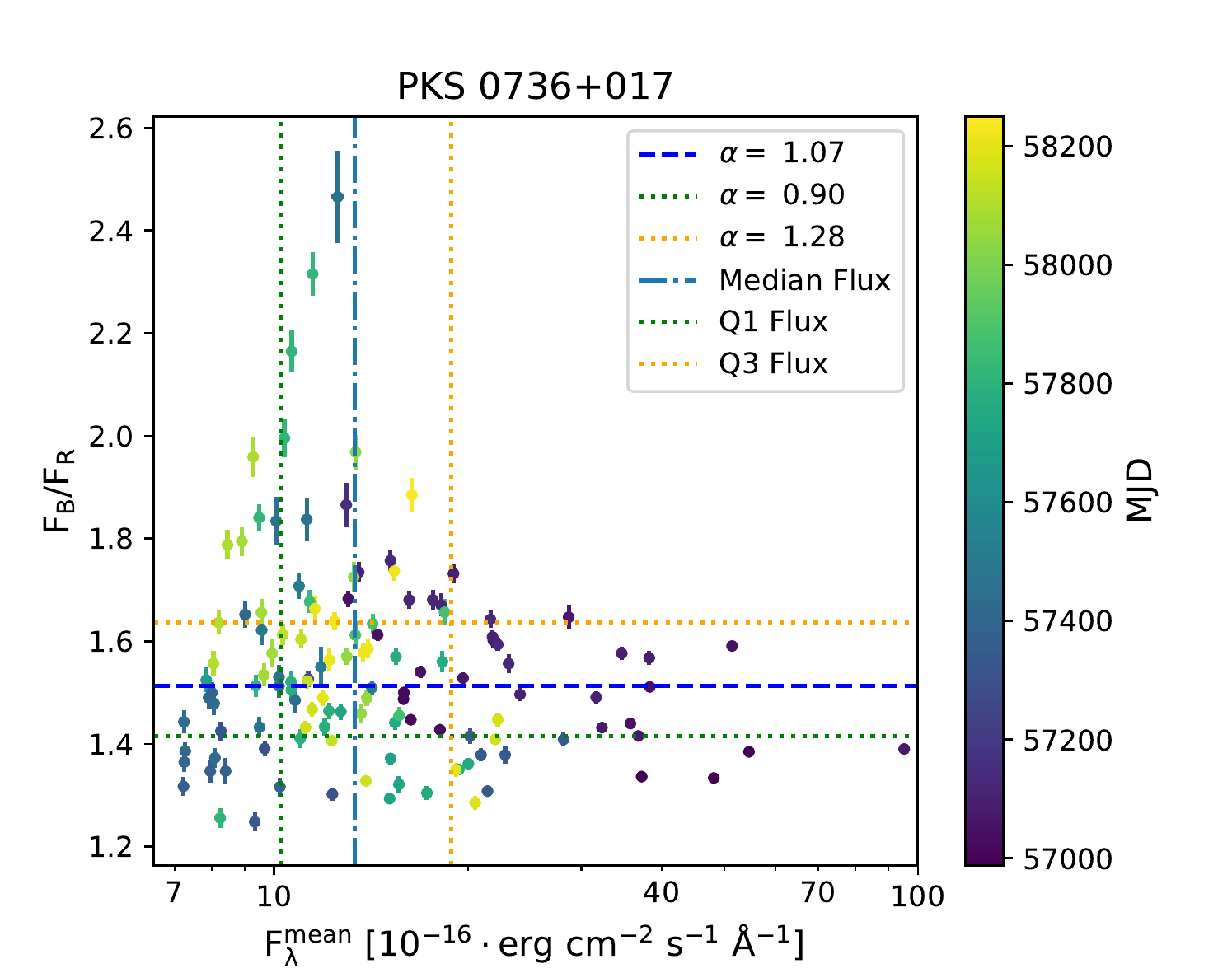}\hfill
    \\[\smallskipamount]
    \includegraphics[width=.5\textwidth]{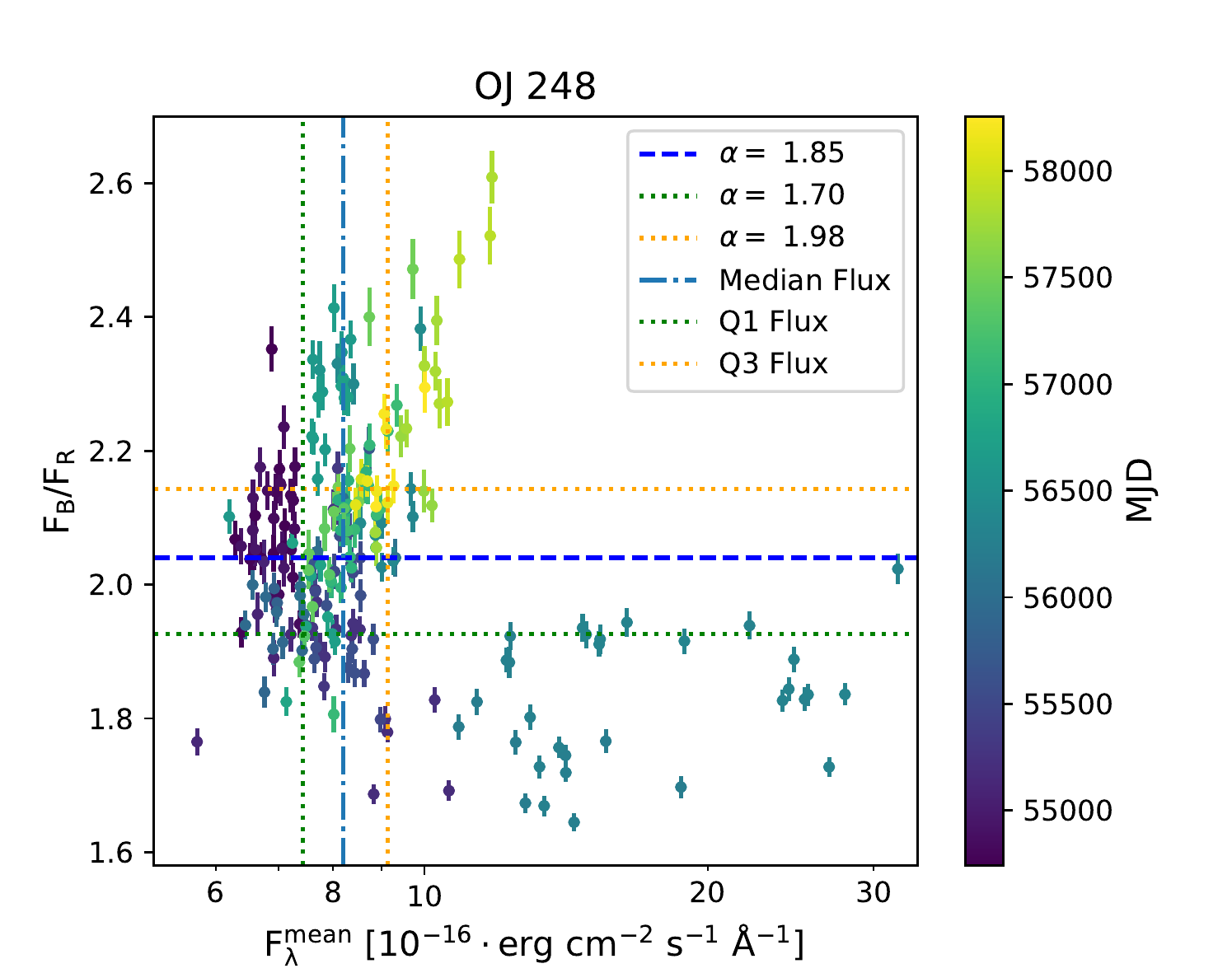}\hfill
    \includegraphics[width=.5\textwidth]{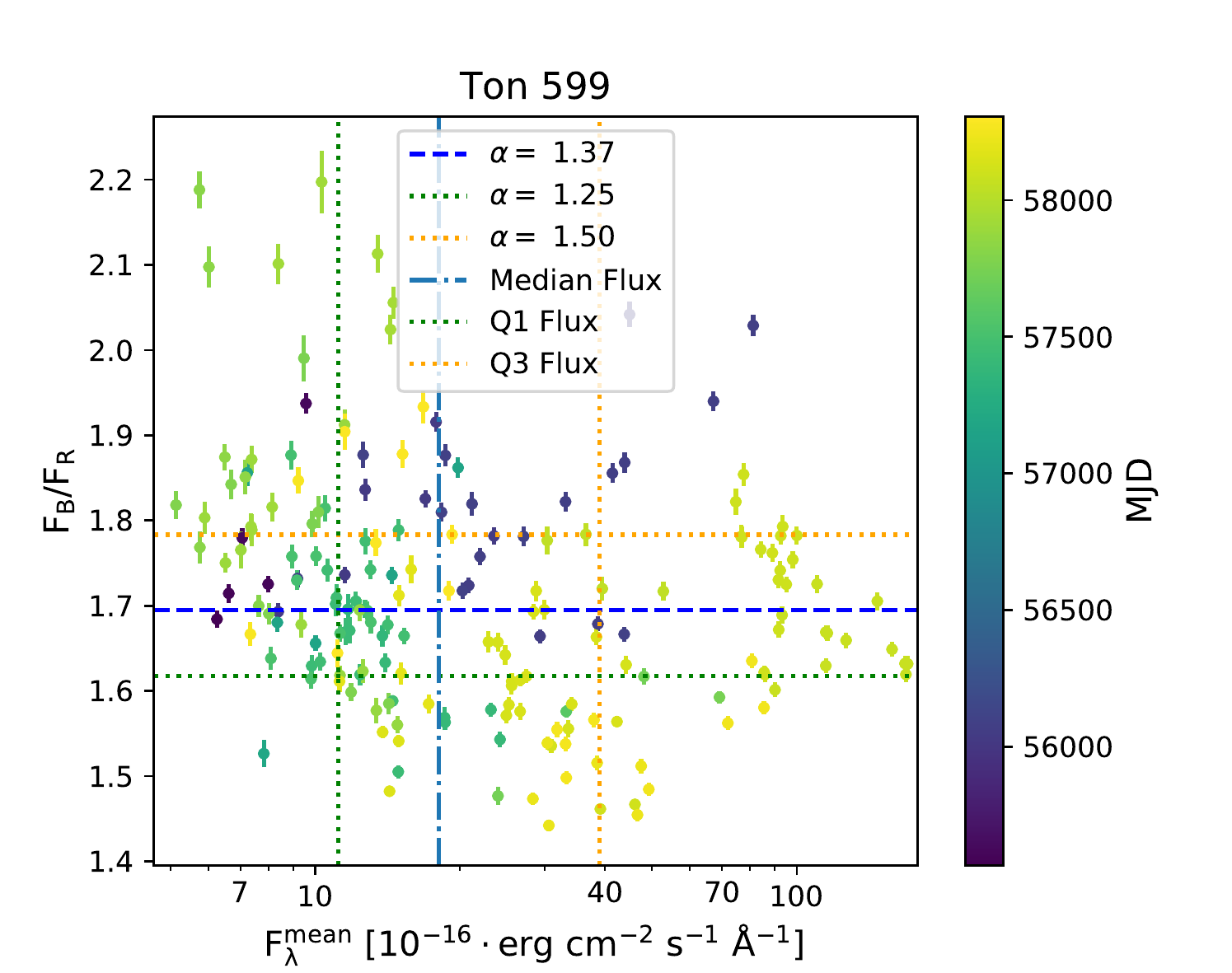}
    \\[\smallskipamount]
    \includegraphics[width=.5\textwidth]{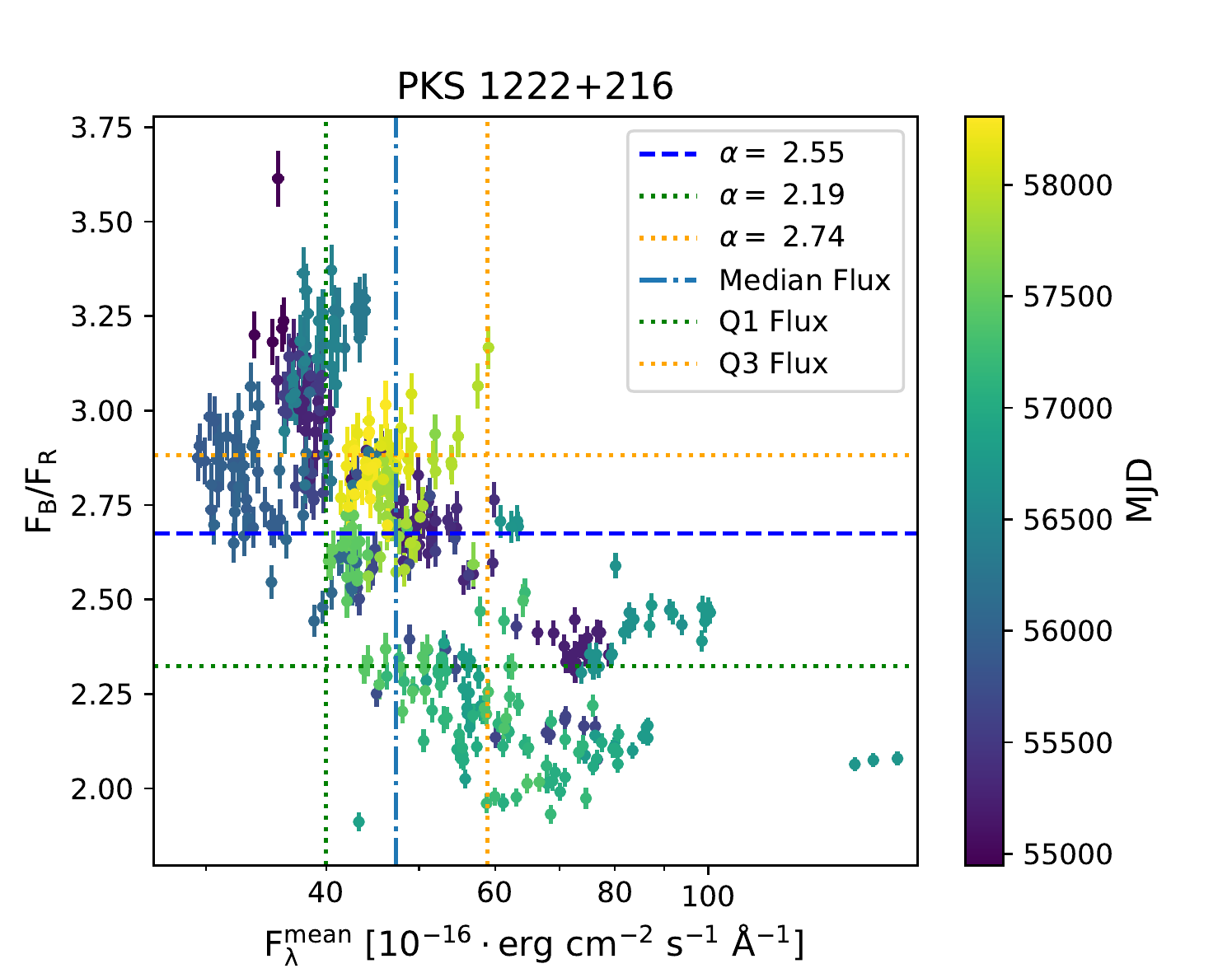}\hfill
    \includegraphics[width=.5\textwidth]{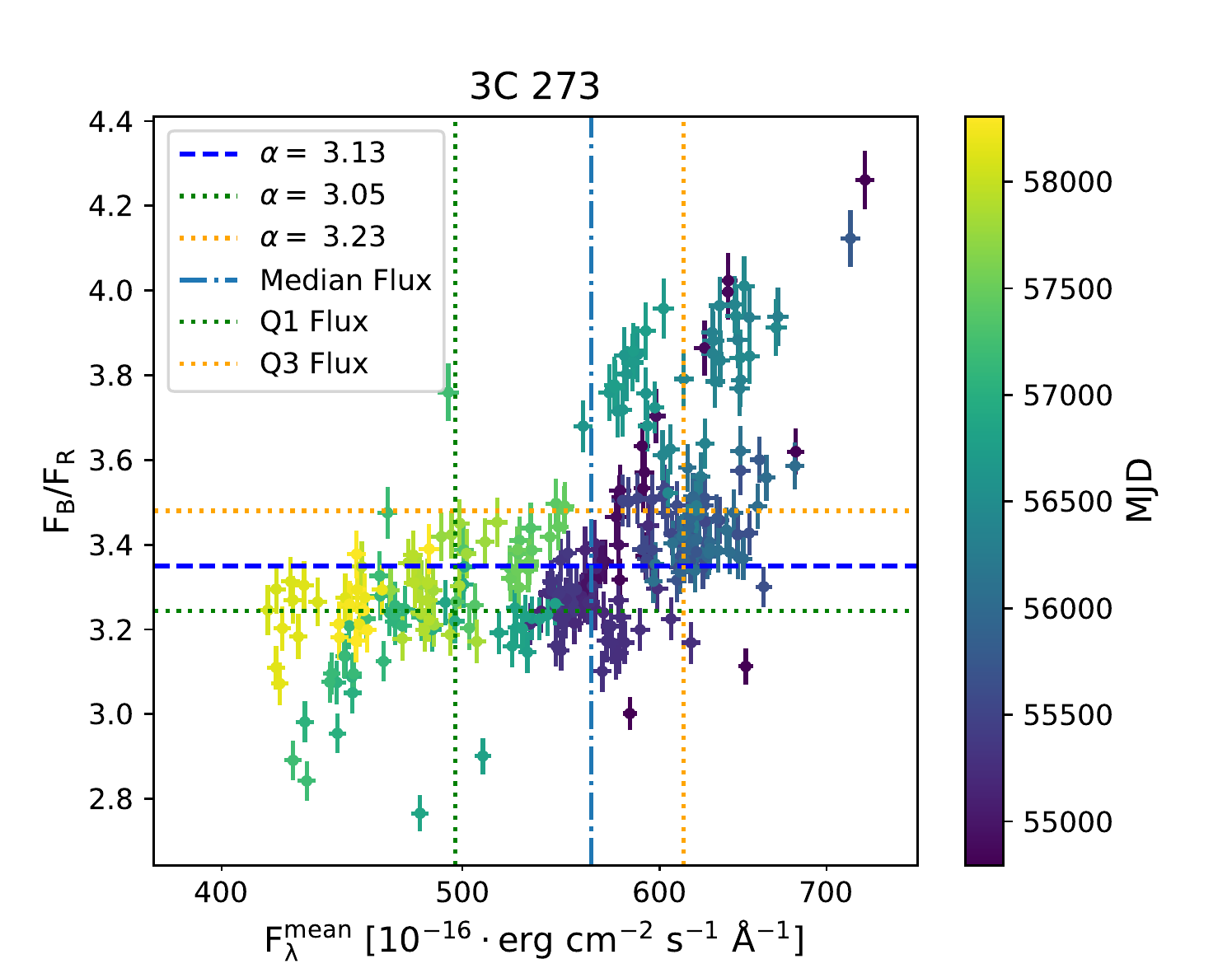}\hfill
    \caption{Colour-flux diagrams for FSRQs. The colour is estimated as the ratio between the blue and red parts of the spectra. Thus, higher colour values represent bluer spectra. The vertical dashed-dotted line corresponds to the median flux. Vertical dotted lines represent the first and third quartiles for the flux values. The horizontal dashed line denotes the median colour value, and its corresponding spectral index is noted in the legend. Horizontal dotted lines represent the first and third quartiles of the colour and their corresponding spectral index are noted in the legend.}
    \label{fig:variability_fsrqs}
\end{figure*}
\begin{figure*}\ContinuedFloat
    \includegraphics[width=0.5\textwidth]{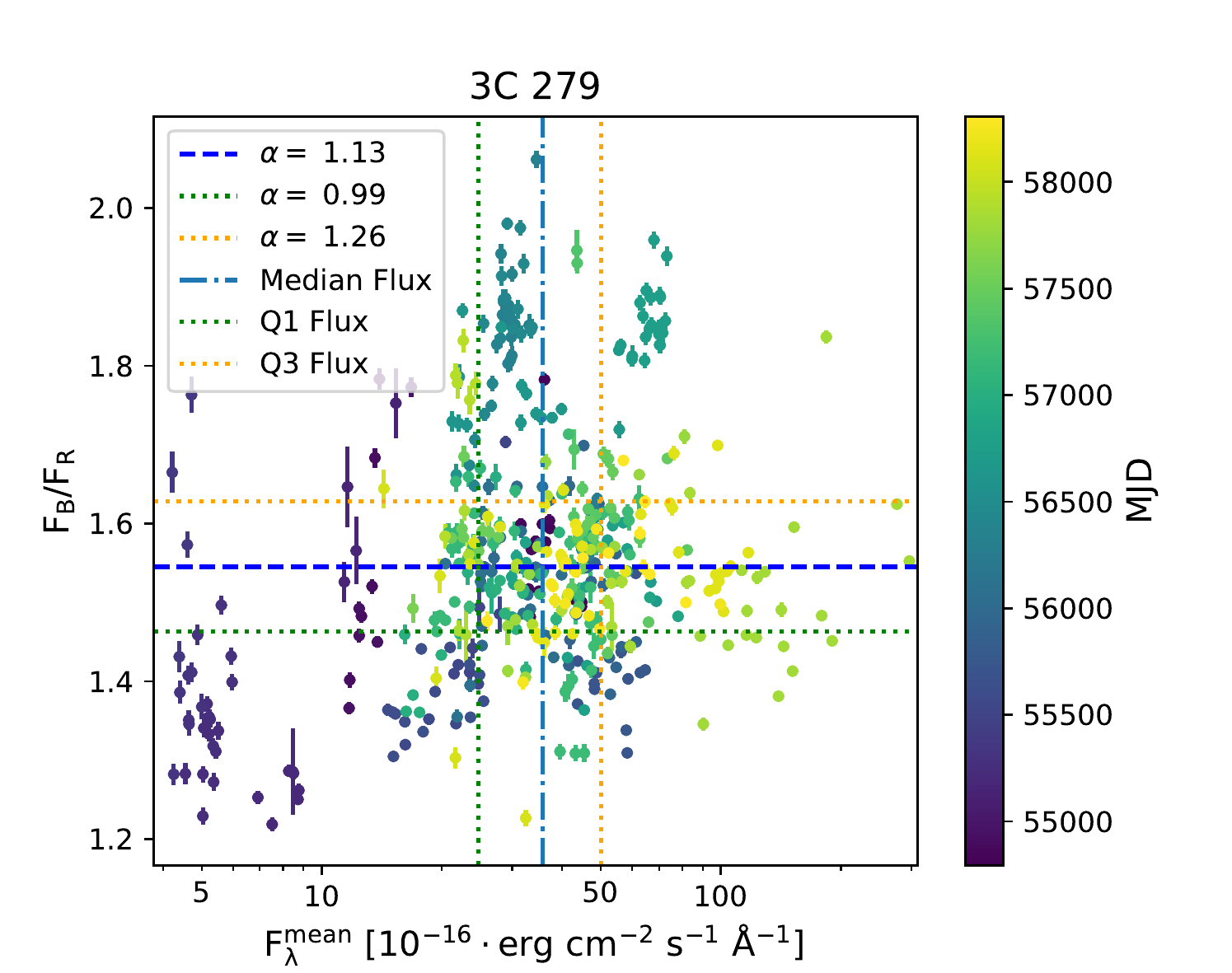}\hfill
    \includegraphics[width=.5\textwidth]{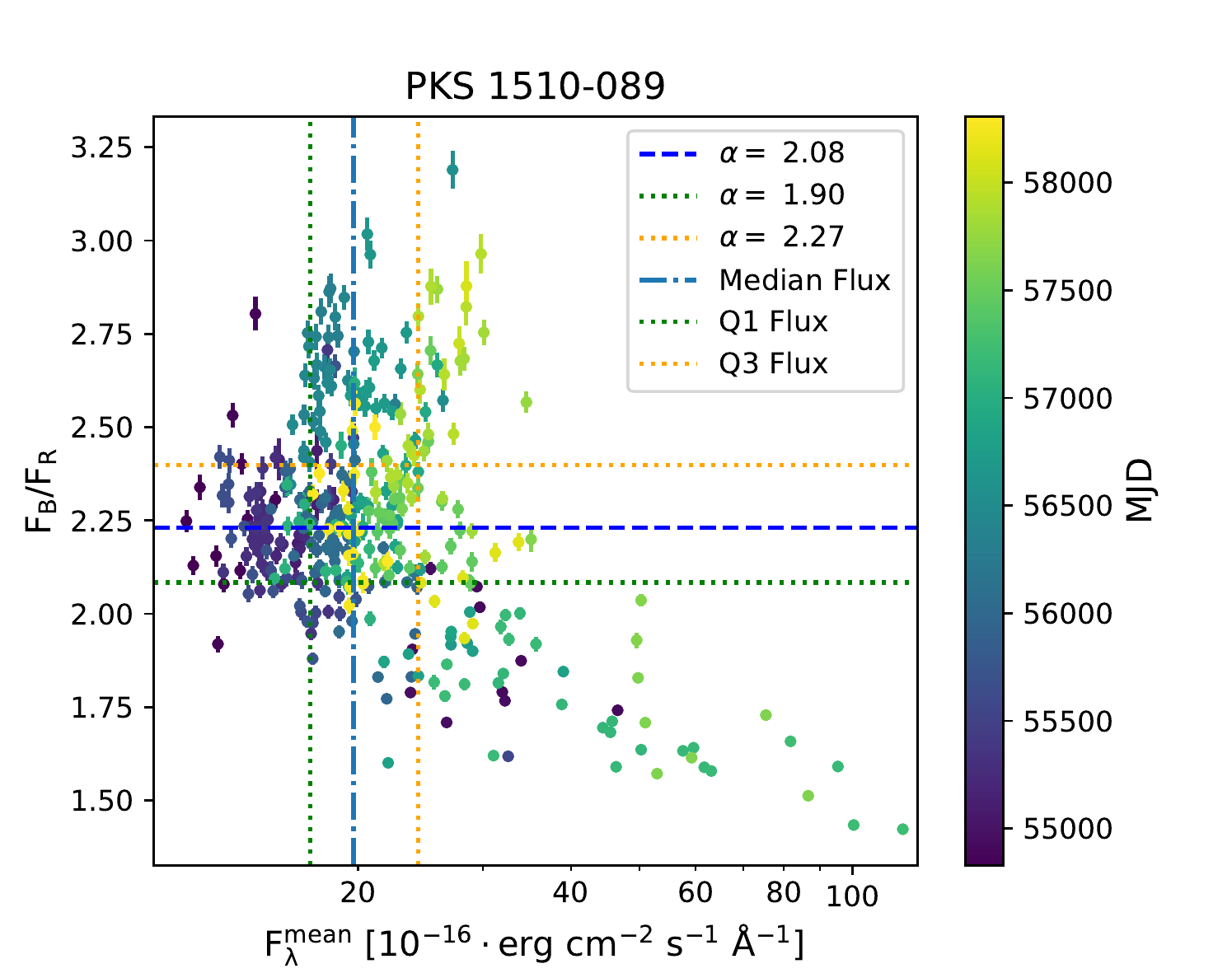}\hfill
    \\[\smallskipamount]
    \includegraphics[width=.5\textwidth]{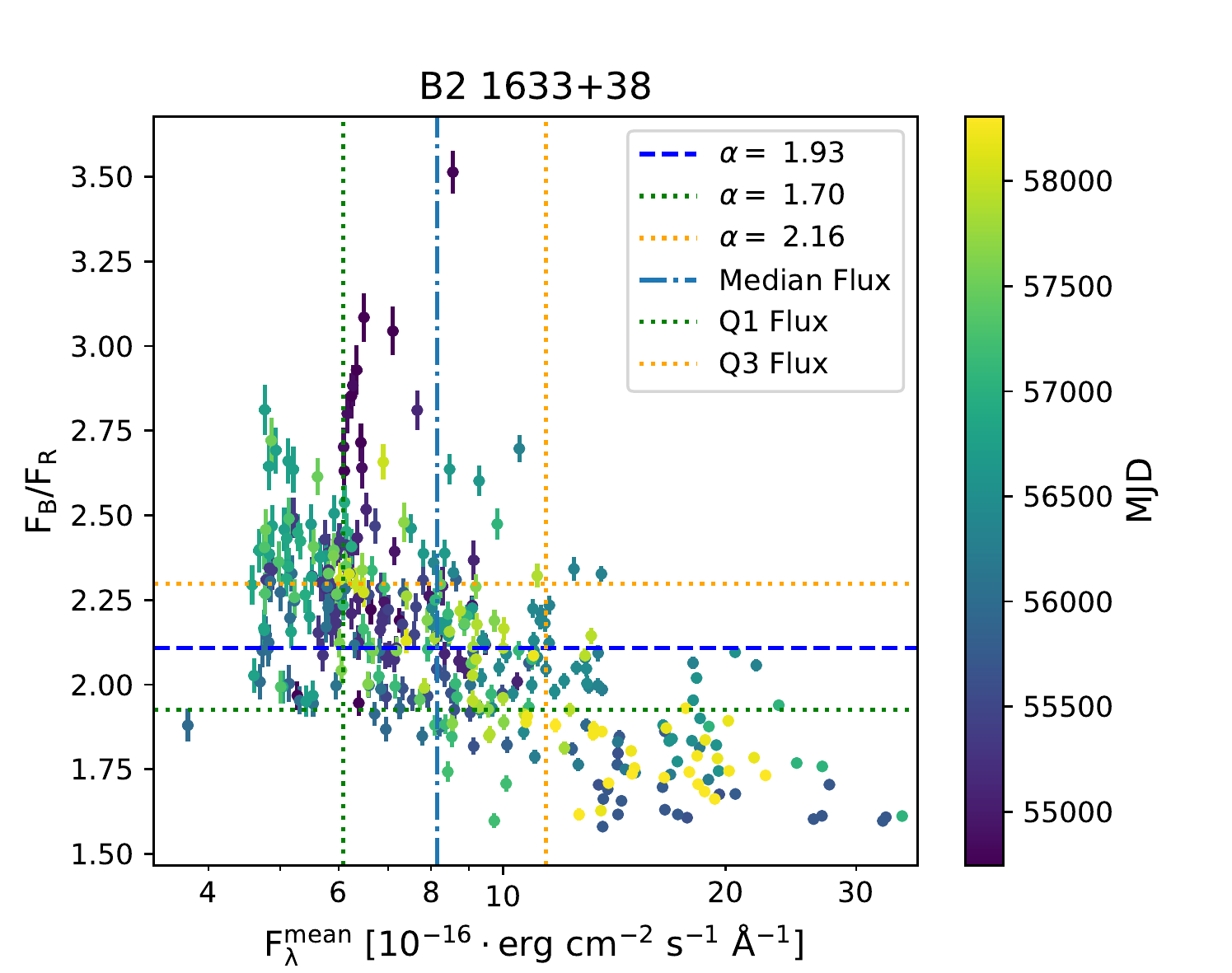}\hfill
    \includegraphics[width=.5\textwidth]{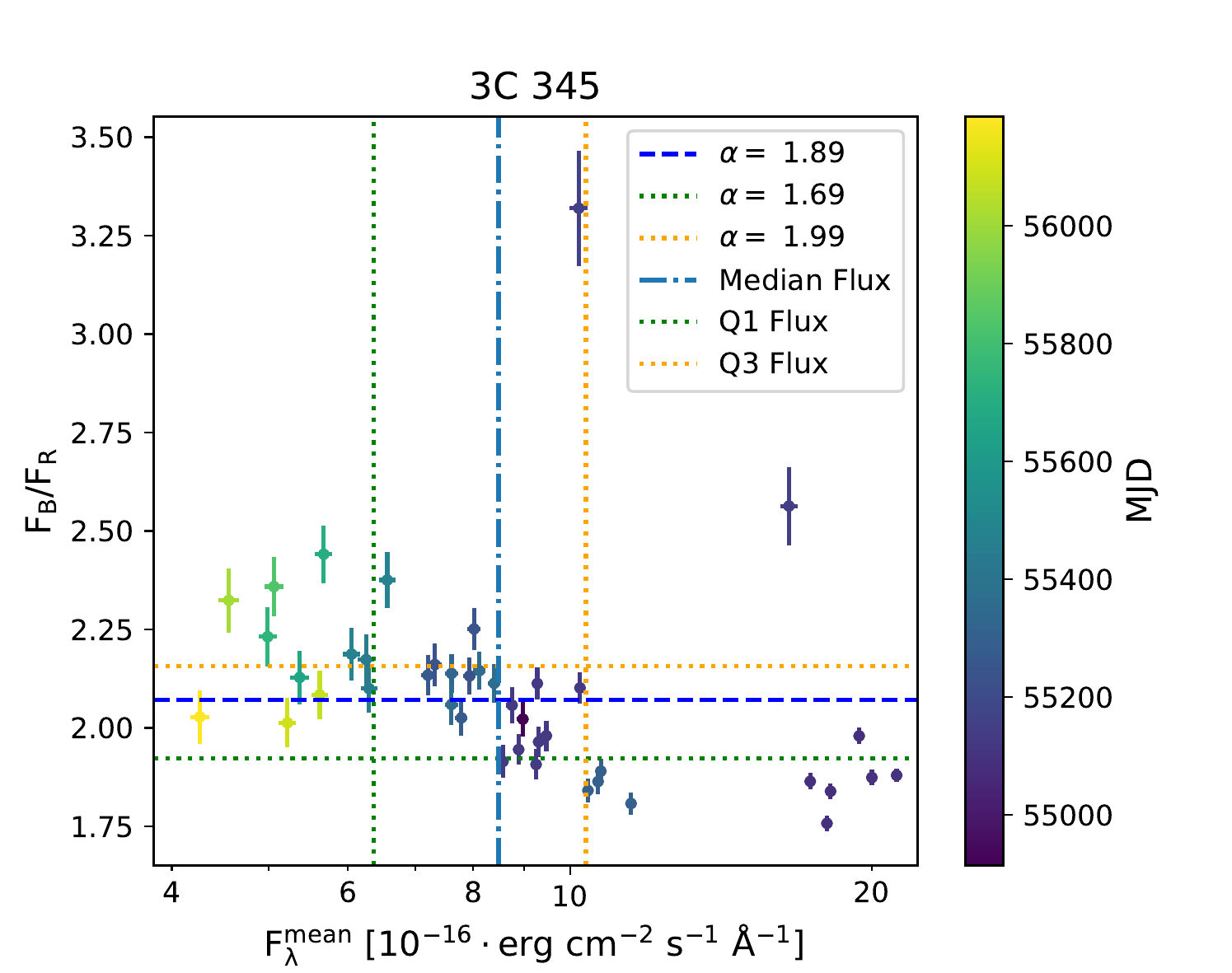}
    \\[\smallskipamount]
    \includegraphics[width=.5\textwidth]{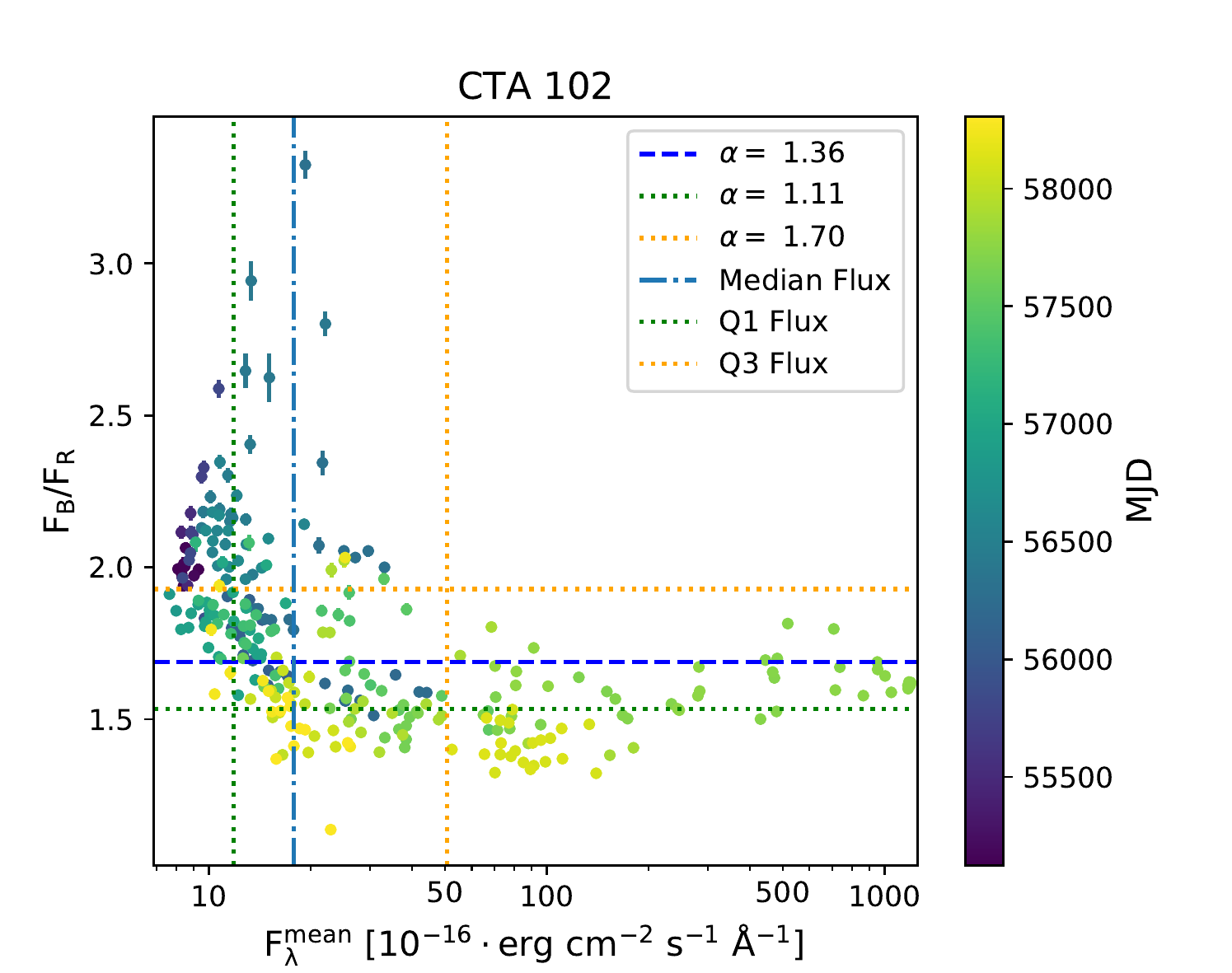}\hfill
    \includegraphics[width=.5\textwidth]{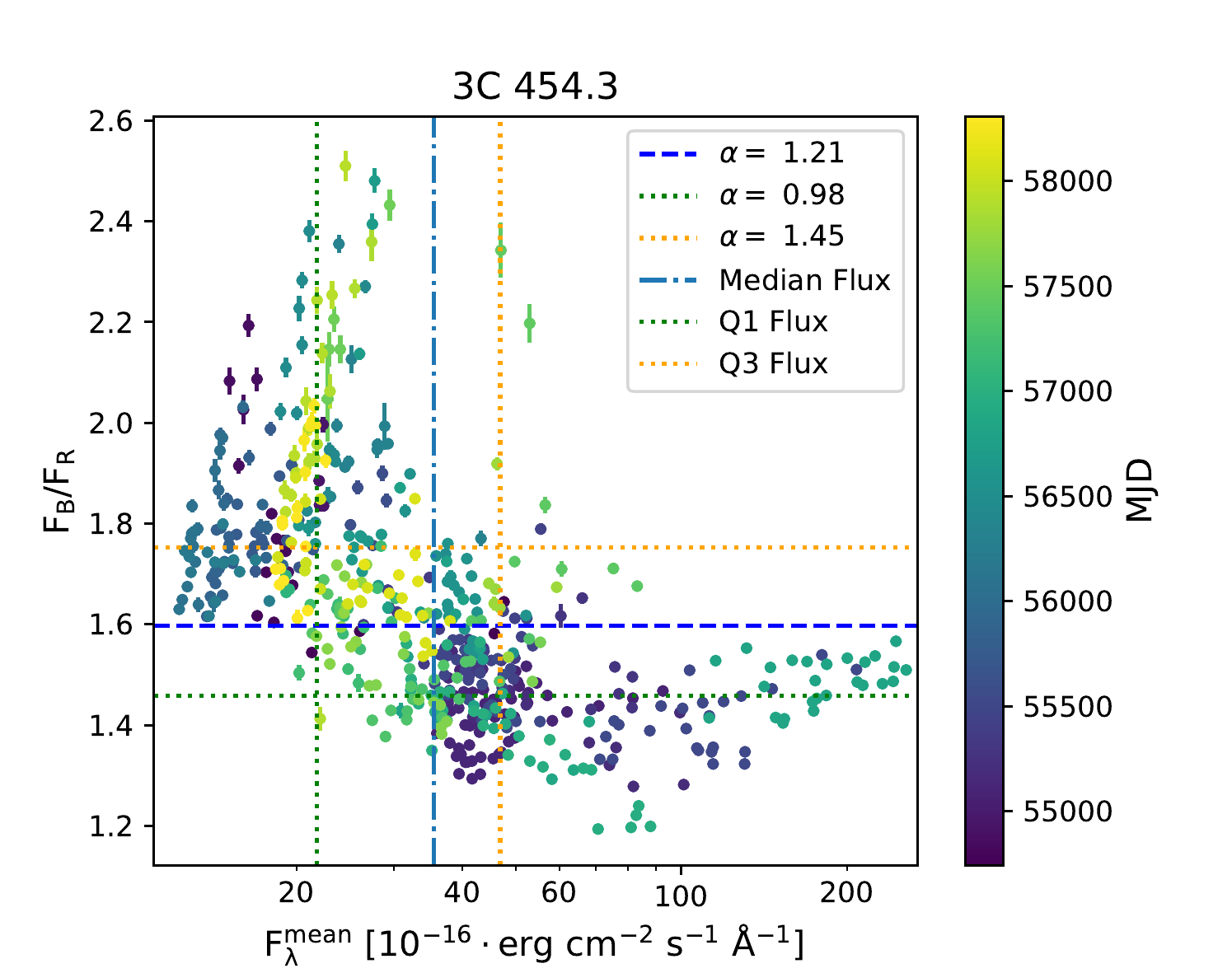}\hfill
    \caption{Colour-flux diagrams for FSRQs (cont.).}
    \label{variability_fsrqs}
\end{figure*}

\section{Methodology}\label{sec4}
\subsection{Non-Negative Matrix Factorization}
The NMF is a statistical tool commonly used for dimensionality reduction \citep{lee1999}. It allows the user to describe a large data set with a reduced number of components or parameters by transforming these high-dimensional data into a low-dimensional space in which it can be easily interpreted \citep{baron2019}. The data are factorized in terms of the different components needed for the reconstruction following 

\begin{equation}
V\approx W\cdot H,
\label{nmf_eq}
\end{equation}

\noindent where $V$ is a \textit{(m $\times$ n)} dimensional matrix containing the data sample. $W$ is a \textit{(m $\times$ p)} matrix containing the coefficients for each component of the reconstruction. $H$ is the vector matrix of the derived components, with a dimension of \textit{(p $\times$ n)}. Each data vector from the matrix $V$ can thus be approximated by a linear combination of the component vectors $H$ and their corresponding weights $W$ \citep{lee2001}. In our case, \textit{m} is the number of spectra introduced as input to be reconstructed, \textit{n} is the number of spectral bins of the data set, and \textit{p} is the number of components used for the reconstruction.

The main advantage of the NMF with respect to other dimensionality reduction algorithms such as the PCA is that the data, weights and components must be non-negative \citep{blanton2007}. While PCA components must be orthogonal and thus, some of them may have negative values, making a physical interpretation more difficult, the NMF leads to an easier physical association due to the non-negative restriction. However, this non-orthogonality condition has also the caveat of resulting in degeneracy of the components. In this work, we make use of the NMF algorithm implemented in the \textit{python} package \texttt{scikit}\footnote{\url{https://scikit-learn.org/stable/modules/generated/sklearn.decomposition.NMF.html}}.

\subsection{Reconstruction of the data sample}
\begin{figure}
	\includegraphics[width=\columnwidth]{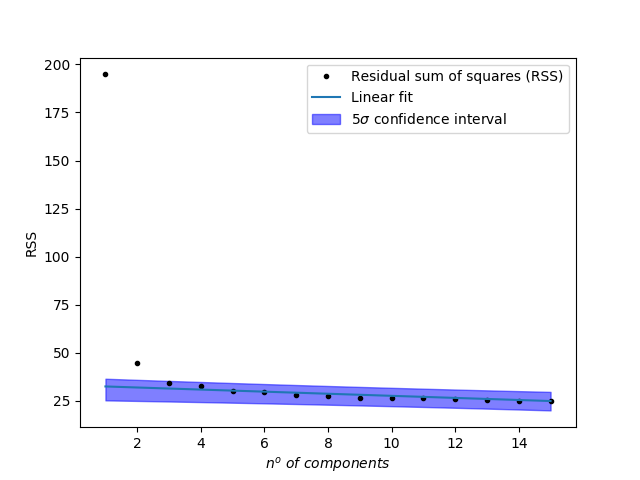}
    \caption{RSS vs. n$^{\circ}$ of components for the BL Lac object PKS~2155-304. The number of components needed for the reconstruction is estimated by a linear fit of the RSS. The amount of components needed is equivalent to the number of points that deviate from the 5$\sigma$ confidence interval of the fit.}
    \label{RSSdiagram}
\end{figure}

Previous works using the NMF are based on a reconstruction with a number of components estimated from the residual sum of squares (RSS) diagram \citep[see e.g.][]{hutchins2008}. The RSS is a measurement of the discrepancy between the data and the model (see Fig. \ref{RSSdiagram}). The RSS of pure random noise follows a decreasing linear trend as the number of components increases. Those points deviating more than $5\sigma$ from this linear trend correspond to the components that represent real variability and consequently, the minimum number of components needed for the reconstruction of the data. No \textit{a priori} information is given to the algorithm. 
This approach results in a total number of 2 components needed for most of the BL Lac and galaxy-dominated objects, and 3 components for almost all of the FSRQs of the sample.

\begin{figure}
	\includegraphics[width=\columnwidth]{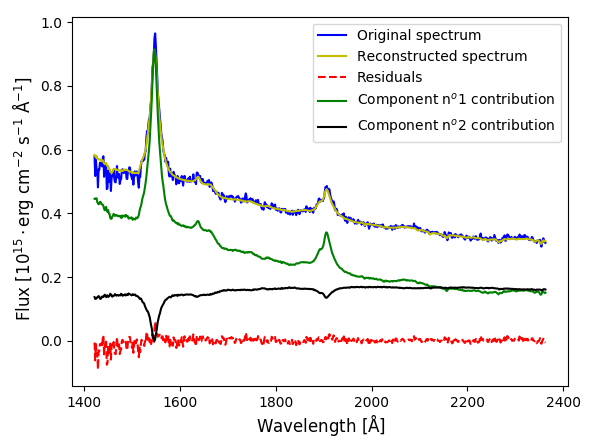}
    \caption{Reconstructed spectrum of the FSRQ B2~1633+38. Two components are used for the reconstruction, without any \textit{a priori} assumption. The first component represents the emission coming from the BLR. The second component shows artificial absorption lines non-existent in the total spectrum, introduced by the NMF to minimize the residual.}
    \label{NMF_coupled_components}
\end{figure}

While this procedure maximizes the variance explained by the reconstruction, the components contain mixed contributions and the physical interpretation of the components is not straightforward. This methodology leads to resulting components in which the emission of different regions of the AGN are mixed into an individual component (e.g. PL type component with characteristic BLR emission lines, see Fig. \ref{NMF_coupled_components}). This effect makes hard to obtain a physical interpretation of the reconstruction and to quantify the amount of variability that is caused by each constituent of the AGN.

In this paper we propose a complementary methodology for the reconstruction of the spectra of the sources. This approach is based on the \textit{a priori} knowledge of the possible components we can expect from the different types of blazars:

\begin{itemize}
    \item BL Lacs: The absence or minor presence of signatures of an accretion disk, dusty torus or a BLR in BL Lac objects makes the relativistic jet the main contribution \citep[e.g.][]{plotkin2012}. The synchrotron emission coming from the jet can be described as a 
    PL function in a narrow spectral range. For larger spectral ranges, other models (e.g. log-parabola) may be more adequate.
    \item FSRQs: Apart from the dominant non-thermal emission from the jet, these sources exhibit broad optical emission lines coming from the BLR \citep[see for instance][]{ghisellini2008} and likely thermal emission from the accretion disk \citep[e.g.][]{shang2005,calderone2013}. In this case, we use a quasi-stellar object (QSO) emission to account for the BLR and a set of PLs to describe the main contribution of the jet, and the accretion disk.
    \item Galaxy-dominated sources: In some cases, the jet of BL Lacs is not bright enough to outshine the host galaxy. In such objects, the emission of the stellar population can be the principal contribution to the optical emission \citep[see e.g.][]{massaro2015}. For these sources, the stellar population and the jet contributions are expected.
\end{itemize}

Under these considerations, we propose an approach that consists in using these \textit{a priori} expected components to reconstruct the spectra data set of each target. These components are fixed and used as input for the reconstruction, and the NMF algorithm estimates the overall contribution of each one for all the spectra. This method also solves the degeneracy caveat that can be introduced by the non-orthogonality condition of the NMF. The initial number of components is estimated with the RSS diagram. 
After a first reconstruction, a residual map is inspected to account for the success of the global reconstruction (see for instance Fig. \ref{residualmap_example}). In case an extra component is needed to explain the variability of a certain source, it can be easily identified in the residual map and added to the reconstruction. Colour/spectral index changes in the data set can be identified as a region of the residual map with a positive (negative) residual in the blue part of the spectrum, and negative (positive) residual in the red part. To quantify if a reconstruction is accurate enough or if an additional component is needed, we estimate the squared errors of each reconstructed spectrum. If more than 5\% of the whole set of spectra is above a 5$\sigma$ threshold w.r.t. the noise of each spectrum in the data set, another component is added to the reconstruction. On the other hand, if more than 95\% the data set is below this limit value, the reconstruction is considered to be satisfactory. It is important to note that, for those sources with a reduced amount of spectra, a visual inspection of the residual map is specially important to evaluate if another component is needed. In those cases, we check if the spectra showing high residuals are clustered within a time period, indicating a true deviation, while isolated cases are less significant.

Once the reconstruction is considered successful, we can estimate the amount of variability caused by each component. This can be done by combining the explained variance of the reconstruction, and the cumulative variance of each component. The explained variance is a measurement that quantifies the proportion of variation or dispersion of the data set that is being accounted by the model, and it is obtained as

\begin{equation}
\sigma^{2}_{\text{exp}} = \left( 1-\frac{\sigma^{2}_{\text{res}}}{\sigma^{2}_{\text{orig}}} \right) \cdot 100,
\label{explained_variance_eq}
\end{equation}

\noindent where $\sigma^{2}_{\text{orig}}$ is the variance of the original data set and $\sigma^{2}_{\text{res}}$ is the variance of the difference between the reconstruction and the original spectra. We compute also the percentage of variability for which each component $n$ is responsible by using the formula

\begin{equation}
\sigma^{2}_{n}(\%) = \frac{\sigma^{2}_{n}}{\sum_{n=1}^{N}\sigma^{2}_{n}} \cdot 100.
\label{variance_comp_eq}
\end{equation}

$\sigma^{2}_{n}$ corresponds to the variance of the $n$ component and $N$ is the total number of components used. Finally, with $\sigma^{2}_{n}$ and the explained variance we can calculate the variance of the original spectral data set for which each component is responsible of (see columns (7) to (10) in Table \ref{table_results}):

\begin{equation}
\sigma^{2}_{\text{exp},n}(\%) = \sigma^{2}_{n}(\%) \cdot \sigma^{2}_{\text{exp}}.
\label{variance_comp_original_eq}
\end{equation}

\subsubsection{Reconstruction of Galaxy-dominated blazars}
The reconstruction of spectra dominated by the host galaxy emission can be used as a test of our methodology, since the derived contribution of the stellar population should remain fairly constant. As established above, in the case of galaxy-dominated sources the optical spectrum is mostly contributed by the host galaxy emission plus a variable part from the jet. The RSS diagram suggests a reconstruction based in two components to reproduce the variability of these objects. The stellar emission is derived using the Penalized Pixel Fitting (pPXF) method \citep[][]{cappellari2004,cappellari2017}. In order to derive the best stellar population template we use the mean spectrum derived for each of the three objects. To better adjust the continuum shape of the spectrum and account for the featureless synchrotron contribution, we include an additive polynomial during the fit. The best fit stellar population templates correspond in all cases to an old stellar population ($\gtrsim 11\, \mathrm{Gyr}$) and metallicity slightly higher than the solar value. Similar results are found when using spectra representing the low state of each object. To model the contribution of the jet we use the polynomial function result of the pPXF to describe this continuum component.

\subsubsection{Reconstruction of BL Lac objects}
BL Lac objects are characterized by the dominance of the jet over any other component, and by an (almost) featureless spectrum. Thus, we use a reconstruction based on different PLs ($F_{\lambda} \propto \lambda^{-\alpha}$) to account for the change of slope of the spectra. These PL functions have a fixed spectral index, but can vary in flux. Thus, combinations of the different PL functions account for the spectral changes, while their flux scales are related to their relative contribution to the emission. We find that, for most of the BL Lac objects, an initial number of two components is needed for the reconstruction. Then if needed, we add more PL components, to explain periods of residual excess due to strong spectral changes.

\begin{figure*}
    \includegraphics[width=0.49\textwidth]{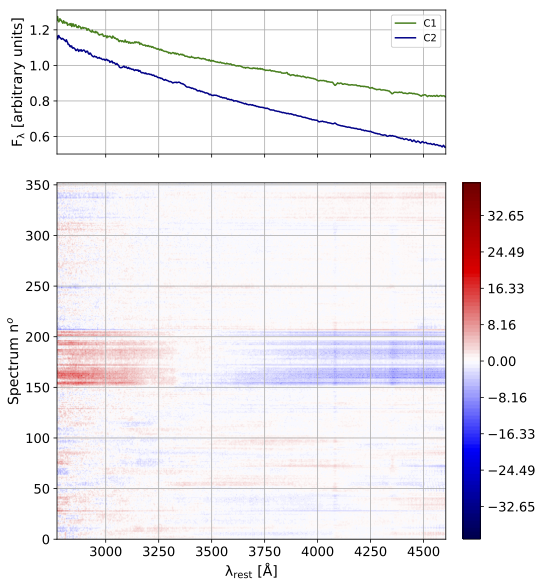}\hfill
    \includegraphics[width=0.49\textwidth]{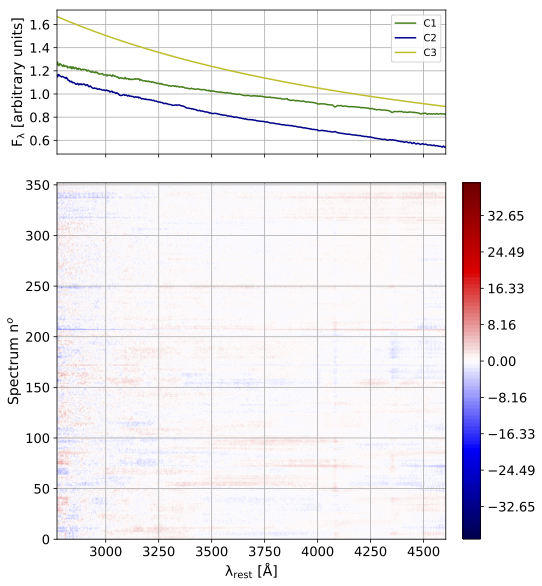}\hfill
    \caption{Residual map for the reconstruction of 3C 66A. \textit{Left:} Reconstruction with 2 components. \textit{Right}: Reconstruction with 3 components. Red and blue colours indicate the spectra for which the reconstruction is not accurate and where an extra component is needed to reproduce the data set. The top panels show the components used for the reconstruction in each case.}
    \label{residualmap_example}
\end{figure*}

The initial PL type components are estimated from the maximum and minimum spectra (computed as the mean of the percentile 90 and percentile 10 spectra, respectively). The minimum spectrum is taken as the first component, while the difference between the maximum and the minimum spectra is used as the second one. Thus, if a BL Lac shows any emission/absorption lines in its spectrum, it will appear associated with the first component. This is due to the fact that these lines will be more visible during low states. If an extra component is required when an excess appears in the residual map of a subset of spectra (usually coincident with colour/spectral index changes as shown in Fig. \ref{residualmap_example}), it is added as the median of this subset.

\subsubsection{Reconstruction of FSRQs}
FSRQs show broad emission lines in their optical spectra due to the presence of the BLR in the surroundings of the black hole. Therefore, we need to add a component that accounts for the presence of these lines in the spectra. Moreover, FSRQs often show an excess in the optical/ultraviolet (UV) part of their spectrum known as the blue bump, introduced by the accretion disk. 
The presence of the BLR and accretion disk makes the number of components  derived from the RSS representation higher than in the BL Lacs case.
The methodology applied in this work leads to a reconstruction that needs 3 or more components to model and reproduce the variability observed in FSRQs. 

In order to define the \textit{a priori} components for the FSRQs, we make use of the QSO template from \cite{vandenberk2001} to account for the emission lines in the spectra. The mean spectrum is fitted to a sum of a PL component that describes the emission of the jet, and the QSO template. Then, the final template, used as input to the NMF, is derived as the difference between the mean spectrum and the fitted PL. In this way, the template derived for each source matches better the intensity/profile of the emission lines of each FSRQ. The derived contribution of this component is expected to remain mostly constant. However, if long term variations are observed, it may indicate intrinsic variation of the BLR+accretion disk emission. The fitted PL is used as the second component. Finally, an additional PL is added to model changes of the spectral index. This third component is estimated from the subset of spectra with significant spectral change as for BL Lac objects, after subtracting the estimated QSO template contribution.

\section{Results}\label{sec5}
In this Section and Section \ref{sec6} we report the general results and the interpretation for all the sources of each type included in the analysis. A detailed discussion on some individual sources of special interest can be found during the following subsections and some notes on the rest of the sample are found in the Appendix \ref{appendix}. The results of the reconstructions are summarized in Table \ref{table_results}.

\subsection{Galaxy-dominated sources}
\begin{figure*}
\centering
\includegraphics[width=\textwidth]{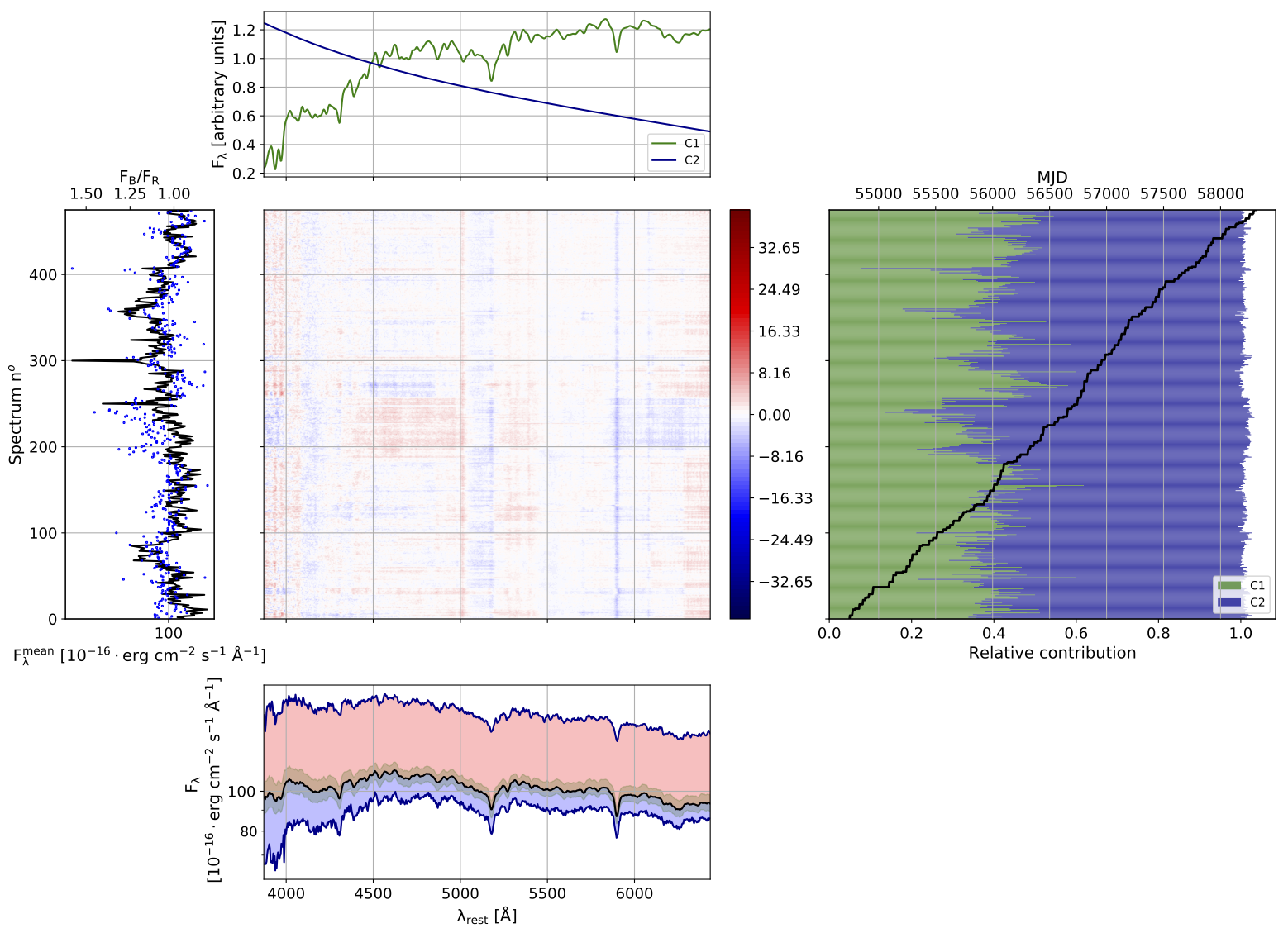}
\caption{Results of Mkn 501. \textit{Top:} Normalized components. \textit{Middle:} Residual map. The colour scale is estimated in terms of the noise of each spectrum. \textit{Bottom:} Median (black), minimum and maximum (dark blue) spectra. The variability range between the maximum (minimum) and the median spectra is represented by the red (light blue) contour, and the variability range between the percentiles 25 and 75 with the green contour. \textit{Right:} Relative contribution of each component. The black line represents the MJD of each spectrum. \textit{Left:} Mean flux and colour of each spectra. Blue dots correspond to the colour of each spectrum.}
\label{mrk501_figures}
\end{figure*}

Galaxy-dominated blazars are characterized by a much smaller variability w.r.t. BL Lac objects and FSRQs. Their fractional variability in the optical band is significantly smaller compared to the latter blazar types (see Fig. \ref{fvar_plot}).
This fact, along with the dominance of the host galaxy over the other parts of the AGN, makes them appropriate targets to test the reliability of the adopted methodology, since the expected stellar contribution should remain constant. The three blazars dominated by the stellar emission of the host galaxy were reconstructed with 2 components: the stellar emission template to account for the bright host galaxy, and the PL component responsible for the synchrotron emission of the jet. We find that we are able to reconstruct accurately the emission and variability of these blazars using these two simple components. These objects show a mild BWB behaviour in their colour evolution with the flux (see Fig. \ref{variability_galdom}). This trend comes from the second component of the reconstruction, associated to the non-thermal synchrotron emission of the jet. In the following subsection we discuss the results of Mkn~501 as an example of this blazar type. The discussion about H~1426+428 and 1ES~2344+514 can be found in the Appendix \ref{appendix_galdom}.


\subsubsection{Mkn 501}\label{sec5.1.1}
The results of the NMF reconstruction of Mkn~501 can be seen in Fig. \ref{mrk501_figures}. This blazar displays a very low variability in optical frequencies  ($F_{var}=10\%$) that is also consistent with the low variability reported in the 4LAC-DR2 \textit{Fermi}-LAT catalog. This is a consequence of Mkn~501 being a high-synchrotron peaked (HSP) blazar, so the variability is stronger after the SED bumps (X rays and VHE $\gamma$ rays). The host galaxy of Mkn 501 contributes with 40\% of the overall optical flux (see Fig. \ref{mrk501_figures}). The component C1 corresponds to a galaxy with an old stellar population and its derived mean flux contribution does not show significant variation. C2 is a PL with a moderate slope and it is responsible for the observed spectral variability. It also accounts for the observed colour variations, in all cases BWB. The colour of this blazar shows two long-term BWB trends due to the same behaviour observed in the second component of the reconstruction, associated to the jet. The presence of different BWB trends due to different states of this sources was also found by other studies \citep[e.g.][]{xiong2016}. We also note by looking at Fig. \ref{mrk501_figures} that in the period between MJD 56400 and MJD 56800 there is a small residual excess. This period is coincident with a rapid bluer trend, not accompanied however by a flux increase. This may indicate that a third component may be needed to reconstruct this small subset. 

\subsection{BL Lac objects}
\begin{figure*}
\centering
\includegraphics[width=\textwidth]{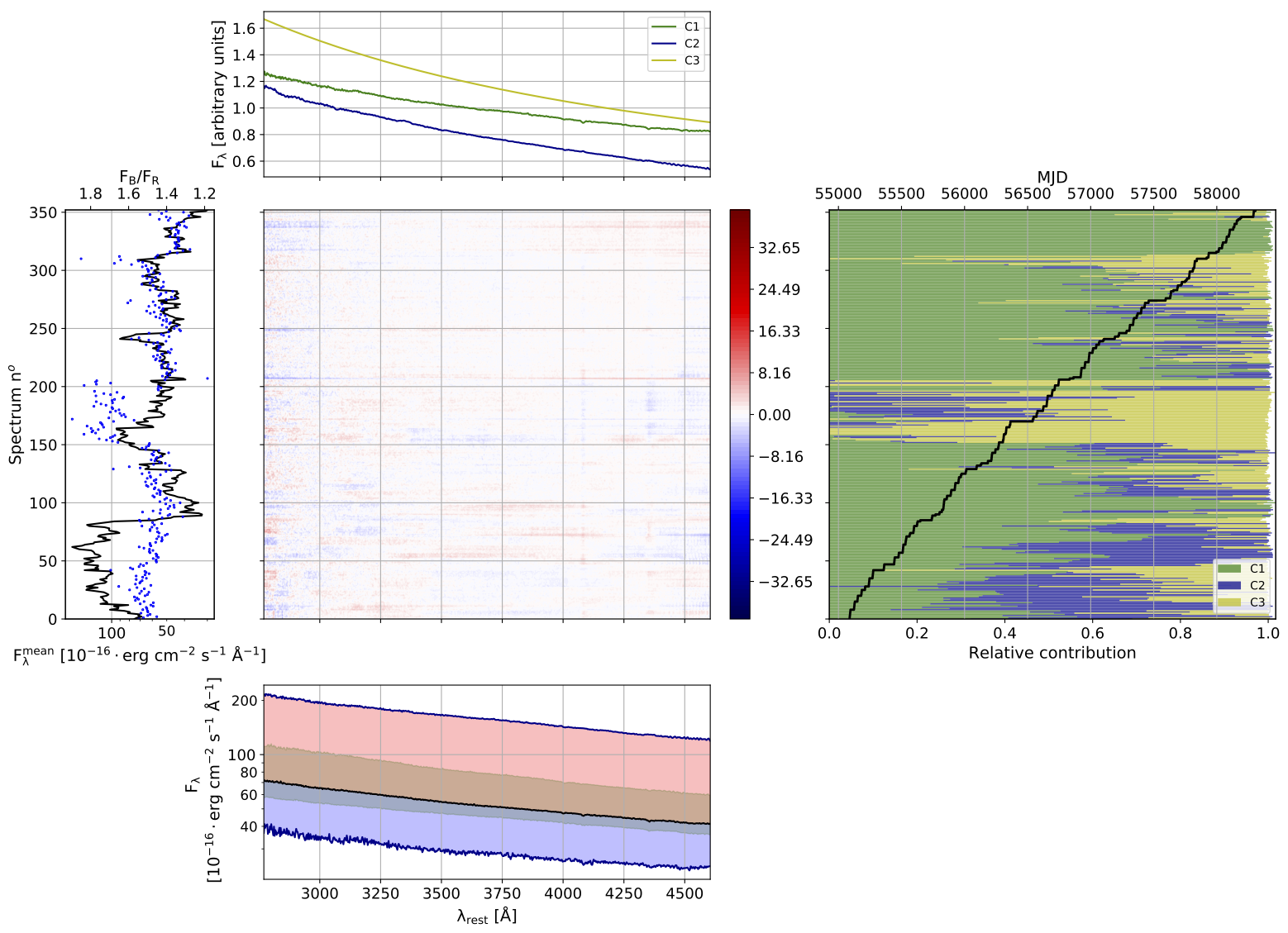}
\caption{Results of 3C 66A (same description as Fig. \ref{mrk501_figures}).}
\label{3c66a_figures}
\end{figure*}

BL Lac objects are reconstructed using 2 to 4 components depending on the amount of spectral variability of the source. These components are able to explain more that 99\% of the variance of the original data set and thus, to model accurately the variability observed in these blazars. Typically the different components obtained from the analysis dominate the emission as the colour changes, reproducing also the spectral evolution of the source. The BL Lac objects used in this analysis are also characterized by a BWB behaviour (as shown in Fig. \ref{variability_bllacs}), followed by almost all the targets of this type (see Fig. \ref{variability_bllacs}). 

Specially remarkable are the cases of the BL Lac objects AO~0235+164, OJ~287 and S5~0716+714. These objects have shown colour behaviours different to the general BWB trend associated to BL Lacs. In addition, OJ~287 is an interesting blazar since it is thought to host a binary system of two supermassive black holes. Furthermore, the appearance of transient components propagating along the jet can be identified in sources such as 3C~66A. A detailed discussion on the results of these sources is included in the following subsections. Mkn~421 is also reported in this section as a prototype of the BL Lac blazar type. An overview of the rest of the BL Lac objects can be found in the Appendix \ref{appendix_bllacs}.

\subsubsection{3C 66A}
The reconstruction of the available spectra is done using 3 PL components (see Fig. \ref{3c66a_figures}). The fraction of explained variance attributed to each component is nearly equal, corresponding to about 33\%. The first component is the flattest one and its contribution to the total flux is above one third for about 80\% of the time, contributing even more when the source is at its low state. The second component has an intermediate slope and appears dominant when the source is at relatively high state, although its contribution is not exceeding 60\%. The third component shows the steepest slope and dominates the spectra ($\sim 90\%$) for the period between MJD 56250--56700 and also shortly in the period MJD 57700--57750. The results of the reconstruction are presented in Fig. \ref{3c66a_figures}.

This source presents an overall BWB behaviour, which has been also reported in previous studies \citep[e.g.][]{meng2018}. This trend can be related mostly to the second component. This component clearly modulates the BWB behaviour, as it increases more during the bluest spectra, and fades by one order of magnitude when the spectra become redder. On the other hand, the first component was found to be roughly constant and achromatic. Moreover, the third component also shows this BWB trend. However, it is representative only for a reduced amount of spectra. This reduced subset is also characterized by a significant change of the spectral index and colour, disconnected from the overall behaviour of the rest of the data set. This can be interpreted as a transient component propagating along the jet that dominates the emission during a short period of time compared to the time span covered by the full data set \citep{wang2021}. 


\begin{figure*}
\centering
\includegraphics[width=\textwidth]{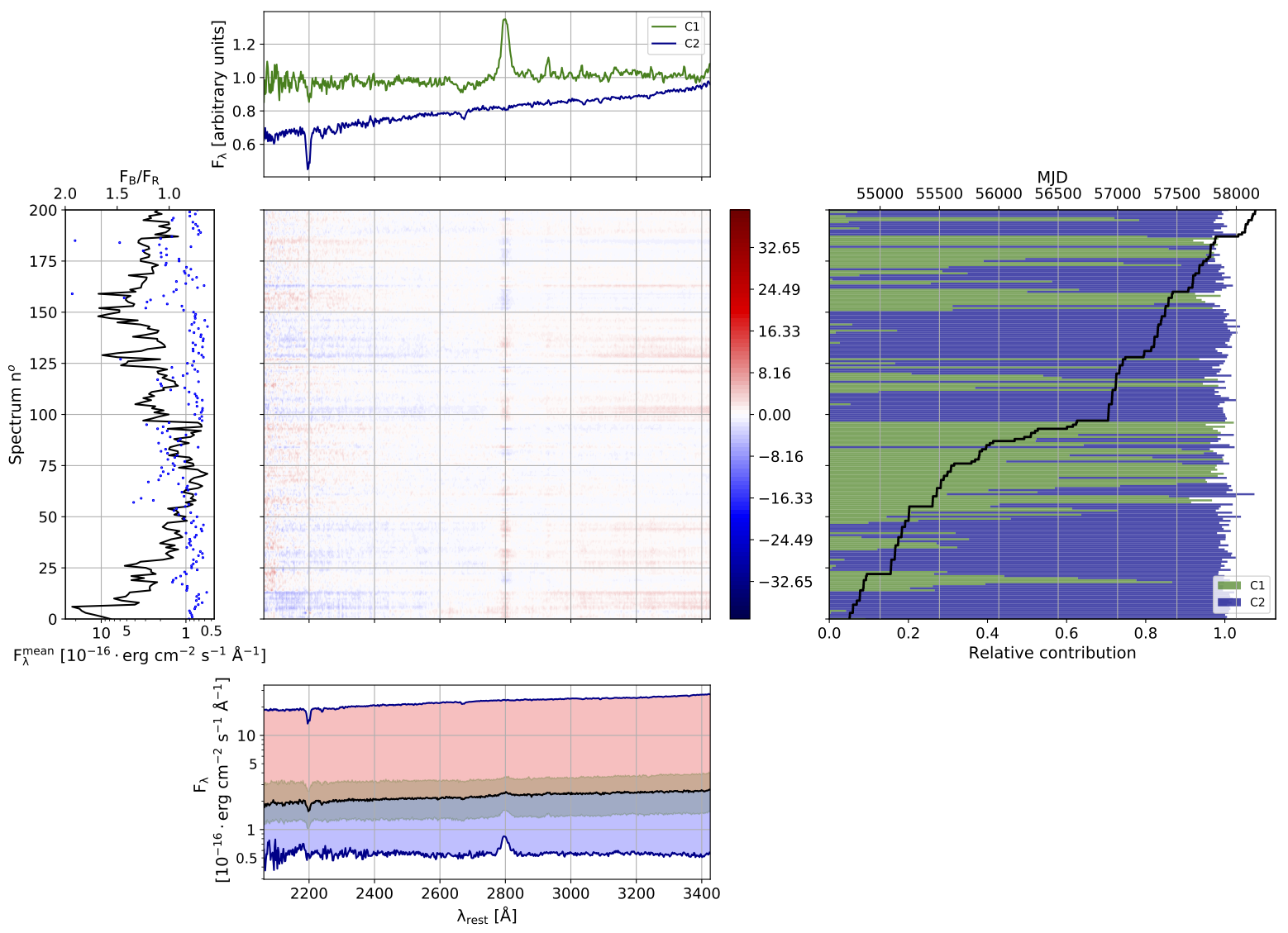}
\caption{Results of AO 0235+164 (same description as Fig. \ref{mrk501_figures}).}
\label{ao0235_figures}
\end{figure*}

\subsubsection{AO 0235+164}
The results of the NMF analysis of AO~0235+164 are shown in Fig. \ref{ao0235_figures}. The RSS diagram indicates that 2 components are enough to explain most of the observed variance. An inspection of the residuals also reveals that two components are sufficient to have a good reconstruction of the spectra with the adopted methodology. The flux of this source is nearly constant during its low state, increasing significantly during two outburst events in which the second component becomes the brightest. The first component is nearly flat and it is dominant when the source is at low state (35\% of the spectra) and accounts for only 15\% of the variance. It also shows the broad MgII$\lambda 2800$ line, that becomes clearly visible during low activity states. The second component is very similar to a red PL, being dominant for about 65\% of the spectra and it accounts for 80\% of the variance. The combination of these two components accounts for the double-trend colour variation with flux (see Fig. \ref{variability_bllacs}), more similar to that observed in FSRQs. 
The brightest spectra show a very mild BWB trend, almost flat, while the low state of the source displays a RWB trend. We note that the PL indices of these components show a negative value, being the only ones in our sample (see Table \ref{table_results}). This is likely due to the extinction produced by the intervening absorber at z=0.524 \citep[A$_{V}$=0.99 derived by][]{junkkarainen2004,raiteri2005}, which has not been corrected for here.

Previous studies have reported for this source a FSRQ-like behaviour, with RWB trends in the colour, contrary to the typical BWB relation observed in BL Lacs \citep{raiteri2014}. \cite{bonning2012} detected a more complicated colour evolution, with two different states depending on the observed $\gamma$-ray intensity. In addition, the optical spectra of AO~0235+164 taken during the low states displayed by this object reveal a prominent and broad emission line, that disappears during the flaring states. These features, more common in FSRQs, may indicate that AO~0235+164 could be a transitional blazar. In this framework, the two components obtained by the NMF reconstruction can be interpreted as a first component with the contribution of a weak BLR, responsible for the FSRQ-like behaviour, and a second component that explains the emission from the jet and it is responsible of the BL Lac-type FWB behaviour and the flaring states of this object \citep{raiteri2006}. Other authors have also proposed the existence of two components to reproduce the variability of this blazar, as well its colour behaviour \citep[see e.g.][]{raiteri2006,cellone2007}. \cite{bonning2012} also consider the possibility of the presence of an accretion disk contributing to the optical-IR emission in this source.


\subsubsection{S5 0716+714}
\begin{figure*}
\centering
\includegraphics[width=\textwidth]{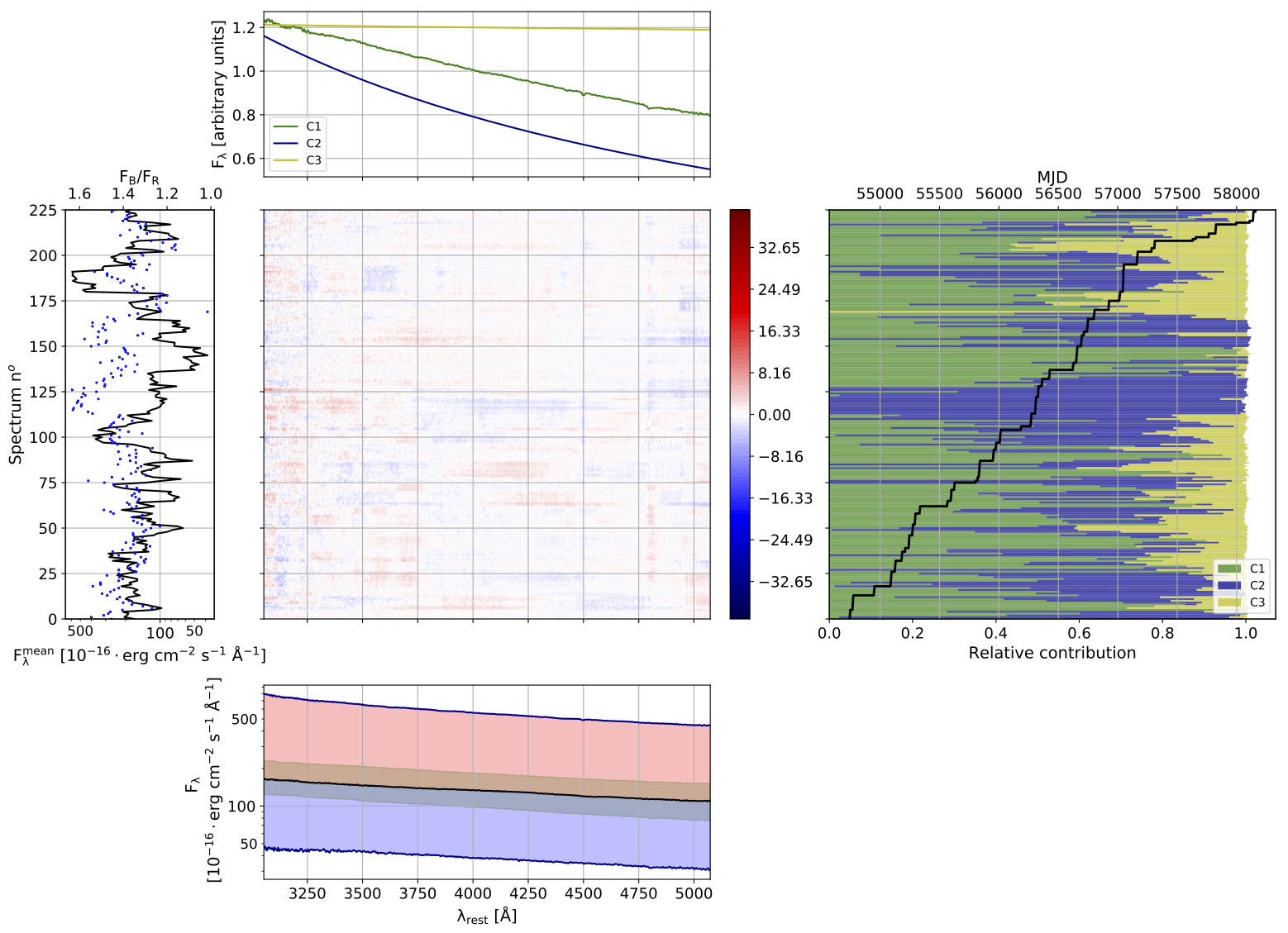}
\caption{Results of S5 0716+714 (same description as Fig. \ref{mrk501_figures}).}
\label{s5_0716_figures}
\end{figure*}

The spectral variability of this blazar is reconstructed with 3 components that reproduce 99.98\% of its variability. The results of this reconstruction can be seen in Fig. \ref{s5_0716_figures}. 
C1 contributes with more than one third of the emission in about 2/3 of the spectra.
Its contribution to the $\sigma^2_{exp}$ is around 30\%. C2 is the bluest component and it is responsible for about 65\% of the $\sigma^2_{exp}$. It shows the larger contribution to the mean flux during the decline of bright flares. C3 is an almost flat PL and it is responsible for about 6\% of the $\sigma^2_{exp}$, indicating that it remains basically constant. Its contribution to the mean flux is large (up to 50\%) when the C2 contribution vanishes, otherwise its contribution stays around 20\%.

S5 0716+714 does not show overall any clear colour trend (see Fig. \ref{variability_bllacs}). However, at shorter time scales it shows the typical BWB trend of BL Lacs, associated with the bright flares observed in the left panel of Fig. \ref{s5_0716_figures}. The peculiarity about its colour evolution is that this BWB displays three different trends with three slopes, associated to three different brightening periods: a very steep BWB behaviour for the faintest spectra; a second trend with higher flux variations and lower colour change, and a third trend with lower colour variability that corresponds to the highest flare observed in this blazar during the 10-year monitoring (MJD 57046). The existence of multiple trends in the colour evolution can be an indicative of different electron populations responsible for the synchrotron emission, or can be associated with different activity levels of the source \citep{xiong2016}. In terms of the individual components, these trends are related to the second component, which is the bluest and showing the different trends identified in the overall behaviour. On the other hand, the first and third component vary uncorrelated with the colour of the spectra.


\subsubsection{OJ 287}
\begin{figure*}
\centering
\includegraphics[width=\textwidth]{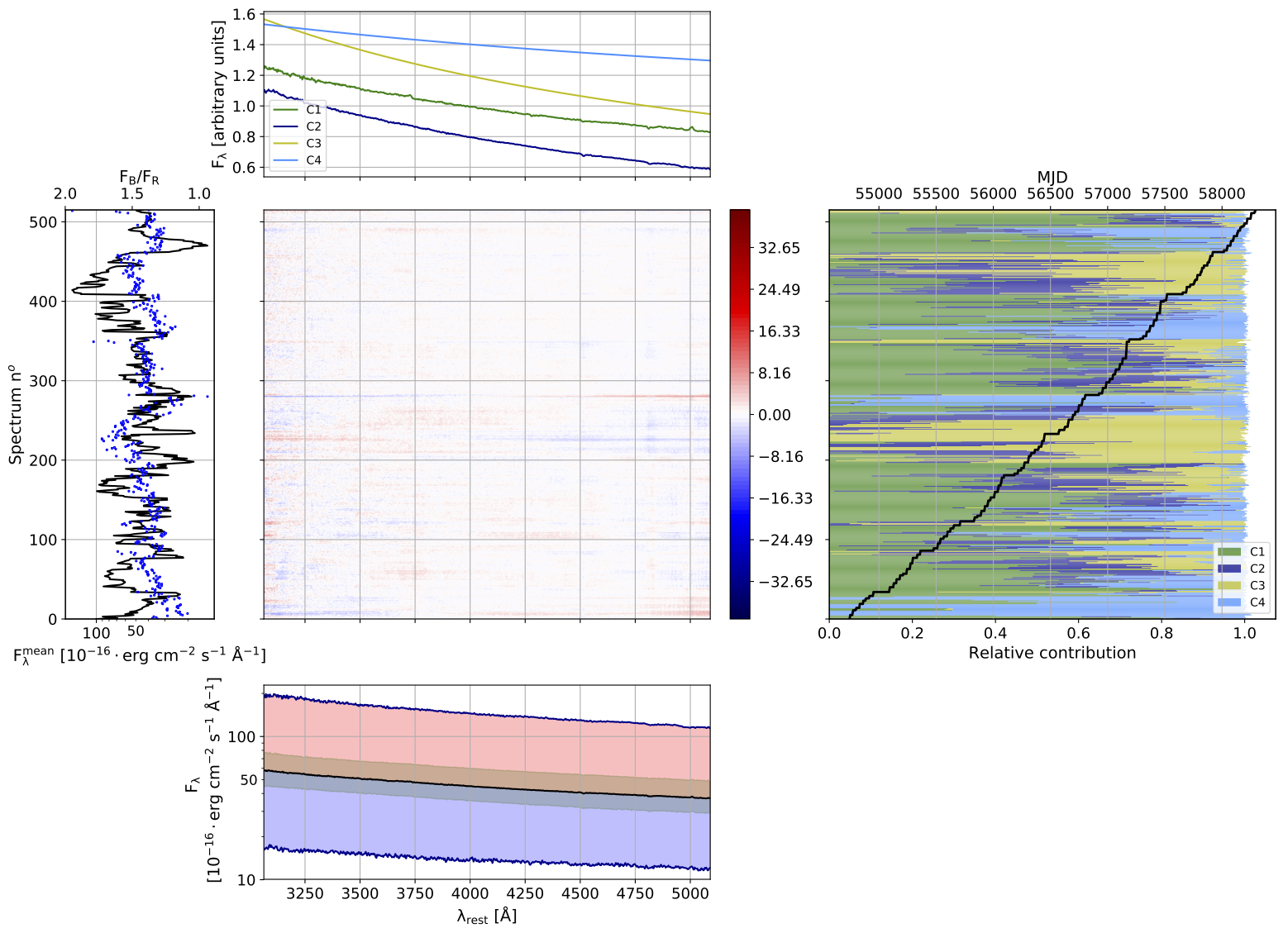}
\caption{Results of OJ 287 (same description as Fig. \ref{mrk501_figures}).}
\label{oj287_figures}
\end{figure*}

The RSS indicates a total of 2 components. However an inspection of the residual map reveals additional components. The best reconstruction with our approach is obtained using 4 components, which are PLs of spectral indices ranging from 0.5 to 1.2. The results of this reconstruction are shown in Fig. \ref{oj287_figures}. All components contribute significantly to the $\sigma^2_{exp}$: C1 and C3 have a higher contribution, reaching about 28\% and 35\% of the $\sigma^2_{exp}$, respectively, whereas C2 and C4 contribute the least, around 15\% of the variance. C1 contributes more than 30\% to the total flux for about one third of the spectrum set and it dominates basically when the source is at lower states. C3 shows the steepest slope and it is dominant within the periods MJD 56400-56650 and 57650-57900. C4 has the shallowest slope, i.e. the reddest color, and it contributes mostly during the periods MJD 54750-55000 and 57250-57400. C2 has a relative contribution of 30\% of the flux associated with the higher states of the source, and coincident when C4 is not contributing significantly. OJ~287 exhibits large colour variability during the monitoring period. However, contrary to the common behaviour observed in BL Lac type blazars, this variability appears uncorrelated with flux changes. This can be understood in terms of the component slopes, which are very similar for C1, C2 and C3, which have the highest contributions. The exception to this behaviour happened during the bright flare occurring between MJD 57650 and MJD 57900, showing a very interesting behaviour in the optical regime. This episode exhibits a FWB colour trend. The activity appears to start around MJD 57250, when the spectrum becomes bluer. During this episode the component C3 contributes almost 100\% and the flux increases roughly achromatically. During this period the spectra can be reconstructed by a combination of C1 and C4. At MJD 57700 the decrease of flux starts and the spectrum is formed by C2 and C3 as basic contributors. After MJD 58000 the flux returns to a low state and the spectra are reconstructed by a large contribution of C1, plus C2 and C4 with lower contributions. The fact that these three components have similar indices and relative contributions may also indicate that they represent the component of the jet, but varying the energy of the electrons responsible for the emission.

OJ~287 is a well known BL Lac object, thought to host a binary supermassive black hole system, showing major outbursts approximately every 12 years \citep{sillanpaa1988}, with a structure of two peaks separated by $\sim1-2$~years. Several models have been proposed to explain the behaviour observed in this object. The flare observed in our data is predicted by the binary model \citep{valtonen2016,kapanadze2018}. A detailed discussion of the different binary black hole models and their limitations can be found in \citep{sillanpaa1996}. \cite{sillanpaa1988} suggested a model where the flares are produced by the secondary black hole crossing the accretion disk of the primary, inducing tidal perturbations in the disk. This variant is able to predict the flares and reproduce their colour stability, as the energy production mechanism is expected to remain unchanged. However this binary BH model is not able to explain the appearance of a double peak during the flares. On the other hand, the variant proposed by \cite{lehto1996} is able to accurately predict the consecutive outbursts every 12 years and reproduce the double-peak structure. However, they fail when explaining the achromatic development of the flares, since in this model, the energy production mechanism is expected to change during these events, as the flares in this model are produced by thermal outburst in the disk. Another plausible variant, suggested by \cite{villata1998}, is based on strongly bent jets emerging from both BHs, in which the periodicity is caused by the beaming effects. This model is able to explain both the double peak structure of the flares and the achromatic brightening, as long as the energy production is the same in both jets. Alternatively, \cite{bonning2012} interpret the achromatic brightenings in blazars as pure geometrical effects, which could explain the behaviour observed in this object. The components found in our NMF decomposition do not indicate the presence of a thermal component, favouring those models in which the emission is coming from the relativistic jet.



\begin{figure*}
\centering
\includegraphics[width=\textwidth]{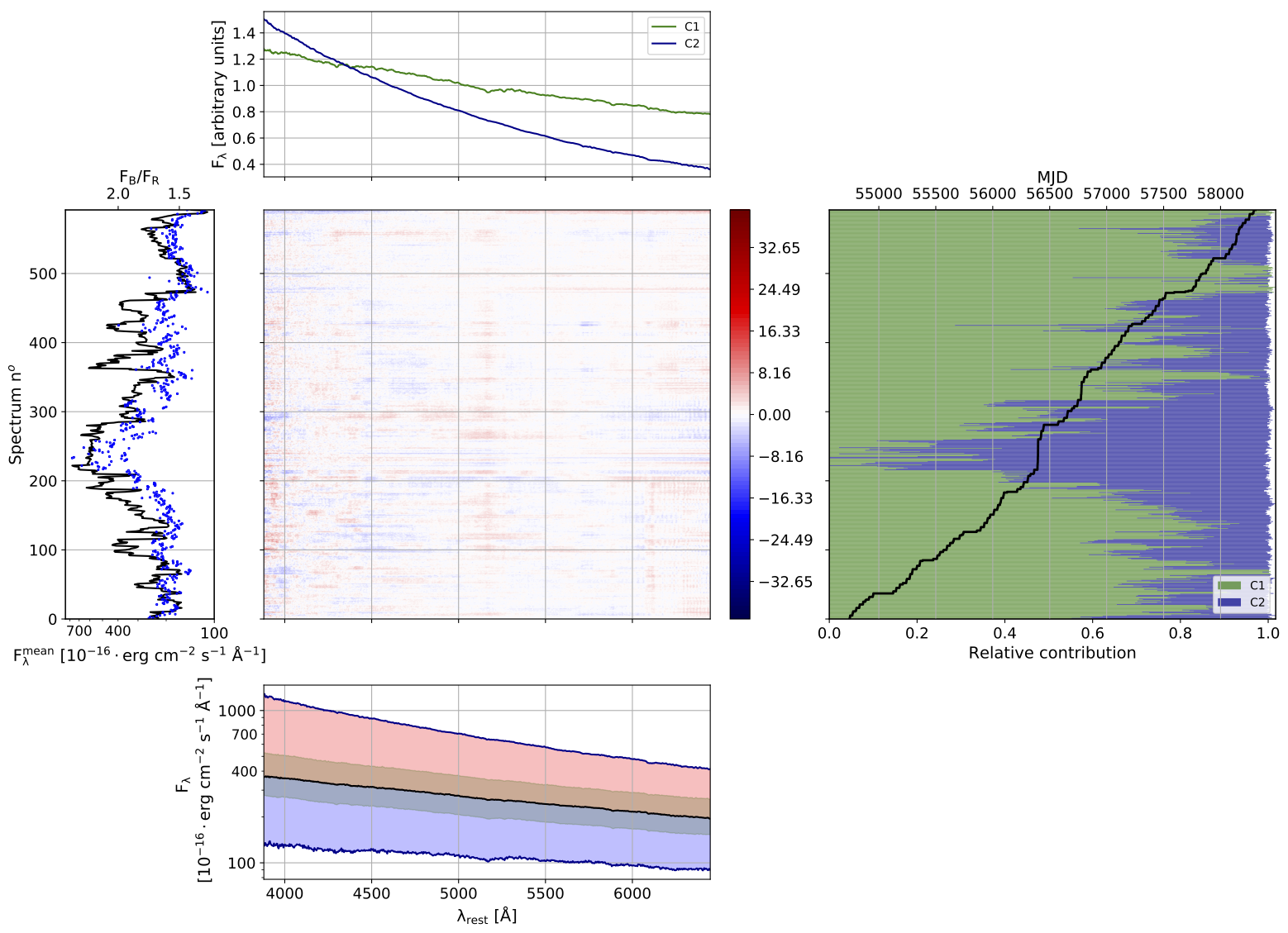}
\caption{Results of Mkn 421 (same description as Fig. \ref{mrk501_figures}).}
\label{mrk421_figures}
\end{figure*}

\subsubsection{Mkn 421}

Mkn 421 is reconstructed by 2 PL like components to account for the emission of the powerful jet, as shown in Fig. \ref{mrk421_figures}. The colour-flux diagram shows a clear BWB trend, and during the brightest state shows a large blue color increases, without almost no flux change, as shown in Fig. \ref{variability_bllacs}. The first component, C1, contributes above 50\% to the mean flux most of the time, except for the period MJD 56250-56750, when the component C2 becomes fully dominant. This corresponds to the bluest spectra shown by Mkn~421, which are consistent with the BWB behaviour, already reported by \cite{carnerero2017}. The first component does not vary dramatically, it accounts for $\sim 15\%\, \sigma^2_{exp}$ and its contribution remains constant even when colour changes. C2 accounts for most of the variability of this source ($\sim 85\%\, \sigma^2_{exp}$). This component is also responsible for the BWB behaviour of this blazar (see Fig. \ref{variability_bllacs}). The two-component decomposition of the data set is supported by a two-zone scenario used in several studies for this blazar in the past \citep[see e.g.][]{aleksic2015}. 


\subsection{FSRQs}
The NMF reconstruction of FSRQs requires at least 3 components to explain more than 99\% of the observed variability. This difference with the BL Lac objects is due to the presence of a bright BLR in FSRQs. 
We observe a RWB trend in most of the FSRQs, historically associated with this blazar type (see Fig. \ref{variability_fsrqs}). We also identify a FWB behaviour during bright outburst events, when the source overcomes a certain flux. For three of the FSRQs, we also detect a BWB following the colour flattening: PKS~0420-014, 3C~454.3 and CTA~102. Furthermore, among the FSRQs studied here there are two objects with a peculiar colour variability: 3C~273 and 3C~279. The former shows a long-term BWB evolution in its colour variations, with different trends at shorter time scales. Such behaviour was already reported in the past for 3C~273 by \cite{dai2009} in both intraday and long term time scales. The latter on the other hand, shows a much more complex behaviour. Furthermore, some of the FSRQs studied here (e.g. PKS~1222+216) show rather steep and less variable components with no correlation with the overall flux that can be associated to a bright accretion disk. The results of 3C~273, 3C~279, 3C~454.3, CTA~102 and PKS~1222+216 are discussed within the following subsections. The rest of the FSRQs of the sample are included in the Appendix \ref{appendix_fsrqs}.


\subsubsection{PKS 1222+216}
\begin{figure*}
\centering
\includegraphics[width=\textwidth]{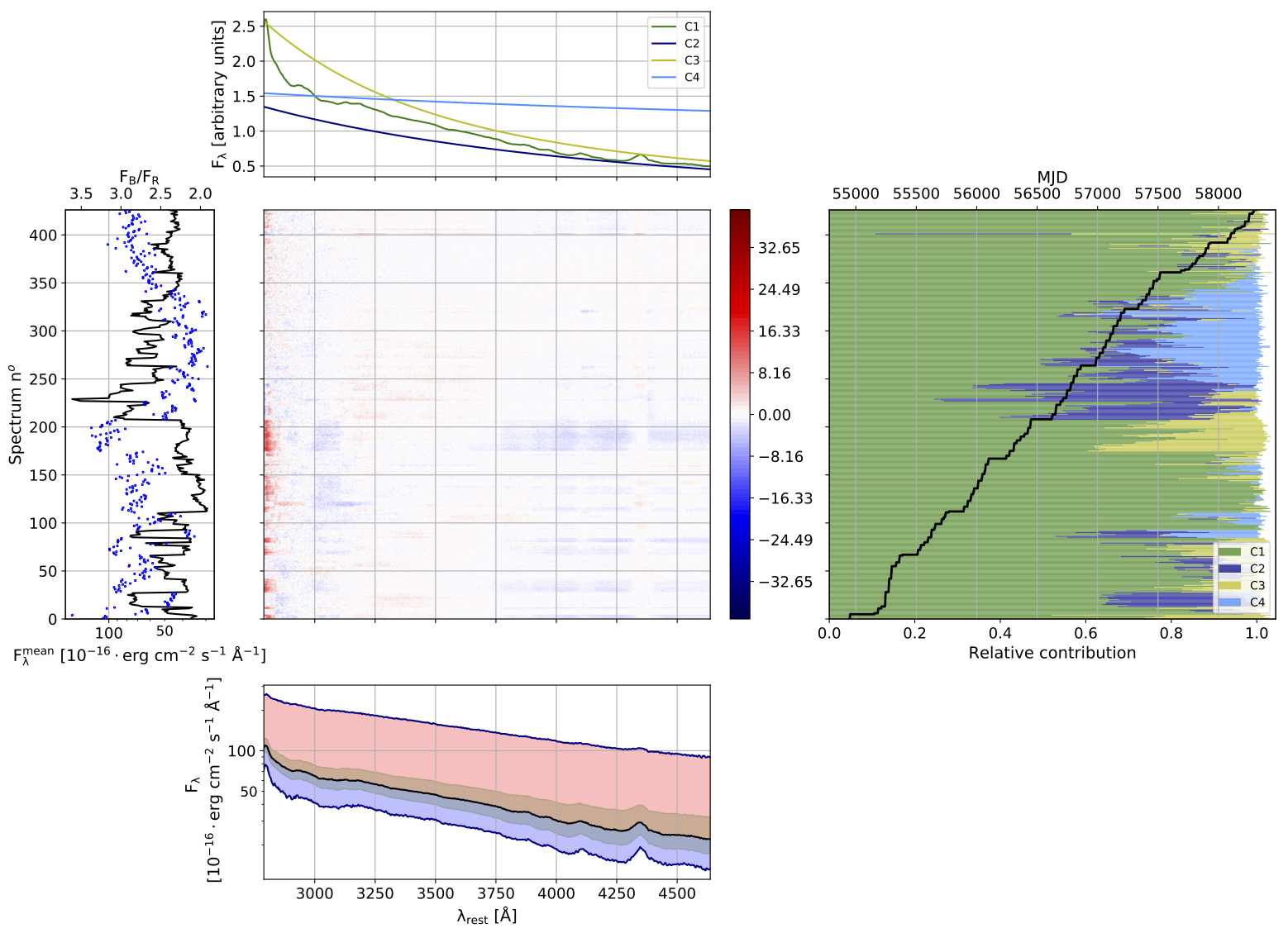}
\caption{Results of PKS 1222+216 (same description as Fig. \ref{mrk501_figures}).}
\label{pks1222_figures}
\end{figure*}

The reconstruction of the FSRQ PKS~1222+216 is based on a total of 4 components. The results obtained for this blazar are shown in Fig. \ref{pks1222_figures}. The first one, associated to the BLR, represents the higher contribution ($\gtrsim 80\%$) to the optical emission of this blazar, except for the period between MJD 56250 to 57250. The component C1 shows some of the HI Balmer lines and at the blue end, the red wing of the MgII 2800. C2 contributes about 50-60\% for the period MJD 56600--56750, coincident with a large flare. It is the most variable component, accounting for 70\% of \explvar. It also has a relatively steep slope ($\alpha=1.70$). On the contrary, C4 shows an almost flat slope ($\alpha=0.52$). This component becomes important during the decline of the big flare, starting around MJD 56760 until MJD 57400. The component C3, much steeper than C2 and C4 ($\alpha=3.65$), shows the largest contribution in the period MJD 56250--56400. It remains relatively constant during this period (4\%~\explvar). This fact and its very steep slope may be indicative of emission coming from the accretion disk in the form of the Big Blue Bump, similarly to the case of 3C~273 (see Section \ref{sec5.3.2} below).

The overall colour behaviour of PKS~1222+216 reveals a RWB trend, followed by a hint of the colour saturation observed in other FSRQs. This RWB evolution is caused by the combination of C2 and C4, also showing this same trend. These components become dominant when the source enters a high state. At low states, C3 dominates over C2 and C4, remaining mostly constant. This is compatible with its association with the bright accretion disk, suggested above. When the synchrotron emission, explained through C2 and C4, equates the blue bump of the accretion disk (C3), a flattening of the RWB trend and posterior colour saturation is observed, as the jet contribution overcomes the accretion disk one.



\subsubsection{3C 273}\label{sec5.3.2}
\begin{figure*}
\centering
\includegraphics[width=\textwidth]{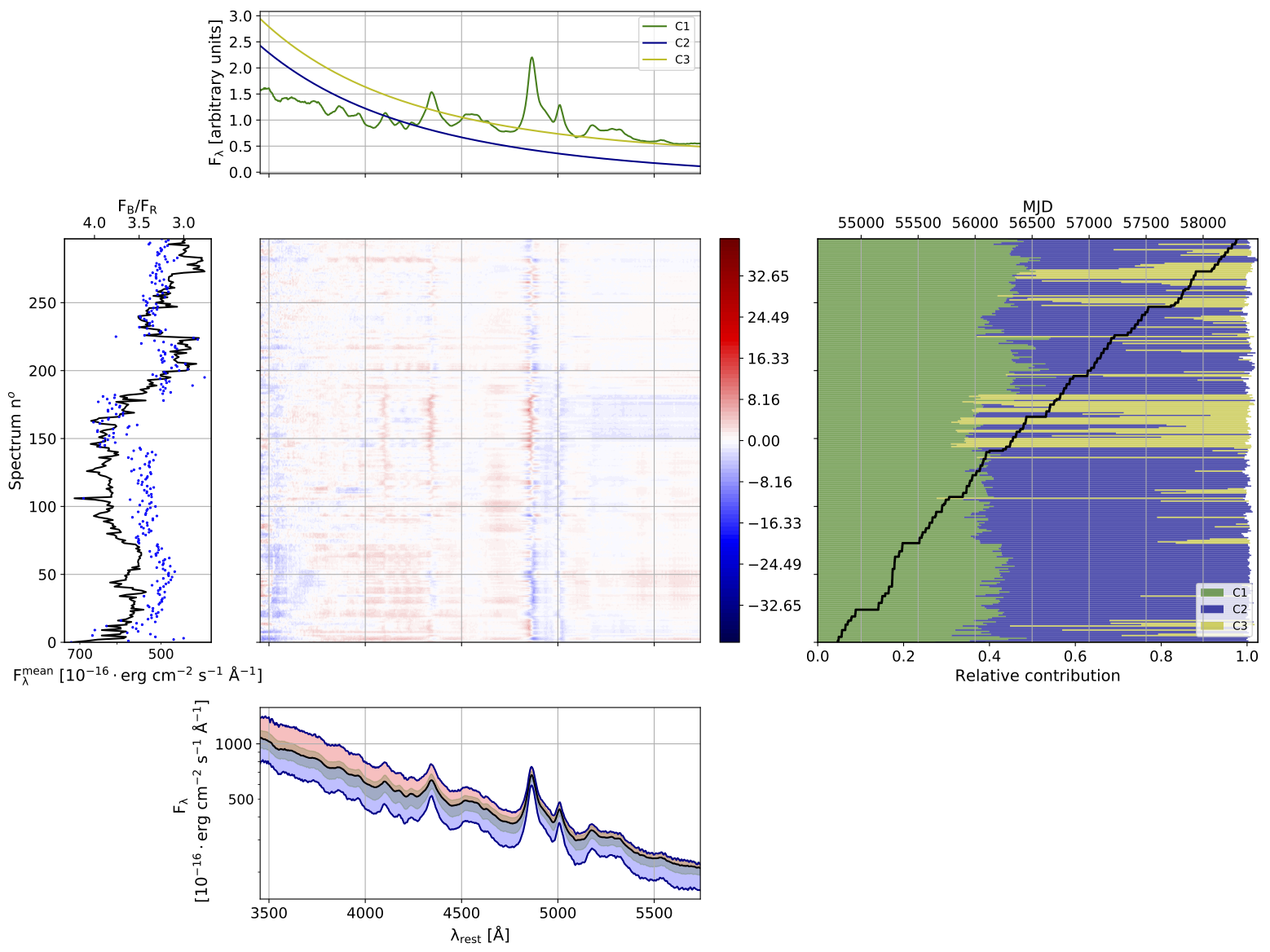}
\caption{Results of 3C 273 (same description as Fig. \ref{mrk501_figures}).}
\label{3c273_figures}
\end{figure*}

The optical spectra of this source do not show much variability ($F_{var} \sim 10\%$). Our NMF decomposition consists of 3 components accounting for about $97\%$ of $\sigma^2_{obs}$. The NMF reconstruction of this source is shown in Fig. \ref{3c273_figures}. The first component, C1, is associated with the BLR and varies very little ($\lesssim 1\%\sigma^2_{exp}$), although its contribution to the optical flux is around 40-50\% over the whole period. The other two components, C2 and C3, contribute almost equally to the $\sigma^2_{exp}$ and are approximated by very steep blue PLs. C2 contributes among 50-60\% to the total flux for about 75\% of the spectrum set. C3 appears only for certain periods, e.g. MJD 56250--56750. During these periods, C3 becomes important when C2 becomes negligible. \cite{raiteri2014} modelled the SED in the optical near-IR range by adding an extra black-body component attributed to a bright accretion disk, apart from the QSO template. Our components C2 and C3 may correspond to this black-body emission coming from the accretion disk. These two components may be associated to two different states of the disk, corresponding to variations in the accretion rate \citep{li2020}. 

Contrary to the general behaviour of the FSRQs studied here, this blazar displays a clear BWB long term evolution. This makes 3C~273 the only blazar of the sample with such behaviour. \cite{dai2009} have also observed this trend in the past, in both intraday and long term time scales. While we do not have the temporal coverage to evaluate the intraday colour variability, we do see different BWB slopes on time scales shorter than a year.



\subsubsection{3C 279}
\begin{figure*}
\centering
\includegraphics[width=\textwidth]{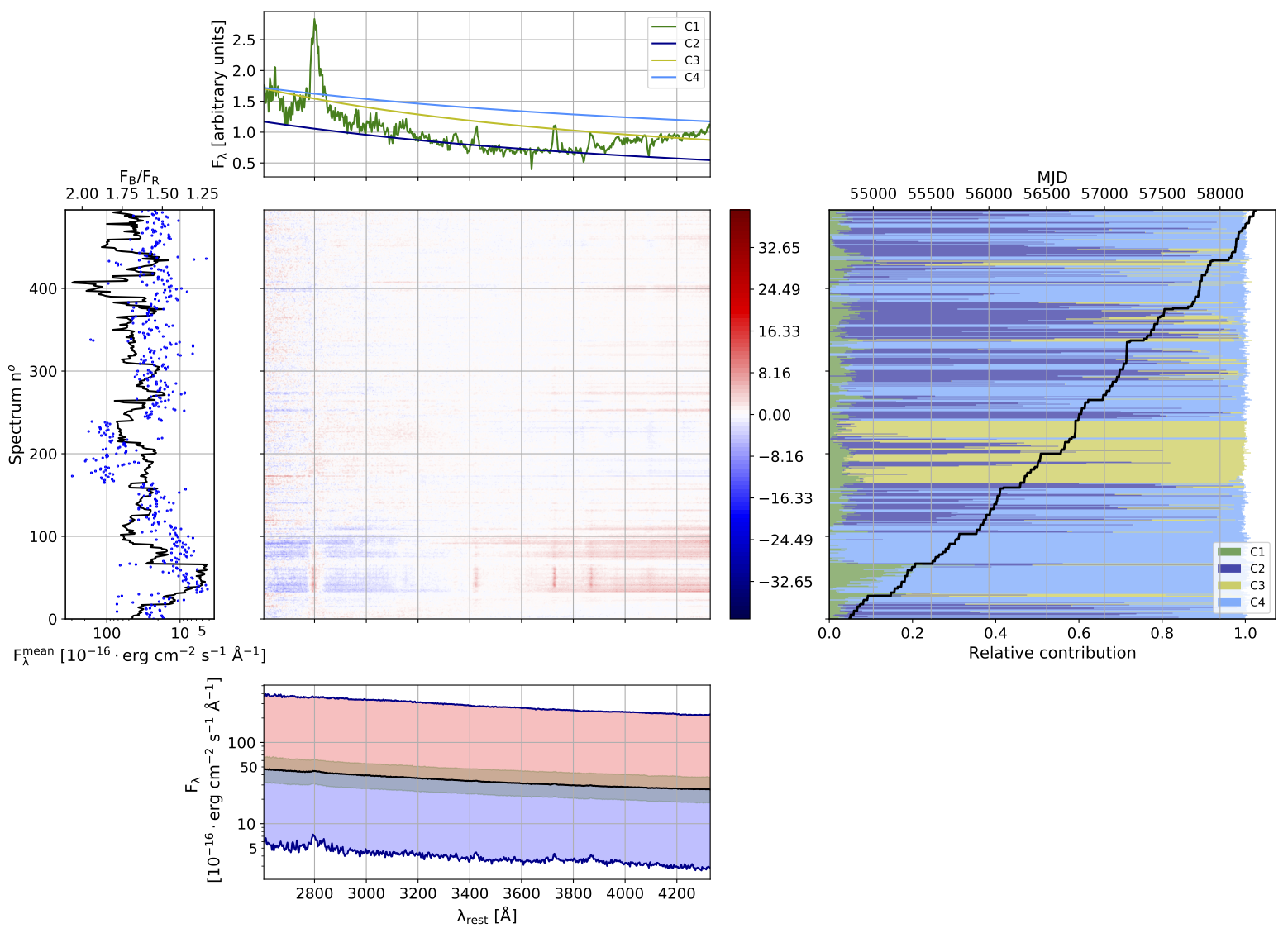}
\caption{Results of 3C 279 (same description as Fig. \ref{mrk501_figures}).}
\label{3c279_figures}
\end{figure*}

The reconstruction of 3C~279 requests 4 components that are able to model 99.91\% of its variability during this 10-year period. The results of the analysis of 3C~279 can be seen in Fig. \ref{3c279_figures}. The NMF analysis on this source reveals a very faint BLR component (C1), which does not contribute more than 5-10\% to the mean flux. Weak emission lines (MgII 2800, [OII]3727 and [NeIII]3869) become evident during the low state of the source. Its contribution to the $\sigma^2_{exp}$ is very low ($\sim0.01$), indicating that it remains constant in the whole spectrum set. When 3C~279 displays an enhanced state, this component remains almost invisible due to the high dominance of the jet. The component C2 is responsible for about 33\% of the $\sigma^2_{exp}$. It is a PL showing an intermediate slope compared to C3 (steepest) and C4 (flattest). Its relative contribution to the mean flux of the spectra ranges from 50\% to 90\%, excluding the period from MJD 55200 to 55900, when C4 contributed  most ($\sim$90-100\%). The component C4 becomes dominant over shorter periods when C2 diminishes. C4 is the most variable, accounting for about 48\% of the $\sigma^2_{exp}$ and shows a slope flatter than C2. 
The component C3 is the PL showing the steepest slope and accounts for about 20\% of $\sigma^2_{exp}$. It is responsible of the bluest spectra, occurring in the period of MJD 56250--56750. We note that despite the goodness of the reconstruction, for the minimum spectra there is still some residual that is not being reconstructed. This may be caused by the higher noise and fluctuations of this subset of spectra.



Contrary to the bulk of the FSRQ population, 3C~279 does not show any clear long-term colour correlation in the optical domain. Moreover, clear colour trends appear at shorter time scales, corresponding to flare events or fast flux variations. This result was already reported by \cite{bottcher2007} using optical data from the WEBT Collaboration from January to April 2006, or \cite{isler2017} with data from the SMARTS telescope. Furthermore, very mild correlations were found in optical wavelengths between the colour index (B-V) and the V magnitude by \cite{kidger1990}, who also reports strong positive correlations between colours and magnitudes in the near-IR bands. The same weak correlation at optical frequencies was also reported by other authors \citep[e.g.][]{agarwal2019}.
In addition, \cite{isler2017} claimed RWB, FWB and BWB trends as the emission of the jet increased and overcame the contributions of the accretion disk and the BLR. A similar behaviour was claimed for this source by \cite{safna2020}, with both RWB and BWB behaviours between the (B-J) colour index and the J magnitude. 
Furthermore, in shorter time scales, 3C~279 shows mild RWB and BWB trends, indicative of different physical processes or activity states in the emission of the jet. 


\subsubsection{CTA 102}
\begin{figure*}
\centering
\includegraphics[width=\textwidth]{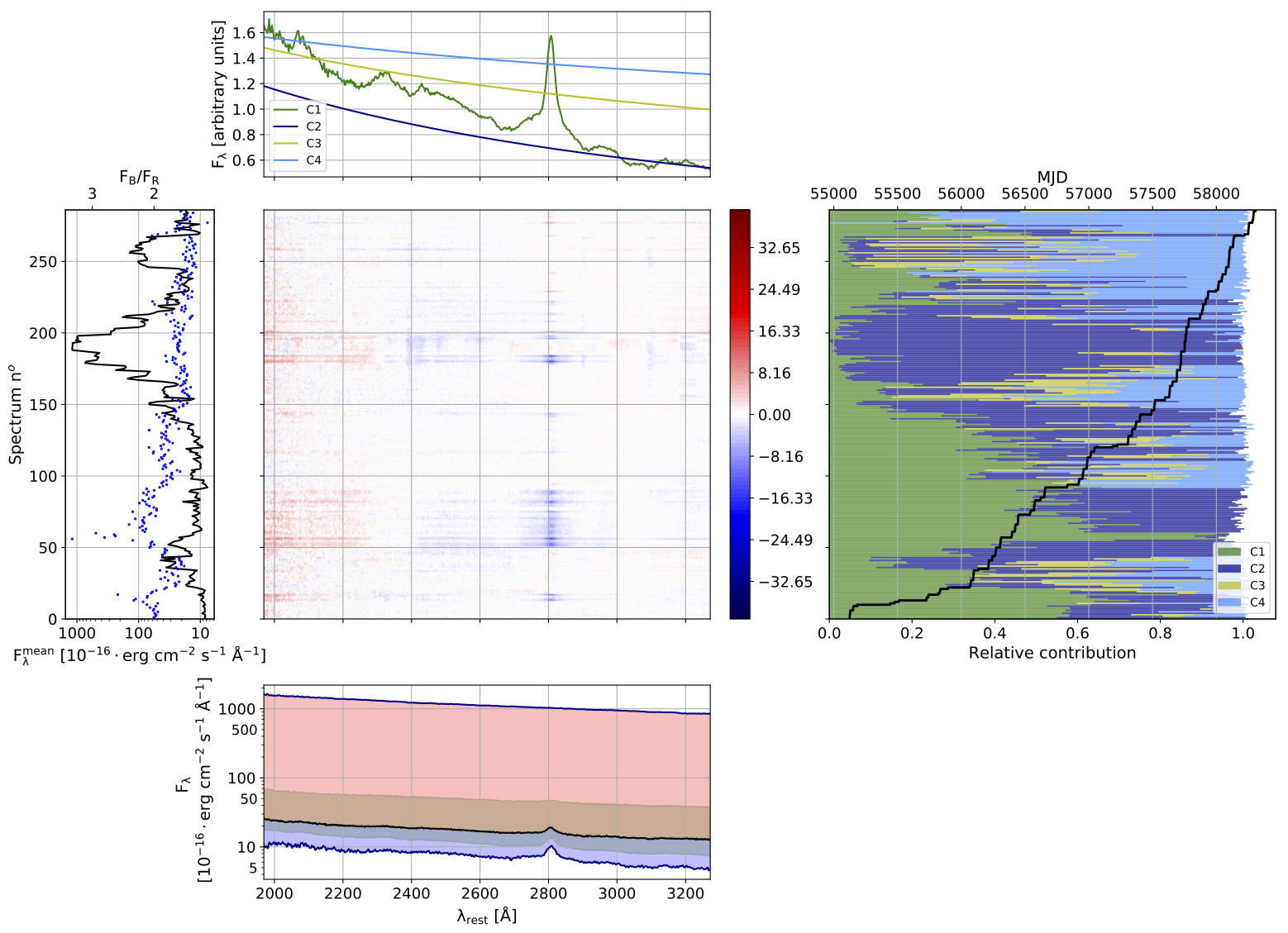}
\caption{Results of CTA 102 (same description as Fig. \ref{mrk501_figures}).}
\label{cta102_figures}
\end{figure*}

CTA 102 is reconstructed with 4 components that explain 99.98\% of its variance (see Fig. \ref{cta102_figures}). This source hosts a bright BLR component that dominates during the observed low emission states. The excess of residuals observed around the emission lines may indicate a variation of these features, including the FeII. Such behaviour has been reported in the past by several authors in other blazars \citep[see e.g.][for 3C~454.3, CTA~102 and 4C~29.45, respectively]{leon-tavares2013,chavushyan2020,hallum2020}.
Almost all the variability is associated to the second component (C2) due to its dominance during the flaring events detected, outshining the rest of the emission by 2 orders of magnitude. C2 is also the one showing the steepest slope ($\alpha=1.2$), while the component C4 has the hardest slope ($\alpha=0.6$). When the source is in low flux state, as the flux increases, the spectra show a RWB trend. During these periods, C3 and C4 are the dominant components. C3 contributes during short periods and for about 20\% to the mean flux, while C4 contributes more when the C2 is less important. For fluxes $\gtrsim$10$^{-14}$~erg~cm$^{-2}$~s$^{-1}$ \AA$^{-1}$, C2 becomes dominant, and the colour trend is reversed, leading to a BWB one.
This change is also visible in the overall colour variability of CTA~102, that shows a mild BWB evolution during the bright flare occurred between MJD 57400 and MJD 58400. 


\subsubsection{3C 454.3}
\begin{figure*}
\centering
\includegraphics[width=\textwidth]{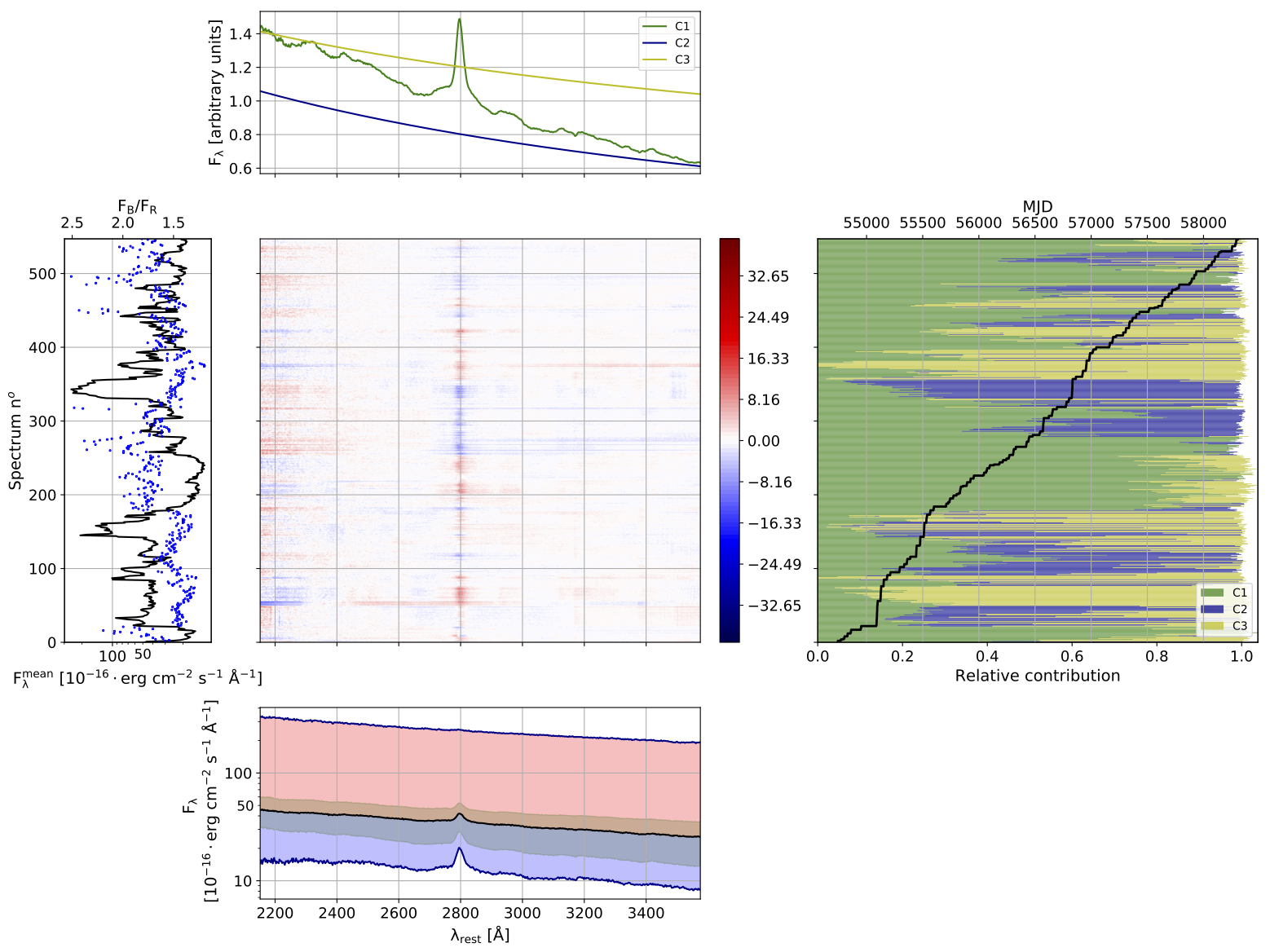}
\caption{Results of 3C 454.3 (same description as Fig. \ref{mrk501_figures}).}
\label{3c454.3_figures}
\end{figure*}

3C 454.3 is one of the most variable sources in the data sample, showing several flares with a flux increase of 1 order of magnitude. Its emission is also reconstructed with 3 components with an accuracy of $\sim$98.96\%. This reconstruction is shown in Fig. \ref{3c454.3_figures}. The first one (C1) models the constant contribution of the BLR (it accounts for about 5\% of the $\sigma^2_{exp}$). The second (C2) and third (C3) components are PLs that account for the variable part of the emission, contributing with the 85\% and 10\% of the $\sigma^2_{exp}$, respectively. 

The component C1 contributes with more than 80\% of the emission during the low states (the periods MJD 55500-56400 and 57800-57900) and only 10\% during high states. C2 is associated with prominent flares, e.g. at MJD 55250, 55500 and 56750. The component C3, a nearly flat PL, shows a large contribution in two periods around MJD 55100 and 56900. The residuals from the reconstruction show positive and negative excesses coincident with the MgII emission line, which may indicates variations in the line intensity, as reported by \cite{leon-tavares2013}.

Overall, the source shows different colour trends with the flux variation. First, a RWB trend typically associated with the FSRQs is seen as the flux increases. A saturation in the colour appears at a flux level of $\sim$ 0.5 $\times$ 10$^{-14}$ erg cm$^{-2}$ s$^{-1}$ \AA$^{-1}$, continuing with an mild BWB evolution, that is not commonly seen in FSRQs. This behaviour was first detected by \cite{villata2006}. \cite{isler2017} also identified different trends by studying the optical/near-infrared (IR) emission and colour variations in 3C~279.  
The colour behaviour can also be related with the two reconstructed components associated to the jet. The third component shows an almost flat spectrum, being partly responsible for the observed RWB trend. Moreover, the second component is also responsible for the RWB behaviour below a certain flux value. For larger contributions of this component, the colour trend is reversed to account for the overall BWB tendency for the brightest spectra. 

\section{Interpretation}\label{sec6}
\subsection{Galaxy-dominated sources}
\begin{figure*}
    \includegraphics[width=0.5\textwidth]{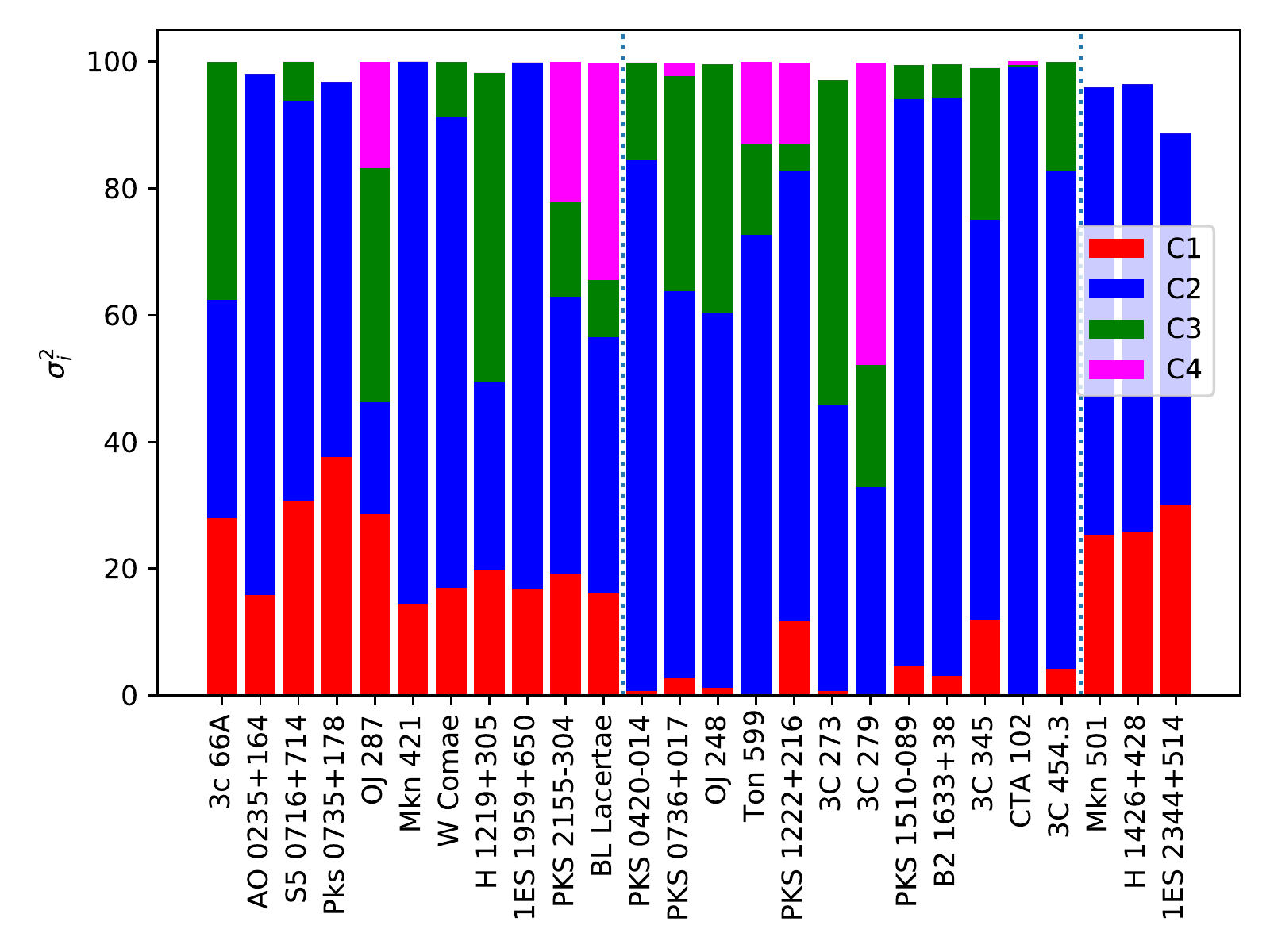}\hfill
    \includegraphics[width=0.5\textwidth]{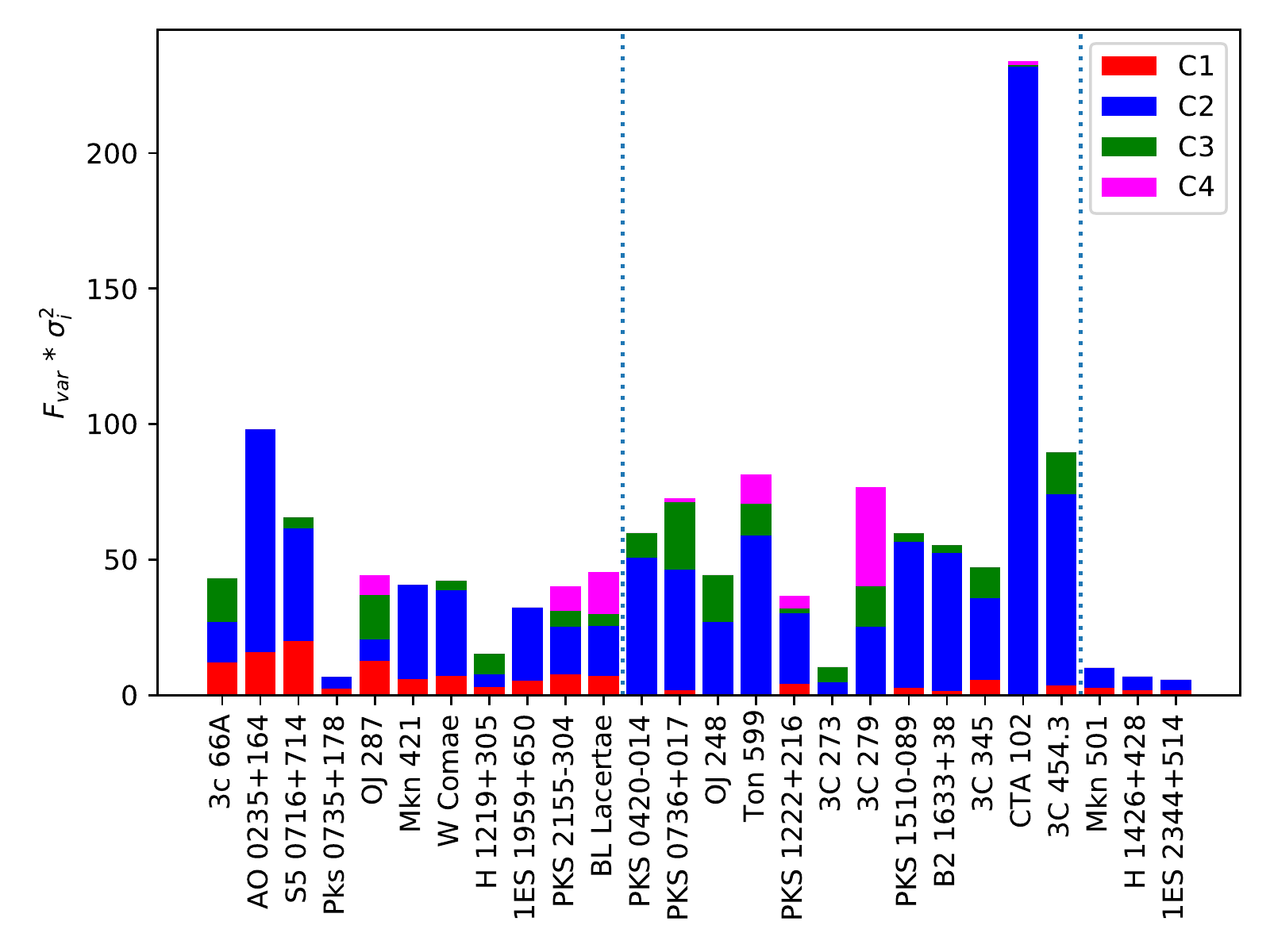}\hfill
    \caption{Explained variance of each component for all the data sample. \textit{Left}: Total explained variance. \textit{Right:} Explained variance scaled with the fractional variability. Vertical dotted lines separate the different types of blazars, classified according to their optical spectra: BL-Lac type at the left; FSRQs in the middle and dominated by the host galaxy at the right.}
    \label{fvar_components}
\end{figure*}

Only three sources of the sample have found to be dominated by the galactic emission. The overall variability of these objects is significantly smaller than that from BL Lacs or FSRQs, as reflected by the low values of the fractional variability in the optical band. This is also shown in the right panel of Fig. \ref{fvar_components}, where the explained variance scaled with the $F_{var}$ of the component associated to the jet is much lower than those of BL Lacs and FSRQs, where the jet is much more variable. As mentioned for Mkn~501, this is due to the fact that these objects are HSP BL Lac objects (see Section \ref{sec5.1.1}). The low variability of these blazars was reproduced using the stellar population plus the synchrotron emission of the jet. The host galaxy of these targets has a relative contribution of the 40-60\% to the total emission each source. Even though these objects show little intrinsic variability both in optical and $\gamma$ rays, our results show that the stellar population component (C1) remains almost constant, as expected (see Fig. \ref{fvar_components}). The variable component (C2) is a PL with spectral indices in the range of other BL Lac objects. Its variation is responsible for the observed BWB trends. These mild BWB trends found suggest that, despite the dominance and brightness of the host galaxy, the spectral evolution of the galaxy-dominated blazars is dominated by the variability of the relativistic jet as expected.

\subsection{BL Lac objects}
The reconstruction of the optical spectra of BL Lac objects requires from 2 to 4 PLs, depending on the spectral variability exhibited by each object. Their indices range mostly from 0.5 to 1.9. This may be indicative of a jet with several emitting regions or substructures, that contribute differently to the emission. The variability in the emission of these regions could lead to both the flux variability and the spectral changes observed over time for these objects.

In some cases (for instance 3C~66A), an extra component appears only for a limited period of time, that may be associated with a transient component in the jet. This behaviour can be caused by the injection of material or instabilities from the accretion disk to the jet, which can lead to the propagation of blobs, shocks or instabilities through the jet, responsible for this emission \citep[][]{marscher1985,marscher2014}. In this scenario, this component will only have a significant contribution to the emission of the source during a relatively short time interval. 

The emission of BL Lacs is expected to have a radiatively inefficient accretion disk and a dominant jet characterized by synchrotron radiation \citep{fragile2009}. This scenario is completely consistent with our reconstruction of BL Lacs, in which all components are PLs associated to synchrotron processes. Most of the BL Lacs (8 out of the 11) studied here show a BWB behaviour in the colour variability. This can be understood in terms of our NMF decomposition by looking at the spectral indices of the components. By construction, C1 is taken to be at a low or quiet state of the source, whereas C2 and the other components that are added to account for the variable term. Our results indicate a prevalence of flatter slopes in the C1 component with respect to the others, which tend to show steeper slopes (see Table \ref{table_results}). For each object the component with the steepest slope coincides with the one explaining  the most of the variance. Thus it is clear that when the flux increases the spectrum becomes bluer. \cite{li2015} claim that chromatic variations may be produced by shock-in-jet mechanisms in flares, while achromatic variability, such as the flare of OJ 287, may be due to geometrical effects. Additionally, the low brightness of their accretion disk is consistent with our reconstruction, where all the components are associated to the synchrotron emission of the jet. This behaviour is likely a result of a less efficient cooling and a very faint accretion disk \citep{isler2017}.

\subsection{FSRQs}
\begin{figure}
\centering
\includegraphics[width=\columnwidth]{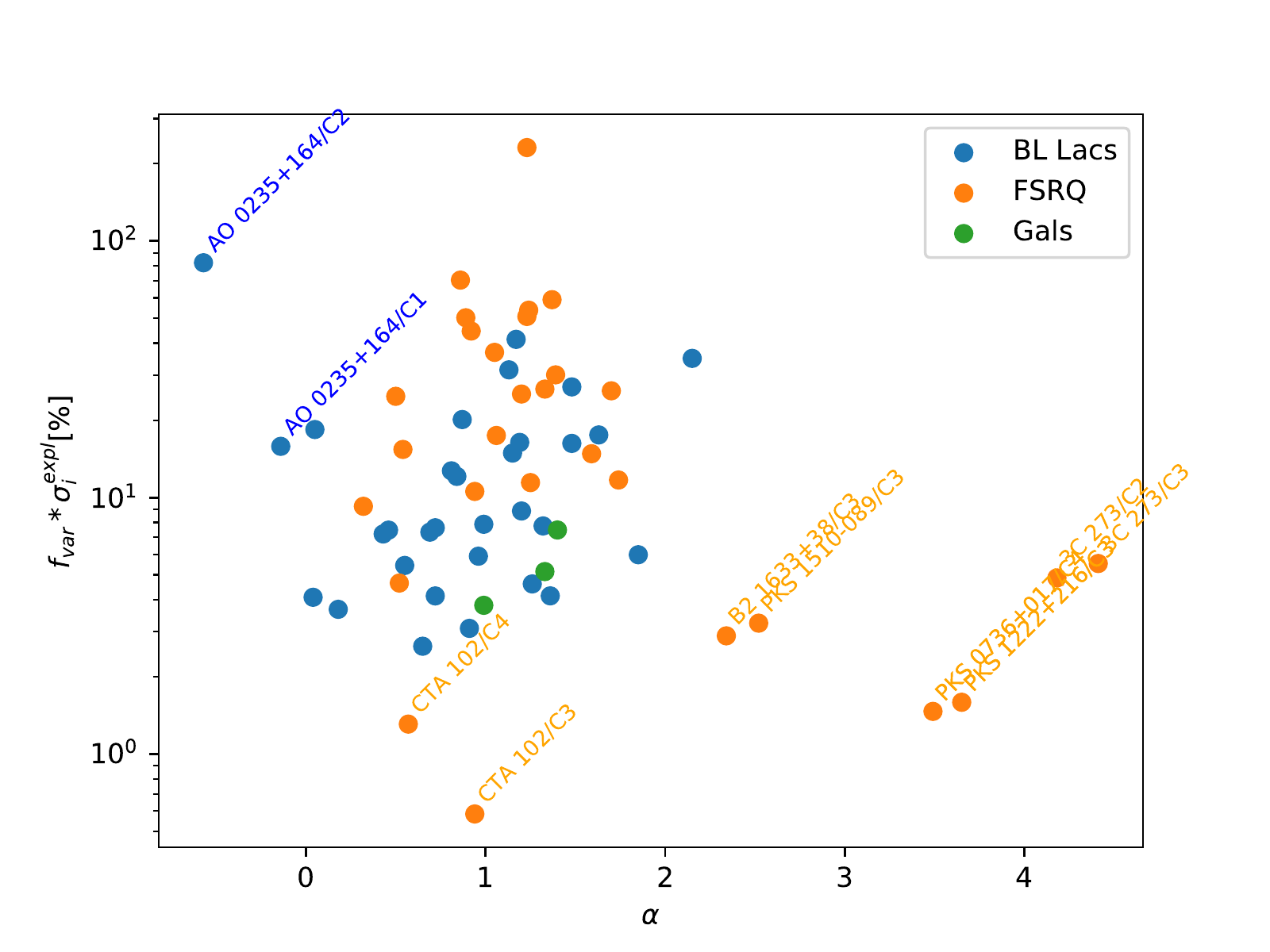}
\caption{Explained variance scaled with the fractional variability for each component vs. the corresponding spectral index.}
\label{fvarcomp_vs_plindex}
\end{figure}

As observed in Fig. \ref{fvar_plot}, these are typically the most variable AGNs of the sample. FSRQs are reproduced with the BLR contribution plus 2 PLs. In some cases another PL is required, giving a total of 4 different contributions. The PL indices are similar to those found in BL Lacs, although the median value is slightly higher. Some of the FSRQs of the sample display steep and less variable components in their reconstruction: 3C~273, PKS~0736+017, PKS~1222+216, PKS~1510-089 and B2~1633+38 (see Fig.\ref{fvarcomp_vs_plindex}). This may be associated with the existence of a bright accretion disk emitting in the UV regime, and extending to optical wavelengths. The slow and low amplitude variation expected from an accretion disk matches the low variability displayed by these components. Moreover, the presence of an enhanced disk was already reported by \cite{raiteri2014} in the modeling of the optical-IR SEDs of B2~1633+38, PKS~1510-089 and 3C~273. Additionally, some sources also display some low amplitude and slow varying trend in the BLR component. This slow varying trend observed in the BLR component can be associated to the variability of the accretion disk coupled with the broad line emission, and the time scales of these trends are compatible with the slow variability of the accretion disk.

The colour variability of these sources does not generally follow a unique RWB behaviour historically as associated to FSRQs, but shows different trends. \cite{isler2017} explains this with the presence of an accretion disk and the different relative contributions of this thermal emission and the non-thermal synchrotron radiation coming from the jet. They identify different regimes that explain the different trends in the colour variability: 

    
    
    

\begin{itemize}
    \item If the jet is in a quiescent state, the accretion disk can be observed in optical to UV wavelengths. In the optical/near-IR regime, the contribution of the disk is bluer, and thus, the spectra tend to bluer colour values. Therefore, as the non-thermal synchrotron contribution increases, it causes a reddening of the spectra as the emission of the jet equals the one from the disk (for instance, B2~1633+38, see Fig. \ref{variability_fsrqs}).
    
    \item Once the jet overcomes the accretion disk, the colour variability flattens, leading to an approximately achromatic flux increase. This effect of ``colour saturation'' has already been detected in the past in several FSRQs \citep[see e.g.][]{villata2006,fan2018}. An example of this can be observed for CTA~102 (see Fig. \ref{variability_fsrqs}).
    
    \item Finally, when the non-thermal emission from the jet fully outshines the thermal contribution from the disk, the colour is determined by the jet. There are two possible behaviours: BWB likely due to the synchrontron electrons reaching higher energies before the radiative cooling (e.g.~3C 454.3, see Fig. \ref{variability_fsrqs}); or RWB if the electrons do not reach such high energies.
    
\end{itemize}

\section{Conclusions}\label{sec7}
NMF is a powerful statistical tool that allows us to reproduce the variability observed in large spectral data sets with high accuracy. We have used an \textit{a priori} knowledge of the expected emission of blazars to model and reproduce their emission in terms of a small number of components. The simplest cases correspond to those objects whose spectra are dominated by the stellar population of the host galaxy. In these cases the spectral variability is explained in terms of the constant contribution of the host galaxy plus a PL component accounting for the synchrotron emission of the jet that is responsible for the variability observed in these sources. BL Lacs are reconstructed with 2 to 4 PL components that account for the spectral variability of the relativistic jet. Furthermore, a QSO-like component is added to analyse FSRQs due to the presence of a bright BLR responsible for the broad emission lines that appear in their optical spectrum, and possibly the contribution of a bright accretion disk.

We have investigated the temporal evolution of the components intervening in the reconstruction of the spectral set of each object. This can result in the identification of transient components that can be associated with blobs or shocks propagating along the jet. Moreover, behaviours such as different colour trends can also be associated with some of the resulting components, revealing different physical processes \citep[see][]{isler2017}. Overall, almost all BL Lacs have found to be BWB (except for OJ~287, whose behaviour may be ruled by a supermassive binary black hole system; and AO~0235+164, that despite being a BL Lac, has some FSRQ-like characteristics, indicating that this blazar may be a transitional object). On the other hand, FSRQs typically display more complicated colour behaviours, with a RWB evolution for low flux states, a colour saturation, and in some cases a final BWB trend for the brightest states.

Some of the sources display a rather steep and low variable PL component in their reconstruction (see e.g. the case of 3C~273). These components can be a signature of an enhanced black body associated to a bright accretion disk. 
The slow and small variability associated to these components is compatible with the behaviour of an accretion disk, and the steepness of the spectra is naturally associated with the optical extension of the UV emission coming from the disk. The sources in which this behaviour was observed are also compatible with those needing an enhanced accretion disk for the optical-IR SED modeling performed by \cite{raiteri2014}. The different colour trends can help to make constrains on the brightness of the accretion disks of FSRQs. 

We propose that an NMF analysis of larger spectral data sets will help to understand the origin of the variability observed in blazars and disentangle the relevant contributions to their emission. In this context, extending the analysis to the IR and UV domains can aid to disentangle the different components present in AGNs and be of great use to study the physical processes taking place in such extreme and violent objects. Moreover, radio data can have a crucial importance in detecting several components propagating through the jet, or substructures such as blobs, shocks or instabilities. Extending this methodology to other AGN types can also shed light in the evolution and unification of AGNs.

\begin{landscape}
\begin{table}
\begin{center}
\caption{Sample of blazars from the Steward Observatory monitoring program included in the NMF analysis. Columns: (1) target name. (2) source type. (3) number of spectra available. (4) number of reconstructed components. (5) fractional variability and its uncertainty. (6) percentage of explained variance. (7-10) percentage of variance explained by each component. (11-14) spectral index of the \textit{n} component. (15) long-term colour trends. When more than one trend is reported, it goes from lower to higher flux states. \label{table_results}}
\begin{tabular}{ccccccccccccccc} \hline
(1) & (2) & (3) & (4) & (5) & (6) & (7) & (8) & (9) & (10) & (11) & (12) & (13) & (14) & (15)\\ 
Source & Type  & n$^o$ spectra & n$^{\circ}$ components & F$_{var}$ & $\sigma^{2}_{\text{exp}}$ (\%) & $\sigma^{2}_{1}$ (\%) & $\sigma^{2}_{2}$ (\%) & $\sigma^{2}_{3}$ (\%) & $\sigma^{2}_{4}$ (\%) & $\alpha_{1}$ & $\alpha_{2}$ & $\alpha_{3}$ & $\alpha_{4}$ & Colour \\ \hline \hline
Mkn 501 & Galaxy dominated  & 476 & 2 & $0.067\pm0.001$ & 95.87 & 25.36 & 70.51 & -- & -- & -- & 1.40 & -- & -- & BWB \\ \hline
H 1426+428 & Galaxy dominated  & 36 & 2 & $0.099\pm0.006$ & 96.46 & 25.86 & 70.60 & -- & -- & -- & 1.33 & -- & -- & Achromatic \\ \hline
1ES 2344+514 & Galaxy dominated  & 81 & 2 & $0.038\pm0.007$ & 88.60 & 30.13 & 58.49 & -- & -- & -- & 0.99 & -- & -- & BWB \\ \hline \hline
3C 66A & BL Lac &  353 & 3 & $0.434\pm0.002$ & 99.91 & 27.93 & 34.43 & 37.55 & -- & 0.84 & 1.15 & 1.48 & -- & BWB \\ \hline
AO 0235+164 & BL Lac  & 207 & 2 & $0.978\pm0.008$ & 98.00 & 15.82 & 82.18 & -- & -- & -0.14 & -0.57 & -- & -- & RWB, FWB \\ \hline
S5 0716+714 & BL Lac  & 226 & 3 & $0.646\pm0.002$ & 99.98 & 30.70 & 63.05 & 6.23 & -- & 0.87 & 1.17 & 0.04 & -- & BWB \\ \hline
PKS 0735+178 & BL Lac  & 34 & 2 & $0.066\pm0.004$ & 96.74 & 37.62 & 59.12 & -- & -- & 0.65 & 1.36 & -- & -- & BWB \\ \hline
OJ 287 & BL Lac  & 516 & 4 & $0.457\pm0.001$ & 99.88 & 28.57 & 17.68 & 36.86 & 16.78 & 0.81 & 0.99 & 1.19 & 0.46 & FWB \\ \hline
Mkn 421 & BL Lac  & 594 & 2 & $0.438\pm0.002$ & 99.95 & 14.48 & 85.47 & -- & -- & 0.96 & 2.15 & -- & -- & BWB \\ \hline
W Comae & BL Lac  & 345 & 3 & $0.432\pm0.002$ & 99.86 & 17.01 & 74.20 & 8.65 & -- & 0.43 & 1.13 & 0.18 & -- & BWB \\ \hline
H 1219+305 & BL Lac  & 34 & 3 & $0.152\pm0.007$ & 98.23 & 19.82 & 29.53 & 48.88 & -- & 0.91 & 1.26 & 0.72 & -- & No trend \\ \hline
1ES 1959+650 & BL Lac  & 147 & 2 & $0.332\pm0.003$ & 99.80 & 16.73 & 83.07 & -- & -- & 0.55 & 1.48 & -- & -- & BWB \\ \hline
PKS 2155-304 & BL Lac  & 299 & 4 & $0.386\pm0.003$ & 99.85 & 19.28 & 43.64 & 14.89 & 22.05 & 1.32 & 1.63 & 1.85 & 1.20 & BWB \\ \hline
BL Lacertae & BL Lac  & 699 & 4 & $0.457\pm0.001$ & 99.71 & 16.08 & 40.39 & 9.07 & 34.17 & 0.25 & 0.69 & 0.05 & 0.72 & BWB \\ \hline \hline
PKS 0420-014 & FSRQ  & 53 & 3 & $0.610\pm0.006$ & 99.83 & 0.72 & 83.69 & 15.41 & -- & -- & 0.89 & 0.32 & -- & RWB, BWB \\ \hline
PKS 0736+017 & FSRQ  & 126 & 4 & $0.682\pm0.006$ & 99.72 & 2.68 & 61.06 & 33.98 & 2.01 & -- & 0.92 & 0.50 & 3.49 & No trend, FWB \\ \hline
OJ 248 & FSRQ  & 224 & 3 & $0.422\pm0.005$ & 99.48 & 1.20 & 59.23 & 39.06 & -- & -- & 1.33 & 1.06 & -- & No trend, FWB \\ \hline
Ton 599 & FSRQ  & 190 & 4 & $1.052\pm0.004$ & 99.96 & 0.05 & 72.55 & 14.38 & 12.98 & -- & 1.37 & 1.74 & 0.94 & RWB, FWB \\ \hline
PKS 1222+216 & FSRQ & 427 & 4 & $0.308\pm0.006$ & 99.75 & 11.72 & 71.03 & 4.34 & 12.65 & -- & 1.70 & 3.65 & 0.52 & RWB, FWB\\ \hline
3C 273 & FSRQ & 298 & 3 & $0.069\pm0.008$ & 97.01 & 0.75 & 45.03 & 51.23 & -- & -- & 4.18 & 4.41 & -- & BWB \\ \hline
3C 279 & FSRQ & 499 & 4 & $0.736\pm0.002$ & 99.90 & 0.01 & 32.85 & 19.22 & 47.74 & -- & 1.20 & 1.59 & 1.05 & No clear trend \\ \hline
PKS 1510-089 & FSRQ  & 371 & 3 & $0.529\pm0.003$ & 99.43 & 4.77 & 89.27 & 5.38 & -- & -- & 1.24 & 2.52 & -- & RWB \\ \hline
B2 1633+38 & FSRQ  & 362 & 3 & $0.519\pm0.005$ & 99.51 & 3.11 & 91.21 & 5.19 & -- & -- & 1.23 & 2.34 & -- & RWB \\ \hline
3C 345 & FSRQ  & 42 & 3 & $0.446\pm0.021$ & 98.96 & 11.93 & 63.06 & 23.98 & -- & -- & 1.39 & 1.25 & -- & RWB\\ \hline
CTA 102 & FSRQ  & 287 & 4 & $2.392\pm0.005$ & 99.98 & 0.25 & 98.93 & 0.25 & 0.56 & -- & 1.23 & 0.94 & 0.57 & RWB, FWB, BWB \\ \hline
3C 454.3 & FSRQ  & 549 & 3 & $0.89\pm0.002$ & 99.90 & 4.25 & 78.48 & 17.17 & -- & -- & 0.86 & 0.54 & -- & RWB, BWB \\ \hline \hline
\end{tabular}
\end{center}
\end{table}
\end{landscape}

\section*{Data availability}
The data underlying this article are available in Steward Observatory Database, at \url{http://james.as.arizona.edu/~psmith/Fermi/DATA/specdata.html}.

\section*{Acknowledgements}

JOS thanks the support from grant FPI-SO from the Spanish Ministry of Economy and Competitiveness (MINECO) (research project SEV-2015-0548-17-3 and predoctoral contract BES-2017-082171).

JOS, JAP and JBG acknowledges financial support from the Spanish Ministry of Science and Innovation (MICINN) through the Spanish State Research Agency, under Severo Ochoa Program 2020-2023 (CEX2019-000920-S) and the project PID2019-107988GB-C22.

OGM acknowledges financial support from the project [IN105720] DGAPA PAPIIT from UNAM. 

Data from the Steward Observatory spectropolarimetric monitoring project were used. This program is supported by Fermi Guest Investigator grants NNX08AW56G, NNX09AU10G, NNX12AO93G, and NNX15AU81G.

We thank the anonymous referee for his/her comments that helped to improve the manuscript.





\bibliographystyle{mnras}
\bibliography{biblio.bib} 

\begin{thebibliography}{}
\makeatletter
\relax
\def\mn@urlcharsother{\let\do\@makeother \do\$\do\&\do\#\do\^\do\_\do\%\do\~}
\def\mn@doi{\begingroup\mn@urlcharsother \@ifnextchar [ {\mn@doi@}
  {\mn@doi@[]}}
\def\mn@doi@[#1]#2{\def\@tempa{#1}\ifx\@tempa\@empty \href
  {http://dx.doi.org/#2} {doi:#2}\else \href {http://dx.doi.org/#2} {#1}\fi
  \endgroup}
\def\mn@eprint#1#2{\mn@eprint@#1:#2::\@nil}
\def\mn@eprint@arXiv#1{\href {http://arxiv.org/abs/#1} {{\tt arXiv:#1}}}
\def\mn@eprint@dblp#1{\href {http://dblp.uni-trier.de/rec/bibtex/#1.xml}
  {dblp:#1}}
\def\mn@eprint@#1:#2:#3:#4\@nil{\def\@tempa {#1}\def\@tempb {#2}\def\@tempc
  {#3}\ifx \@tempc \@empty \let \@tempc \@tempb \let \@tempb \@tempa \fi \ifx
  \@tempb \@empty \def\@tempb {arXiv}\fi \@ifundefined
  {mn@eprint@\@tempb}{\@tempb:\@tempc}{\expandafter \expandafter \csname
  mn@eprint@\@tempb\endcsname \expandafter{\@tempc}}}

\bibitem[\protect\citeauthoryear{{Abdo} et~al.,}{{Abdo}
  et~al.}{2010a}]{abdo2010}
{Abdo} A.~A.,  et~al., 2010a, \mn@doi [\apj] {10.1088/0004-637X/716/1/30},
  \href {https://ui.adsabs.harvard.edu/abs/2010ApJ...716...30A} {716, 30}

\bibitem[\protect\citeauthoryear{{Abdo} et~al.,}{{Abdo}
  et~al.}{2010b}]{abdo2010b}
{Abdo} A.~A.,  et~al., 2010b, \mn@doi [\apj] {10.1088/0004-637X/722/1/520},
  \href {https://ui.adsabs.harvard.edu/abs/2010ApJ...722..520A} {722, 520}

\bibitem[\protect\citeauthoryear{{Ackermann} et~al.,}{{Ackermann}
  et~al.}{2015}]{ackermann2015}
{Ackermann} M.,  et~al., 2015, \mn@doi [\apjl] {10.1088/2041-8205/813/2/L41},
  \href {https://ui.adsabs.harvard.edu/abs/2015ApJ...813L..41A} {813, L41}

\bibitem[\protect\citeauthoryear{{Acosta-Pulido}, {Castro Segura}, {Carnerero}
  \& {Raiteri}}{{Acosta-Pulido} et~al.}{2017}]{acosta2017}
{Acosta-Pulido} J.,  {Castro Segura} N.,  {Carnerero} M.,   {Raiteri} C.,
  2017, \mn@doi [Galaxies] {10.3390/galaxies5010001}, \href
  {https://ui.adsabs.harvard.edu/abs/2017Galax...5....1A} {5, 1}

\bibitem[\protect\citeauthoryear{{Agarwal} et~al.,}{{Agarwal}
  et~al.}{2019}]{agarwal2019}
{Agarwal} A.,  et~al., 2019, \mn@doi [\mnras] {10.1093/mnras/stz1981}, \href
  {https://ui.adsabs.harvard.edu/abs/2019MNRAS.488.4093A} {488, 4093}

\bibitem[\protect\citeauthoryear{{Ajello} et~al.,}{{Ajello}
  et~al.}{2020}]{ajello2020}
{Ajello} M.,  et~al., 2020, \mn@doi [\apj] {10.3847/1538-4357/ab791e}, \href
  {https://ui.adsabs.harvard.edu/abs/2020ApJ...892..105A} {892, 105}

\bibitem[\protect\citeauthoryear{{Aleksi{\'c}} et~al.,}{{Aleksi{\'c}}
  et~al.}{2015}]{aleksic2015}
{Aleksi{\'c}} J.,  et~al., 2015, \mn@doi [\aap] {10.1051/0004-6361/201424811},
  \href {https://ui.adsabs.harvard.edu/abs/2015A&A...578A..22A} {578, A22}

\bibitem[\protect\citeauthoryear{{Baron}}{{Baron}}{2019}]{baron2019}
{Baron} D.,  2019, arXiv e-prints, \href
  {https://ui.adsabs.harvard.edu/abs/2019arXiv190407248B} {p. arXiv:1904.07248}

\bibitem[\protect\citeauthoryear{{Blandford} \& {Rees}}{{Blandford} \&
  {Rees}}{1978}]{blandford1978}
{Blandford} R.~D.,  {Rees} M.~J.,  1978, in {Wolfe} A.~M.,  ed., BL Lac
  Objects. pp 328--341

\bibitem[\protect\citeauthoryear{{Blanton} \& {Roweis}}{{Blanton} \&
  {Roweis}}{2007}]{blanton2007}
{Blanton} M.~R.,  {Roweis} S.,  2007, \mn@doi [\aj] {10.1086/510127}, \href
  {https://ui.adsabs.harvard.edu/abs/2007AJ....133..734B} {133, 734}

\bibitem[\protect\citeauthoryear{{Bonning} et~al.,}{{Bonning}
  et~al.}{2012}]{bonning2012}
{Bonning} E.,  et~al., 2012, \mn@doi [\apj] {10.1088/0004-637X/756/1/13}, \href
  {https://ui.adsabs.harvard.edu/abs/2012ApJ...756...13B} {756, 13}

\bibitem[\protect\citeauthoryear{{B{\"o}ttcher} et~al.,}{{B{\"o}ttcher}
  et~al.}{2007}]{bottcher2007}
{B{\"o}ttcher} M.,  et~al., 2007, \mn@doi [\apj] {10.1086/522583}, \href
  {https://ui.adsabs.harvard.edu/abs/2007ApJ...670..968B} {670, 968}

\bibitem[\protect\citeauthoryear{{Calderone}, {Ghisellini}, {Colpi}  \&
  {Dotti}}{{Calderone} et~al.}{2013}]{calderone2013}
{Calderone} G.,  {Ghisellini} G.,  {Colpi} M.,   {Dotti} M.,  2013, \mn@doi
  [\mnras] {10.1093/mnras/stt157}, \href
  {https://ui.adsabs.harvard.edu/abs/2013MNRAS.431..210C} {431, 210}

\bibitem[\protect\citeauthoryear{{Camenzind} \& {Krockenberger}}{{Camenzind} \&
  {Krockenberger}}{1992}]{camenzind1992}
{Camenzind} M.,  {Krockenberger} M.,  1992, \aap, \href
  {https://ui.adsabs.harvard.edu/abs/1992A&A...255...59C} {255, 59}

\bibitem[\protect\citeauthoryear{{Cappellari}}{{Cappellari}}{2017}]{cappellari2017}
{Cappellari} M.,  2017, \mn@doi [MNRAS] {10.1093/mnras/stw3020}, 466, 798

\bibitem[\protect\citeauthoryear{{Cappellari} \& {Emsellem}}{{Cappellari} \&
  {Emsellem}}{2004}]{cappellari2004}
{Cappellari} M.,  {Emsellem} E.,  2004, \mn@doi [\pasp] {10.1086/381875}, \href
  {https://ui.adsabs.harvard.edu/abs/2004PASP..116..138C} {116, 138}

\bibitem[\protect\citeauthoryear{{Carnerero} et~al.,}{{Carnerero}
  et~al.}{2017}]{carnerero2017}
{Carnerero} M.~I.,  et~al., 2017, \mn@doi [\mnras] {10.1093/mnras/stx2185},
  \href {https://ui.adsabs.harvard.edu/abs/2017MNRAS.472.3789C} {472, 3789}

\bibitem[\protect\citeauthoryear{{Cellone}, {Romero}, {Combi}  \&
  {Mart{\'\i}}}{{Cellone} et~al.}{2007}]{cellone2007}
{Cellone} S.~A.,  {Romero} G.~E.,  {Combi} J.~A.,   {Mart{\'\i}} J.,  2007,
  \mn@doi [\mnras] {10.1111/j.1745-3933.2007.00366.x}, \href
  {https://ui.adsabs.harvard.edu/abs/2007MNRAS.381L..60C} {381, L60}

\bibitem[\protect\citeauthoryear{{Cerruti}, {Zech}, {Boisson}  \&
  {Inoue}}{{Cerruti} et~al.}{2015}]{cerruti2015}
{Cerruti} M.,  {Zech} A.,  {Boisson} C.,   {Inoue} S.,  2015, \mn@doi [\mnras]
  {10.1093/mnras/stu2691}, \href
  {https://ui.adsabs.harvard.edu/abs/2015MNRAS.448..910C} {448, 910}

\bibitem[\protect\citeauthoryear{{Chavushyan}, {Pati{\~n}o-{\'A}lvarez},
  {Amaya-Almaz{\'a}n}  \& {Carrasco}}{{Chavushyan}
  et~al.}{2020}]{chavushyan2020}
{Chavushyan} V.,  {Pati{\~n}o-{\'A}lvarez} V.~M.,  {Amaya-Almaz{\'a}n} R.~A.,
  {Carrasco} L.,  2020, \mn@doi [\apj] {10.3847/1538-4357/ab6ef6}, \href
  {https://ui.adsabs.harvard.edu/abs/2020ApJ...891...68C} {891, 68}

\bibitem[\protect\citeauthoryear{{Covino}, {Sandrinelli}  \& {Treves}}{{Covino}
  et~al.}{2019}]{covino2019}
{Covino} S.,  {Sandrinelli} A.,   {Treves} A.,  2019, \mn@doi [\mnras]
  {10.1093/mnras/sty2720}, \href
  {https://ui.adsabs.harvard.edu/abs/2019MNRAS.482.1270C} {482, 1270}

\bibitem[\protect\citeauthoryear{{D'Ammando} et~al.,}{{D'Ammando}
  et~al.}{2009}]{d'ammando2009}
{D'Ammando} F.,  et~al., 2009, \mn@doi [\aap] {10.1051/0004-6361/200912560},
  \href {https://ui.adsabs.harvard.edu/abs/2009A&A...508..181D} {508, 181}

\bibitem[\protect\citeauthoryear{{Dai} et~al.,}{{Dai} et~al.}{2009}]{dai2009}
{Dai} B.~Z.,  et~al., 2009, \mn@doi [\mnras]
  {10.1111/j.1365-2966.2008.14137.x}, \href
  {https://ui.adsabs.harvard.edu/abs/2009MNRAS.392.1181D} {392, 1181}

\bibitem[\protect\citeauthoryear{{Danforth}, {Nalewajko}, {France}  \&
  {Keeney}}{{Danforth} et~al.}{2013}]{danforth2013}
{Danforth} C.~W.,  {Nalewajko} K.,  {France} K.,   {Keeney} B.~A.,  2013,
  \mn@doi [\apj] {10.1088/0004-637X/764/1/57}, \href
  {https://ui.adsabs.harvard.edu/abs/2013ApJ...764...57D} {764, 57}

\bibitem[\protect\citeauthoryear{{Dermer} \& {Schlickeiser}}{{Dermer} \&
  {Schlickeiser}}{1994}]{dermer1994}
{Dermer} C.~D.,  {Schlickeiser} R.,  1994, \mn@doi [\apjs] {10.1086/191929},
  \href {https://ui.adsabs.harvard.edu/abs/1994ApJS...90..945D} {90, 945}

\bibitem[\protect\citeauthoryear{{Fan}}{{Fan}}{2005}]{fan2005}
{Fan} J.~H.,  2005, Chinese Journal of Astronomy and Astrophysics Supplement,
  \href {https://ui.adsabs.harvard.edu/abs/2005ChJAS...5..213F} {5, 213}

\bibitem[\protect\citeauthoryear{{Fan} et~al.,}{{Fan} et~al.}{2018a}]{fan2018a}
{Fan} J.~H.,  et~al., 2018a, \mn@doi [\aj] {10.3847/1538-3881/aaa547}, \href
  {https://ui.adsabs.harvard.edu/abs/2018AJ....155...90F} {155, 90}

\bibitem[\protect\citeauthoryear{{Fan} et~al.,}{{Fan} et~al.}{2018b}]{fan2018}
{Fan} X.,  et~al., 2018b, \mn@doi [\apj] {10.3847/1538-4357/aab09d}, \href
  {https://ui.adsabs.harvard.edu/abs/2018ApJ...856...80F} {856, 80}

\bibitem[\protect\citeauthoryear{{Finke}}{{Finke}}{2013}]{finke2013}
{Finke} J.~D.,  2013, \mn@doi [\apj] {10.1088/0004-637X/763/2/134}, \href
  {https://ui.adsabs.harvard.edu/abs/2013ApJ...763..134F} {763, 134}

\bibitem[\protect\citeauthoryear{{Fitzpatrick}}{{Fitzpatrick}}{1999}]{fitzpatrick1999}
{Fitzpatrick} E.~L.,  1999, \mn@doi [\pasp] {10.1086/316293}, \href
  {https://ui.adsabs.harvard.edu/abs/1999PASP..111...63F} {111, 63}

\bibitem[\protect\citeauthoryear{{Fragile} \& {Meier}}{{Fragile} \&
  {Meier}}{2009}]{fragile2009}
{Fragile} P.~C.,  {Meier} D.~L.,  2009, \mn@doi [\apj]
  {10.1088/0004-637X/693/1/771}, \href
  {https://ui.adsabs.harvard.edu/abs/2009ApJ...693..771F} {693, 771}

\bibitem[\protect\citeauthoryear{{Gallant}, {Gallo}  \& {Parker}}{{Gallant}
  et~al.}{2018}]{gallant2018}
{Gallant} D.,  {Gallo} L.~C.,   {Parker} M.~L.,  2018, \mn@doi [\mnras]
  {10.1093/mnras/sty1987}, \href
  {https://ui.adsabs.harvard.edu/abs/2018MNRAS.480.1999G} {480, 1999}

\bibitem[\protect\citeauthoryear{{Ghisellini} \& {Tavecchio}}{{Ghisellini} \&
  {Tavecchio}}{2008}]{ghisellini2008}
{Ghisellini} G.,  {Tavecchio} F.,  2008, \mn@doi [\mnras]
  {10.1111/j.1365-2966.2008.13360.x}, \href
  {https://ui.adsabs.harvard.edu/abs/2008MNRAS.387.1669G} {387, 1669}

\bibitem[\protect\citeauthoryear{{Ghisellini}, {Tavecchio}, {Foschini},
  {Ghirland a}, {Maraschi}  \& {Celotti}}{{Ghisellini}
  et~al.}{2010}]{ghisellini2010}
{Ghisellini} G.,  {Tavecchio} F.,  {Foschini} L.,  {Ghirland a} G.,  {Maraschi}
  L.,   {Celotti} A.,  2010, \mn@doi [\mnras]
  {10.1111/j.1365-2966.2009.15898.x}, \href
  {https://ui.adsabs.harvard.edu/abs/2010MNRAS.402..497G} {402, 497}

\bibitem[\protect\citeauthoryear{{Hallum}, {Jorstad}, {Marscher}  \&
  {Larionov}}{{Hallum} et~al.}{2020}]{hallum2020}
{Hallum} M.~K.,  {Jorstad} S.~G.,  {Marscher} A.~P.,   {Larionov} V.~M.,  2020,
  in American Astronomical Society Meeting Abstracts \#235. p. 151.04

\bibitem[\protect\citeauthoryear{Hotelling}{Hotelling}{1933}]{hotelling1933}
Hotelling H.,  1933, Journal of Educational Psychology, 24, 417

\bibitem[\protect\citeauthoryear{{Hurley}, {Oliver}, {Farrah}, {Lebouteiller}
  \& {Spoon}}{{Hurley} et~al.}{2014}]{hurley2014}
{Hurley} P.~D.,  {Oliver} S.,  {Farrah} D.,  {Lebouteiller} V.,   {Spoon}
  H.~W.~W.,  2014, \mn@doi [\mnras] {10.1093/mnras/stt1875}, \href
  {https://ui.adsabs.harvard.edu/abs/2014MNRAS.437..241H} {437, 241}

\bibitem[\protect\citeauthoryear{Hutchins, Murphy, Singh  \& Graber}{Hutchins
  et~al.}{2008}]{hutchins2008}
Hutchins L.,  Murphy S.,  Singh P.,   Graber J.,  2008, \mn@doi [Bioinformatics
  (Oxford, England)] {10.1093/bioinformatics/btn526}, 24, 2684

\bibitem[\protect\citeauthoryear{{Isler}, {Urry}, {Coppi}, {Bailyn}, {Brady},
  {MacPherson}, {Buxton}  \& {Hasan}}{{Isler} et~al.}{2017}]{isler2017}
{Isler} J.~C.,  {Urry} C.~M.,  {Coppi} P.,  {Bailyn} C.,  {Brady} M.,
  {MacPherson} E.,  {Buxton} M.,   {Hasan} I.,  2017, \mn@doi [\apj]
  {10.3847/1538-4357/aa79fc}, \href
  {https://ui.adsabs.harvard.edu/abs/2017ApJ...844..107I} {844, 107}

\bibitem[\protect\citeauthoryear{{Ivezi{\'c}}, {Connelly}, {Vand erPlas}  \&
  {Gray}}{{Ivezi{\'c}} et~al.}{2014}]{ivezic2014}
{Ivezi{\'c}} {\v{Z}}.,  {Connelly} A.~J.,  {Vand erPlas} J.~T.,   {Gray} A.,
  2014, {Statistics, Data Mining, and Machine Learning in Astronomy}

\bibitem[\protect\citeauthoryear{{Junkkarinen}, {Cohen}, {Beaver}, {Burbidge},
  {Lyons}  \& {Madejski}}{{Junkkarinen} et~al.}{2004}]{junkkarainen2004}
{Junkkarinen} V.~T.,  {Cohen} R.~D.,  {Beaver} E.~A.,  {Burbidge} E.~M.,
  {Lyons} R.~W.,   {Madejski} G.,  2004, \mn@doi [\apj] {10.1086/423777}, \href
  {https://ui.adsabs.harvard.edu/abs/2004ApJ...614..658J} {614, 658}

\bibitem[\protect\citeauthoryear{{Kapanadze}, {Vercellone}, {Romano}, {Hughes},
  {Aller}, {Aller}, {Kapanadze}  \& {Tabagari}}{{Kapanadze}
  et~al.}{2018}]{kapanadze2018}
{Kapanadze} B.,  {Vercellone} S.,  {Romano} P.,  {Hughes} P.,  {Aller} M.,
  {Aller} H.,  {Kapanadze} S.,   {Tabagari} L.,  2018, \mn@doi [\mnras]
  {10.1093/mnras/sty1803}, \href
  {https://ui.adsabs.harvard.edu/abs/2018MNRAS.480..407K} {480, 407}

\bibitem[\protect\citeauthoryear{{Kidger}, {Garcia Lario}  \& {de
  Diego}}{{Kidger} et~al.}{1990}]{kidger1990}
{Kidger} M.~R.,  {Garcia Lario} P.,   {de Diego} J.~A.,  1990, \mn@doi [\apss]
  {10.1007/BF00646813}, \href
  {https://ui.adsabs.harvard.edu/abs/1990Ap&SS.171...13K} {171, 13}

\bibitem[\protect\citeauthoryear{{Lee} \& {Seung}}{{Lee} \&
  {Seung}}{1999}]{lee1999}
{Lee} D.~D.,  {Seung} H.~S.,  1999, \mn@doi [\nat] {10.1038/44565}, \href
  {https://ui.adsabs.harvard.edu/abs/1999Natur.401..788L} {401, 788}

\bibitem[\protect\citeauthoryear{Lee \& Seung}{Lee \& Seung}{2001}]{lee2001}
Lee D.,  Seung H.,  2001, Adv. Neural Inform. Process. Syst., 13

\bibitem[\protect\citeauthoryear{{Lehto} \& {Valtonen}}{{Lehto} \&
  {Valtonen}}{1996}]{lehto1996}
{Lehto} H.~J.,  {Valtonen} M.~J.,  1996, \mn@doi [\apj] {10.1086/176962}, \href
  {https://ui.adsabs.harvard.edu/abs/1996ApJ...460..207L} {460, 207}

\bibitem[\protect\citeauthoryear{{Le{\'o}n-Tavares} et~al.,}{{Le{\'o}n-Tavares}
  et~al.}{2013}]{leon-tavares2013}
{Le{\'o}n-Tavares} J.,  et~al., 2013, \mn@doi [\apjl]
  {10.1088/2041-8205/763/2/L36}, \href
  {https://ui.adsabs.harvard.edu/abs/2013ApJ...763L..36L} {763, L36}

\bibitem[\protect\citeauthoryear{{Li}, {Zhang}, {Luo}, {Wang}  \& {Zhou}}{{Li}
  et~al.}{2015}]{li2015}
{Li} X.,  {Zhang} L.,  {Luo} Y.,  {Wang} L.,   {Zhou} L.,  2015, \mn@doi
  [\mnras] {10.1093/mnras/stv563}, \href
  {https://ui.adsabs.harvard.edu/abs/2015MNRAS.449.2750L} {449, 2750}

\bibitem[\protect\citeauthoryear{{Li}, {Luo}, {Yang}, {Yang}, {Yang}  \&
  {Cai}}{{Li} et~al.}{2018}]{li2018}
{Li} X.-P.,  {Luo} Y.-H.,  {Yang} H.-T.,  {Yang} H.-Y.,  {Yang} C.,   {Cai} Y.,
   2018, \mn@doi [Research in Astronomy and Astrophysics]
  {10.1088/1674-4527/18/12/150}, \href
  {https://ui.adsabs.harvard.edu/abs/2018RAA....18..150L} {18, 150}

\bibitem[\protect\citeauthoryear{{Li}, {Zhang}, {Jin}, {Du}, {Cui}, {Liu}  \&
  {Wang}}{{Li} et~al.}{2020}]{li2020}
{Li} Y.,  {Zhang} Z.,  {Jin} C.,  {Du} P.,  {Cui} L.,  {Liu} X.,   {Wang} J.,
  2020, \mn@doi [\apj] {10.3847/1538-4357/ab95a3}, \href
  {https://ui.adsabs.harvard.edu/abs/2020ApJ...897...18L} {897, 18}

\bibitem[\protect\citeauthoryear{{Mangalam} \& {Wiita}}{{Mangalam} \&
  {Wiita}}{1993}]{mangalam1993}
{Mangalam} A.~V.,  {Wiita} P.~J.,  1993, \mn@doi [\apj] {10.1086/172453}, \href
  {https://ui.adsabs.harvard.edu/abs/1993ApJ...406..420M} {406, 420}

\bibitem[\protect\citeauthoryear{{Maraschi}, {Ghisellini}  \&
  {Celotti}}{{Maraschi} et~al.}{1992}]{maraschi1992}
{Maraschi} L.,  {Ghisellini} G.,   {Celotti} A.,  1992, \mn@doi [\apjl]
  {10.1086/186531}, \href
  {https://ui.adsabs.harvard.edu/abs/1992ApJ...397L...5M} {397, L5}

\bibitem[\protect\citeauthoryear{{Marscher}}{{Marscher}}{2014}]{marscher2014}
{Marscher} A.~P.,  2014, \mn@doi [\apj] {10.1088/0004-637X/780/1/87}, \href
  {https://ui.adsabs.harvard.edu/abs/2014ApJ...780...87M} {780, 87}

\bibitem[\protect\citeauthoryear{{Marscher} \& {Gear}}{{Marscher} \&
  {Gear}}{1985}]{marscher1985}
{Marscher} A.~P.,  {Gear} W.~K.,  1985, \mn@doi [\apj] {10.1086/163592}, \href
  {https://ui.adsabs.harvard.edu/abs/1985ApJ...298..114M} {298, 114}

\bibitem[\protect\citeauthoryear{{Massaro}, {Maselli}, {Leto}, {Marchegiani},
  {Perri}, {Giommi}  \& {Piranomonte}}{{Massaro} et~al.}{2015}]{massaro2015}
{Massaro} E.,  {Maselli} A.,  {Leto} C.,  {Marchegiani} P.,  {Perri} M.,
  {Giommi} P.,   {Piranomonte} S.,  2015, \mn@doi [\apss]
  {10.1007/s10509-015-2254-2}, \href
  {https://ui.adsabs.harvard.edu/abs/2015Ap&SS.357...75M} {357, 75}

\bibitem[\protect\citeauthoryear{{Meng}, {Zhang}, {Wu}, {Ma}  \& {Zhou}}{{Meng}
  et~al.}{2018}]{meng2018}
{Meng} N.,  {Zhang} X.,  {Wu} J.,  {Ma} J.,   {Zhou} X.,  2018, \mn@doi [\apjs]
  {10.3847/1538-4365/aacffe}, \href
  {https://ui.adsabs.harvard.edu/abs/2018ApJS..237...30M} {237, 30}

\bibitem[\protect\citeauthoryear{{Otero-Santos} et~al.,}{{Otero-Santos}
  et~al.}{2020}]{otero2020}
{Otero-Santos} J.,  et~al., 2020, \mn@doi [\mnras] {10.1093/mnras/staa134},
  \href {https://ui.adsabs.harvard.edu/abs/2020MNRAS.492.5524O} {492, 5524}

\bibitem[\protect\citeauthoryear{Paatero \& Tapper}{Paatero \&
  Tapper}{1994}]{paatero1994}
Paatero P.,  Tapper U.,  1994, \mn@doi [Environmetrics]
  {10.1002/env.3170050203}, 5, 111

\bibitem[\protect\citeauthoryear{{Parker} et~al.,}{{Parker}
  et~al.}{2015}]{parker2015}
{Parker} M.~L.,  et~al., 2015, \mn@doi [\mnras] {10.1093/mnras/stu2424}, \href
  {https://ui.adsabs.harvard.edu/abs/2015MNRAS.447...72P} {447, 72}

\bibitem[\protect\citeauthoryear{{Pe{\~n}il} et~al.,}{{Pe{\~n}il}
  et~al.}{2020}]{penil2020}
{Pe{\~n}il} P.,  et~al., 2020, \mn@doi [\apj] {10.3847/1538-4357/ab910d}, \href
  {https://ui.adsabs.harvard.edu/abs/2020ApJ...896..134P} {896, 134}

\bibitem[\protect\citeauthoryear{{Pearson}}{{Pearson}}{1901}]{pearson1901}
{Pearson} K.,  1901, \mn@doi [The London, Edinburgh, and Dublin Philosophical
  Magazine and Journal of Science] {10.1080/14786440109462720}, 2, 559

\bibitem[\protect\citeauthoryear{{Plotkin}, {Anderson}, {Brandt}, {Markoff},
  {Shemmer}  \& {Wu}}{{Plotkin} et~al.}{2012}]{plotkin2012}
{Plotkin} R.~M.,  {Anderson} S.~F.,  {Brandt} W.~N.,  {Markoff} S.,  {Shemmer}
  O.,   {Wu} J.,  2012, \mn@doi [\apjl] {10.1088/2041-8205/745/2/L27}, \href
  {https://ui.adsabs.harvard.edu/abs/2012ApJ...745L..27P} {745, L27}

\bibitem[\protect\citeauthoryear{{Raiteri} et~al.,}{{Raiteri}
  et~al.}{2005}]{raiteri2005}
{Raiteri} C.~M.,  et~al., 2005, \mn@doi [\aap] {10.1051/0004-6361:20042567},
  \href {https://ui.adsabs.harvard.edu/abs/2005A&A...438...39R} {438, 39}

\bibitem[\protect\citeauthoryear{{Raiteri} et~al.,}{{Raiteri}
  et~al.}{2006}]{raiteri2006}
{Raiteri} C.~M.,  et~al., 2006, \mn@doi [\aap] {10.1051/0004-6361:20065744},
  \href {https://ui.adsabs.harvard.edu/abs/2006A&A...459..731R} {459, 731}

\bibitem[\protect\citeauthoryear{{Raiteri} et~al.,}{{Raiteri}
  et~al.}{2010}]{raiteri2010}
{Raiteri} C.~M.,  et~al., 2010, \mn@doi [\aap] {10.1051/0004-6361/201015191},
  \href {https://ui.adsabs.harvard.edu/abs/2010A&A...524A..43R} {524, A43}

\bibitem[\protect\citeauthoryear{{Raiteri} et~al.,}{{Raiteri}
  et~al.}{2014}]{raiteri2014}
{Raiteri} C.~M.,  et~al., 2014, \mn@doi [\mnras] {10.1093/mnras/stu886}, \href
  {https://ui.adsabs.harvard.edu/abs/2014MNRAS.442..629R} {442, 629}

\bibitem[\protect\citeauthoryear{{Raiteri} et~al.,}{{Raiteri}
  et~al.}{2017}]{raiteri2017}
{Raiteri} C.~M.,  et~al., 2017, \mn@doi [\nat] {10.1038/nature24623}, \href
  {https://ui.adsabs.harvard.edu/abs/2017Natur.552..374R} {552, 374}

\bibitem[\protect\citeauthoryear{{Raiteri} et~al.,}{{Raiteri}
  et~al.}{2019}]{raiteri2019}
{Raiteri} C.~M.,  et~al., 2019, \mn@doi [\mnras] {10.1093/mnras/stz2264}, \href
  {https://ui.adsabs.harvard.edu/abs/2019MNRAS.489.1837R} {489, 1837}

\bibitem[\protect\citeauthoryear{{Raiteri} et~al.,}{{Raiteri}
  et~al.}{2021a}]{raiteri2021a}
{Raiteri} C.~M.,  et~al., 2021a, \mn@doi [\mnras] {10.1093/mnras/staa3561},
  \href {https://ui.adsabs.harvard.edu/abs/2021MNRAS.501.1100R} {501, 1100}

\bibitem[\protect\citeauthoryear{{Raiteri} et~al.,}{{Raiteri}
  et~al.}{2021b}]{raiteri2021b}
{Raiteri} C.~M.,  et~al., 2021b, \mn@doi [\mnras] {10.1093/mnras/stab1268},
  \href {https://ui.adsabs.harvard.edu/abs/2021MNRAS.504.5629R} {504, 5629}

\bibitem[\protect\citeauthoryear{{Rani} et~al.,}{{Rani}
  et~al.}{2010}]{rani2010}
{Rani} B.,  et~al., 2010, \mn@doi [\mnras] {10.1111/j.1365-2966.2010.16419.x},
  \href {https://ui.adsabs.harvard.edu/abs/2010MNRAS.404.1992R} {404, 1992}

\bibitem[\protect\citeauthoryear{{Safna}, {Stalin}, {Rakshit}  \&
  {Mathew}}{{Safna} et~al.}{2020}]{safna2020}
{Safna} P.~Z.,  {Stalin} C.~S.,  {Rakshit} S.,   {Mathew} B.,  2020, \mn@doi
  [\mnras] {10.1093/mnras/staa2622}, \href
  {https://ui.adsabs.harvard.edu/abs/2020MNRAS.498.3578S} {498, 3578}

\bibitem[\protect\citeauthoryear{{Sandrinelli}, {Covino}  \&
  {Treves}}{{Sandrinelli} et~al.}{2014}]{sandrinelli2014}
{Sandrinelli} A.,  {Covino} S.,   {Treves} A.,  2014, \mn@doi [\aap]
  {10.1051/0004-6361/201321558}, \href
  {https://ui.adsabs.harvard.edu/abs/2014A&A...562A..79S} {562, A79}

\bibitem[\protect\citeauthoryear{{Sandrinelli}, {Covino}, {Treves}, {Holgado},
  {Sesana}, {Lindfors}  \& {Ramazani}}{{Sandrinelli}
  et~al.}{2018}]{sandrinelli2018}
{Sandrinelli} A.,  {Covino} S.,  {Treves} A.,  {Holgado} A.~M.,  {Sesana} A.,
  {Lindfors} E.,   {Ramazani} V.~F.,  2018, \mn@doi [\aap]
  {10.1051/0004-6361/201732550}, \href
  {https://ui.adsabs.harvard.edu/abs/2018A&A...615A.118S} {615, A118}

\bibitem[\protect\citeauthoryear{{Schlafly} \& {Finkbeiner}}{{Schlafly} \&
  {Finkbeiner}}{2011}]{schlafly2011}
{Schlafly} E.~F.,  {Finkbeiner} D.~P.,  2011, \mn@doi [\apj]
  {10.1088/0004-637X/737/2/103}, \href
  {https://ui.adsabs.harvard.edu/abs/2011ApJ...737..103S} {737, 103}

\bibitem[\protect\citeauthoryear{{Shang} et~al.,}{{Shang}
  et~al.}{2005}]{shang2005}
{Shang} Z.,  et~al., 2005, \mn@doi [\apj] {10.1086/426134}, \href
  {https://ui.adsabs.harvard.edu/abs/2005ApJ...619...41S} {619, 41}

\bibitem[\protect\citeauthoryear{{Sillanpaa}, {Haarala}, {Valtonen},
  {Sundelius}  \& {Byrd}}{{Sillanpaa} et~al.}{1988}]{sillanpaa1988}
{Sillanpaa} A.,  {Haarala} S.,  {Valtonen} M.~J.,  {Sundelius} B.,   {Byrd}
  G.~G.,  1988, \mn@doi [\apj] {10.1086/166033}, \href
  {https://ui.adsabs.harvard.edu/abs/1988ApJ...325..628S} {325, 628}

\bibitem[\protect\citeauthoryear{{Sillanpaa} et~al.,}{{Sillanpaa}
  et~al.}{1996}]{sillanpaa1996}
{Sillanpaa} A.,  et~al., 1996, \aap, \href
  {https://ui.adsabs.harvard.edu/abs/1996A&A...315L..13S} {315, L13}

\bibitem[\protect\citeauthoryear{{Smith}, {Montiel}, {Rightley}, {Turner},
  {Schmidt}  \& {Jannuzi}}{{Smith} et~al.}{2009}]{smith2009}
{Smith} P.~S.,  {Montiel} E.,  {Rightley} S.,  {Turner} J.,  {Schmidt} G.~D.,
  {Jannuzi} B.~T.,  2009, arXiv e-prints, \href
  {https://ui.adsabs.harvard.edu/abs/2009arXiv0912.3621S} {p. arXiv:0912.3621}

\bibitem[\protect\citeauthoryear{{Stickel}, {Padovani}, {Urry}, {Fried}  \&
  {Kuehr}}{{Stickel} et~al.}{1991}]{stickel1991}
{Stickel} M.,  {Padovani} P.,  {Urry} C.~M.,  {Fried} J.~W.,   {Kuehr} H.,
  1991, \mn@doi [\apj] {10.1086/170133}, \href
  {https://ui.adsabs.harvard.edu/abs/1991ApJ...374..431S} {374, 431}

\bibitem[\protect\citeauthoryear{{Stocke}, {Morris}, {Gioia}, {Maccacaro},
  {Schild}, {Wolter}, {Fleming}  \& {Henry}}{{Stocke}
  et~al.}{1991}]{stocke1991}
{Stocke} J.~T.,  {Morris} S.~L.,  {Gioia} I.~M.,  {Maccacaro} T.,  {Schild} R.,
   {Wolter} A.,  {Fleming} T.~A.,   {Henry} J.~P.,  1991, \mn@doi [\apjs]
  {10.1086/191582}, \href
  {https://ui.adsabs.harvard.edu/abs/1991ApJS...76..813S} {76, 813}

\bibitem[\protect\citeauthoryear{{Torres-Zafra}, {Cellone}, {Buzzoni},
  {Andruchow}  \& {Portilla}}{{Torres-Zafra} et~al.}{2018}]{torres-zafra2018}
{Torres-Zafra} J.,  {Cellone} S.~A.,  {Buzzoni} A.,  {Andruchow} I.,
  {Portilla} J.~G.,  2018, \mn@doi [\mnras] {10.1093/mnras/stx2561}, \href
  {https://ui.adsabs.harvard.edu/abs/2018MNRAS.474.3162T} {474, 3162}

\bibitem[\protect\citeauthoryear{{Urry} \& {Padovani}}{{Urry} \&
  {Padovani}}{1995}]{urry1995}
{Urry} C.~M.,  {Padovani} P.,  1995, \mn@doi [\pasp] {10.1086/133630}, \href
  {https://ui.adsabs.harvard.edu/abs/1995PASP..107..803U} {107, 803}

\bibitem[\protect\citeauthoryear{{Valtonen} et~al.,}{{Valtonen}
  et~al.}{2016}]{valtonen2016}
{Valtonen} M.~J.,  et~al., 2016, \mn@doi [\apjl] {10.3847/2041-8205/819/2/L37},
  \href {https://ui.adsabs.harvard.edu/abs/2016ApJ...819L..37V} {819, L37}

\bibitem[\protect\citeauthoryear{{Vanden Berk} et~al.,}{{Vanden Berk}
  et~al.}{2001}]{vandenberk2001}
{Vanden Berk} D.~E.,  et~al., 2001, \mn@doi [\aj] {10.1086/321167}, \href
  {https://ui.adsabs.harvard.edu/abs/2001AJ....122..549V} {122, 549}

\bibitem[\protect\citeauthoryear{{Vaughan}, {Edelson}, {Warwick}  \&
  {Uttley}}{{Vaughan} et~al.}{2003}]{vaughan2003}
{Vaughan} S.,  {Edelson} R.,  {Warwick} R.~S.,   {Uttley} P.,  2003, \mn@doi
  [\mnras] {10.1046/j.1365-2966.2003.07042.x}, \href
  {https://ui.adsabs.harvard.edu/abs/2003MNRAS.345.1271V} {345, 1271}

\bibitem[\protect\citeauthoryear{{Villata}, {Raiteri}, {Sillanpaa}  \&
  {Takalo}}{{Villata} et~al.}{1998}]{villata1998}
{Villata} M.,  {Raiteri} C.~M.,  {Sillanpaa} A.,   {Takalo} L.~O.,  1998,
  \mn@doi [\mnras] {10.1046/j.1365-8711.1998.01244.x}, \href
  {https://ui.adsabs.harvard.edu/abs/1998MNRAS.293L..13V} {293, L13}

\bibitem[\protect\citeauthoryear{{Villata} et~al.,}{{Villata}
  et~al.}{2006}]{villata2006}
{Villata} M.,  et~al., 2006, \mn@doi [\aap] {10.1051/0004-6361:20064817}, \href
  {https://ui.adsabs.harvard.edu/abs/2006A&A...453..817V} {453, 817}

\bibitem[\protect\citeauthoryear{{Vovk} \& {Babi{\'c}}}{{Vovk} \&
  {Babi{\'c}}}{2015}]{vovk2015}
{Vovk} I.,  {Babi{\'c}} A.,  2015, \mn@doi [\aap]
  {10.1051/0004-6361/201526004}, \href
  {https://ui.adsabs.harvard.edu/abs/2015A&A...578A..92V} {578, A92}

\bibitem[\protect\citeauthoryear{Wang, Liu, Petropoulou, Oikonomou, Xue  \&
  Wang}{Wang et~al.}{2021}]{wang2021}
Wang Z.,  Liu R.,  Petropoulou M.,  Oikonomou F.,  Xue R.,   Wang X.,  2021,
  \mn@doi [PoS] {10.22323/1.395.0984}, ICRC2021, 984

\bibitem[\protect\citeauthoryear{{Xiong}, {Zhang}, {Zhang}, {Yi}, {Bai},
  {Wang}, {Liu}  \& {Zheng}}{{Xiong} et~al.}{2016}]{xiong2016}
{Xiong} D.,  {Zhang} H.,  {Zhang} X.,  {Yi} T.,  {Bai} J.,  {Wang} F.,  {Liu}
  H.,   {Zheng} Y.,  2016, \mn@doi [\apjs] {10.3847/0067-0049/222/2/24}, \href
  {https://ui.adsabs.harvard.edu/abs/2016ApJS..222...24X} {222, 24}

\bibitem[\protect\citeauthoryear{{Zhang}, {Zhou}, {Zhao}  \& {Dai}}{{Zhang}
  et~al.}{2015}]{zhang2015}
{Zhang} B.-K.,  {Zhou} X.-S.,  {Zhao} X.-Y.,   {Dai} B.-Z.,  2015, \mn@doi
  [Research in Astronomy and Astrophysics] {10.1088/1674-4527/15/11/002}, \href
  {https://ui.adsabs.harvard.edu/abs/2015RAA....15.1784Z} {15, 1784}

\bibitem[\protect\citeauthoryear{{Zhang}, {Yan}, {Zhou}, {Wang}  \&
  {Zhang}}{{Zhang} et~al.}{2020}]{zhang2020}
{Zhang} P.,  {Yan} D.,  {Zhou} J.,  {Wang} J.,   {Zhang} L.,  2020, \mn@doi
  [\apj] {10.3847/1538-4357/ab71fe}, \href
  {https://ui.adsabs.harvard.edu/abs/2020ApJ...891..163Z} {891, 163}

\bibitem[\protect\citeauthoryear{{Zhu}}{{Zhu}}{2016}]{zhu2016}
{Zhu} G.,  2016, arXiv e-prints, \href
  {https://ui.adsabs.harvard.edu/abs/2016arXiv161206037Z} {p. arXiv:1612.06037}

\makeatother
\end{thebibliography}




\section*{Supplementary material}
\noindent\textbf{Figure A1\labtext{A1}{h1426_figures}}. Results of H 1426+428 (same description as Fig. \ref{mrk501_figures}, available in the online version).

\noindent\textbf{Figure A2\labtext{A2}{1es2344_figures}}. Results of 1ES 2344+514 (same description as Fig. \ref{mrk501_figures}, available in the online version).

\noindent\textbf{Figure A3\labtext{A3}{pks0735_figures}}. Results of PKS 0735+178 (same description as Fig. \ref{mrk501_figures}, available in the online version).

\noindent\textbf{Figure A4\labtext{A4}{wcom_figures}}. Results of W Comae (same description as Fig. \ref{mrk501_figures}, available in the online version).

\noindent\textbf{Figure A5\labtext{A5}{h1219_figures}}. Results of H 1219+305 (same description as Fig. \ref{mrk501_figures}, available in the online version).

\noindent\textbf{Figure A6\labtext{A6}{1es1959_figures}}. Results of 1ES 1959+650 (same description as Fig. \ref{mrk501_figures}, available in the online version).

\noindent\textbf{Figure A7\labtext{A7}{pks2155_figures}}. Results of PKS 2155-304 (same description as Fig. \ref{mrk501_figures}, available in the online version).

\noindent\textbf{Figure A8\labtext{A8}{bllac_figures}}. Results of BL Lacertae (same description as Fig. \ref{mrk501_figures}, available in the online version).

\noindent\textbf{Figure A9\labtext{A9}{pks0420_figures}}. Results of PKS 0420-014 (same description as Fig. \ref{mrk501_figures}, available in the online version).

\noindent\textbf{Figure A10\labtext{A10}{pks0736_figures}}. Results of PKS 0736+017 (same description as Fig. \ref{mrk501_figures}, available in the online version).

\noindent\textbf{Figure A11\labtext{A11}{oj248_figures}}. Results of OJ 248 (same description as Fig. \ref{mrk501_figures}, available in the online version).

\noindent\textbf{Figure A12\labtext{A12}{ton599_figures}}. Results of Ton 599 (same description as Fig. \ref{mrk501_figures}, available in the online version).

\noindent\textbf{Figure A13\labtext{A13}{pks1510_figures}}. Results of PKS 1510-089 (same description as Fig. \ref{mrk501_figures}, available in the online version).

\noindent\textbf{Figure A14\labtext{A14}{b2_1633_figures}}. Results of B2 1633+38 (same description as Fig. \ref{mrk501_figures}, available in the online version).

\noindent\textbf{Figure A15\labtext{A15}{3c345_figures}}. Results of 3C 345 (same description as Fig. \ref{mrk501_figures}, available in the online version).

\appendix

\section{Remarks on individual sources}\label{appendix}
Here we describe the results of the NMF analyses for the rest of blazars not presented in section \ref{sec5}. The figures are available in the online version of the paper.

\subsection{Galaxy-dominated sources}\label{appendix_galdom}
\subsubsection{H 1426+428}
Similarly to 1ES 2344+514 and Mkn~501, this blazar emission displays a bright stellar emission, that contributes with approximately 30\% of the total optical emission detected and shows little variability ($F_{var}\simeq9\%$). The results of this reconstruction are displayed in Fig. \ref{h1426_figures}. The second component, associated to the jet, accounts for the rest of the emission and the low variability observed on this blazar. The spectral shape on H~1426+428 remains roughly constant during the monitoring due to its low variability. Thus, the colour of this source does not show large variations or correlation with the flux.

\subsubsection{1ES 2344+514}
The optical spectral variability of this blazar is very low ($F_{var}\simeq4\%$). 1ES 2344+514 has a bright host galaxy that contributes with 40-60\% of the optical emission of this source. The rest of the emission is associated to the jet, represented by the second component of the reconstruction. Fig. \ref{1es2344_figures} displays the results of the reconstruction of this blazar. Despite its very small variability, it shows a mild overall BWB behaviour that comes from the component C2, associated to the jet, which has a not very steep slope. The colour trend is explained by the variable contribution of this component relative to the stellar population one.

\subsection{BL Lac objects}\label{appendix_bllacs}
\subsubsection{PKS 0735+178}
The mean flux of the spectra shows very little variation ($F_{var}=7\%$). The source is well reconstructed by 2 PL-like components C1 and C2, with spectral indices 0.63 and 1.45, respectively. Fig. \ref{pks0735_figures} shows the results of the reconstruction of this source. The first component is always the brightest one. Its contribution to the mean flux is always above 80\%, and its mean flux remains nearly constant over time. It also displays a constant behaviour with respect to the colour. The largest flux contribution of the component C2 is about 20\%. Despite its low contribution, it accounts for more than half of the $\sigma^2_{exp}$. Moreover, the second PL shows a BWB behaviour and it explains the observed overall BWB trend, despite the low number of spectra. This BWB trend has also been observed in the past for this object \citep{meng2018}. 

\subsubsection{W Comae}
The variability of the BL Lac object W~Comae is reproduced using 3 components associated to the synchrotron emission of the relativistic jet. These components are able to model the 99.86\% of the variability of this blazar. The results of this blazar can be seen in Fig. \ref{wcom_figures}. The component C1 is the dominant when the mean flux is low. Its contribution to the $\sigma^2_{exp}$ is around 20\%. The component C2 is the steepest one and its contribution to the $\sigma^2_{exp}$ is above 70\%. It is the dominant component when the source is at high state. C3 is showing an almost flat slope ($\alpha=0.18$). It remains rather constant according to its contribution to the $\sigma^2_{exp}$, around 10\%. It shows a large contribution after MJD 56750, generally associated to low states, and it coincides with a reddening of the spectra. As for almost all the objects of its type, a long-term BWB evolution is observed for this object and it can be attributed to the variations of the component C2.

\subsubsection{H 1219+305}
Overall the spectra of H~1219+305 displays a relatively small variability ($F_{var}\sim 15\,\%$). The flux of this blazar remains approximately stable during the whole observing period, although the colour F$_{\text{B}}$/F$_{\text{R}}$ varies considerably in two epochs: from bluer in the period MJD 54800--54950 to redder in the period MJD 55200--55350. This BL Lac is reconstructed with 3 PL components, whose indices range from 0.72 to 1.26. The results of the NMF analysis are displayed in Fig. \ref{h1219_figures}. The components C1 ($\alpha=0.91$) and C2 ($\alpha=1.26$) contribute almost equally to the mean flux during the first epoch when the spectra are bluer. The component C3 ($\alpha=0.72$) is dominant during the second epoch, when the spectra become redder. Since the average flux of the source remains nearly constant, no (anti)correlated trend with the colour is observed.

\subsubsection{1ES 1959+650}
The reconstruction of the spectra of this source is achieved with 2 components (see Fig. \ref{1es1959_figures}): the component C1 is rather stable ($\sim 15\%\,\sigma^2_{exp}$) and the component C2 is the one responsible of most of the variability. C1 resembles a rather flat PL, which also contains very faint absorption (Mg\textit{b}) and emission ([OIII]5007~\AA) features. Its contribution to the spectrum flux is almost unique when the target is at low state and moderate ($\sim 50\%$) at intermediate states. C2 is a blue PL that becomes very dominant when the source is on flare states. This object exhibits the typical BWB trend observed in BL Lacs. This behaviour was previously observed for this blazar \citep[see for instance][]{meng2018}. 

\subsubsection{PKS 2155-304}
A total of 4 PL components were used to model the variability of the BL Lac object PKS~2155-304 (see Fig. \ref{pks2155_figures}). This reconstruction accounts for the 99.85\% of the variance of the original spectra.
The component C1 contributes to the mean flux with  30-50\% of the emission for about 3/4 of the spectra, accounting for about 20\% of the $\sigma^2_{exp}$. The component C2 has its largest contribution associated to short flares, as shown by its higher contribution (40\%) to the $\sigma^2_{exp}$. C3 shows a contribution $\gtrsim 60\,\%$ to the emission during the period MJD 56400--56600, being the main component when the colour is bluest. The component C4 is the less steep one. It contributes appreciably for short periods, being the main component when the colour is the reddest. A BWB trend is observed in the long-term variability of this blazar. 

\subsubsection{BL Lacertae}
BL Lacertae was reconstructed using 4 PL components due to the high spectral variability displayed. The reconstruction accounts for the 99.71\% of the variance of the data set, and it is shown in Fig. \ref{bllac_figures}. The PL indices range from 0.05 to 0.7. The component C1 shows a relatively flat slope ($\alpha=0.27$). Its spectrum also exhibits weak emission lines such as the HI recombination lines and [OIII]5007~\AA. It has a moderate contribution, $\lesssim 50\%$ to the total flux during the lower states, and accounts for $\sim 15\%$ of the $\sigma^2_{exp}.$  The components C2 and C4 are the most variable ones, being responsible for $\sim 40\%$ and $\sim 35\%$ of the $\sigma^2_{exp}$, respectively. Their slopes are very similar and may represent the evolution of the same physical component. The component C3 is almost flat and remains basically constant  ($10\%\,\sigma^2_{exp}$). Its contribution to the total flux is as large as 60\%, not coincident with the high states. The overall evolution of this object follows the traditional BWB behaviour observed in BL Lacs, explained by the variable contribution of the blue components C2 and C4 ($\alpha \sim 0.7$), and expected from this dominant synchrotron emission.

\subsection{FSRQs}\label{appendix_fsrqs}
\subsubsection{PKS 0420-014}
The variability of PKS~0420-014 is accurately reproduced with 3 different components. The results are displayed in Fig. \ref{pks0420_figures}. The percentage of variance explained by the reconstruction of this target is 99.82\%. Despite its low time coverage, PKS~0420-014 is one of the few FSRQ showing the different trends reported by \cite{isler2017}. During its low flux state, it displays a very clear RWB evolution. At a flux of $\sim$10$^{-15}$ erg cm$^{-2}$ s$^{-1}$ \AA$^{-1}$, this trend flattens, and the source starts becoming bluer as it brightens. The reddening evolution with increasing flux was reported in the past by \cite{rani2010}. However, the authors do not claim for the BWB trend that is also observed in this work.

The component C1 of the reconstruction (showing a prominent MgII line) is associated to the BLR. As expected, the variability associated to this component is very low (less than 1\%). Its contribution to the total flux is about 20\% most of the time, except for the period MJD 55300-55650. The components C2 and C3 account for the non-thermal synchrotron emission emerging from the jet. The component C2 is bluer, with a slope around 1.0 and it accounts for $\sim 85\%$ of the $\sigma^2_{exp}$, becoming dominant in flux during high states. The component C3 is flatter and accounts for less than $15\% \sigma^2_{exp}$, but it contributes about 80\% of the flux when the source is at intermediate states. Each one is responsible of one of the two colour trends observed in PKS~0420-014. C2 accounts for the BWB trends during the flaring states. Furthermore, the  component C3  explains the RWB trend.

\subsubsection{PKS 0736+017}
PKS 0736+017 is reconstructed with 4 components that explain 99.72\% of the variance of the data set. The results derived for this FSRQ are shown in Fig. \ref{pks0736_figures}. The first component, associated to the BLR, dominates up to a flux of $\sim$10$^{-15}$ erg cm$^{-2}$ s$^{-1}$ \AA$^{-1}$. For brighter spectra, the jet components overcome the BLR emission. The component C1 contributes almost 80\% in most of the spectra, although accounts only for a very small fraction of $\sigma^2_{exp}$. The component C2 is the one accounting for most of the \explvar, and its contribution to the mean flux reaches 70\% in coincidence with bright flares. The component C3 has the less steep slope, appearing during short periods associated with less intense flares. The component C4 has a very steep slope and appears only in about 10\% of the spectra, generally when the source is at low state. It accounts for a tiny fraction of \explvar, indicating a rather small variability. This low variability and its steep PL are compatible with being the optical extension of the UV disk emission (similarly to the cases of 3C~273, PKS~1222+216 or B2~1633+38). Overall, the low state of the source does not reveal any clear colour trend. However, for the brightest spectra, a hint of the colour saturation already seen in other FSRQs is observed. 

\subsubsection{OJ 248}
The reconstruction of OJ~248 was performed with a total of 3 components. This reconstruction is shown in Fig. \ref{oj248_figures}. The BLR component observed in this FSRQ is always outshined by the PL components associated to the relativistic jet, and contributes with $\sim$25\% of the optical emission, except during the enhanced state observed at MJD 56300. Despite its low contribution to \explvar, the mean flux of component C1 shows a slow monotonic increase of about 20\% in around 10 years. The component C2 ($\alpha = 1.33$) is dominant in about half of the spectra, especially when the colour of the source is the bluest. The component C3 ($\alpha=1.06$) contributes about 80\% to the mean flux, and its maximum coincides when the colour is the reddest, just before the big flare at MJD 56250. The low emission state of OJ~248 does not reveal any clear spectral evolution, with no trend between the colour and the flux, and large colour variation when the flux remains roughly constant, which coincides with an increase of the relative contribution of C1 at MJD $\sim$57800. A RWB trend is observed for OJ 248 during the low states between MJD 54700 and MJD 56250, approximately. Moreover, a FWB brightening is detected for the outburst observed during the flaring event at MJD 56300. The rising of the PL components is responsible for the FWB. Additionally, a very mild BWB increase is observed during the increase experienced at MJD 57750.

\subsubsection{Ton 599}
The spectra of this blazar show a broad line emission that is typically weaker than the emission of the jet. The reconstruction of the spectra was done using 4 components (see Fig. \ref{ton599_figures}). The component C1 corresponds to the BLR emission. It shows a strong emission line corresponding to MgII~2800 \AA, plus a fainter [OII]3727~\AA. This component contributes at most by $\sim 50\%$ to the mean flux during low states and remains basically constant. The component C2 contributes up to 50\% when the source is at high states and carries out most of the \explvar, around 75\%. The components C3 and C4 show the most and the less steep slopes ($\alpha=1.7, 0.9$), respectively. They contribute more within the periods when the colour becomes bluer (component C3) and redder (component C4), in both cases with contributions above 25\% for $\sim 30\% $ of the whole spectrum set. As it happened to other FSRQs, for the brightest spectra, a colour saturation, followed by an achromatic brightening, can be seen for this blazar. 

\subsubsection{PKS 1510-089}
The NMF reconstruction of PKS~1510-089 is obtained using 3 components. The results of the NMF analysis performed on this object can be seen in Fig. \ref{pks1510_figures}. PKS 1510-089 displays a bright BLR represented by the first component of the reconstruction. This is the dominant contribution for about 80\% of the spectra, except for three short outburst events at MJD 55000, MJD 57150 and MJD 57600, respectively. As expected, the BLR contribution is approximately constant, and no trends or correlations are seen with the colour. The second component is a simple PL accounting for the emission of the jet and 91\% of the \explvar. It is usually less bright than the BLR, except for the outbursts already mentioned. The contribution of this component varies correlated with the colour, displaying a RWB trend. This result is consistent with the one presented by \cite{d'ammando2009}, and follows the general behaviour of FSRQs. Overall, this blazar displays the classical RWB trend observed in FSRQs, and the colour saturation already discussed in other sources. Moreover, the component C3 shows a blue contribution, less bright than the BLR and the PL from the jet. It appears at low states of the source, shows little variation and has a steep slope. These characteristics are compatible with the disk emission, which has been already suggested by \cite{raiteri2014}. 

\subsubsection{B2 1633+38}
The spectra of this source have been reconstructed using 3 components (see Fig. \ref{b2_1633_figures}): the first one represents the disk and BLR emission (showing the C[IV] and C[III] features) that dominates up to a flux of $\sim$0.6 $\times$ 10$^{-15}$ erg cm$^{-2}$ s$^{-1}$ \AA$^{-1}$. The second and third ones are PLs, with indices $\alpha=1.2$ and $\alpha=2.3$, respectively. The main contribution to the mean flux comes from the component C1, except when flares appear in which component C2 takes over, or component C3 in the period between MJD 56250 and 56400. The mean flux of C1 varies smoothly in scales of hundreds of days. Moreover, most of the variability (92\% of \explvar) is due to the component C2. As it happened for 3C~273, the steep and less variable ($\sigma^{2}_{exp} \simeq 5\%$) component C3 may be associated with an enhanced black body explaining the contribution of the accretion disk. This would be in agreement with the need for a bright accretion disk in the optical-IR SED modeling claimed by \cite{raiteri2014}. A loose RWB trend is observed for most of the spectra, except for the high flux regime, in which a flattening in the colour is seen (see Fig \ref{variability_fsrqs}). This flattening coincides with the period when the component C2 is highest. A flattening in the colour-flux trend is also observed for the most intense spectra, sign of the colour saturation. 

\subsubsection{3C 345}
The spectra of 3C 345 can be reconstructed using 3 components. The results can be seen in Fig. \ref{3c345_figures}. 3C 345 has an almost constant contribution $\sigma^{2}_{exp} \simeq 10\%$ from the BLR that outshines the other two components used to explain the variability in this source for almost all the data set. This component has a constant contribution to the emission. It shows a relative contribution to the mean flux above 50\% for about 4/5 of the spectra. The mean flux associated to this component shows a clear slow decline trend, which may indicate a disk emission real variation. It does not show variations associated to changes in the colour of the source. The second and third components correspond to PL functions. The component C2 has a steep index ($\alpha=1.72$) and accounts for about 55\% of \explvar. It contributes by about 60\% of the mean flux in the period around MJD 55100. The component C3 has a less steep index and accounts for about 30\% of \explvar. It contributes with 30-40\% of the mean flux during the periods around MJD 55150 and 55300. The residual excess observed around the MgII emission line might indicate a variation of the line. A clear RWB colour evolution is show by this object, which can be explained by the increase of C2 and C3 relative to C1, which is slightly bluer. Apart from this trend, a flattening of the RWB trend is observed, leading to an almost achromatic variability when the source is in a high emission state.


\bsp	
\label{lastpage}
\end{document}